\documentclass[12pt]{article}
\usepackage{epsf}
\usepackage{amsmath}
\usepackage{amsfonts}
\usepackage{latexsym}
\usepackage{graphicx}
\usepackage{color}
\usepackage{xy}
\usepackage{cancel,slashed}
\usepackage[linktocpage]{hyperref}

\usepackage{ifpdf}

\newcommand{\bmat}{\left(\begin{array}}
\newcommand{\emat}{\end{array}\right)}

\def\yzero{\smash{\hbox{$y\kern-4pt\raise1pt\hbox{${}^\circ$}$}}}
\def\p{\partial}
\def\a{\alpha}
\def\b{\beta}
\def\g{\gamma}
\def\d{\delta}
\def\sig{\sigma}
\def\beq{\begin{equation}}
\def\eeq{\end{equation}}
\def\beqa{\begin{eqnarray}}
\def\eeqa{\end{eqnarray}}
\def\Om{\Omega}
\def\om{\omega}

\def\vphi{\varphi}
\def\-{\hphantom{-}}
\def\ov{\overline}
\def\s2{\frac{1}{\sqrt2}}

\def\oh{\frac{1}{2}}
\def\tr{{\rm tr \,}}

\def\IF{\relax{\rm I\kern-.18em F}}
\def\II{\relax{\rm I\kern-.18em I}}

\def\cc{{\cal C}}

\def\cn{{\cal N}}
\def\cl{{\cal L}}
\def\cam{{\cal M}}

\def\cv{{\cal V}}

\def\ch{{\cal H}}
\def\cf{{\cal F}}

\def\cb{{\cal B}}

\def\Dsl{\,\raise.15ex\hbox{/}\mkern-13.5mu D} 

\def\IC{\mathbb{C}}

\def\IR{\mathbb{R}}
\def\IZ{\mathbb{Z}}
\def\IN{\mathbb{N}}
\def\Id{{\mathbb{I}}}

\def\vphi{\varphi}
\def\eps{\epsilon}

\def\im{{\rm Im}\,}
\def\re{{\rm Re}\,}

\def\R{{\cal R}}

\def\lam{\lambda}
\def\raw{\rightarrow}
\def\Raw{\Rightarrow}
\def\G{\Gamma}

\newcommand{\ul}{\underline}
\def\bes{\begin{subequations}}
\def\ees{\end{subequations}}

\catcode`\@=11
\newdimen\@rotdimen
\newbox\@rotbox

\def\@vspec#1{\special{ps:#1}}
\def\@rotstart#1{\@vspec{gsave currentpoint currentpoint translate
   #1 neg exch neg exch translate}}
\def\@rotfinish{\@vspec{currentpoint grestore moveto}}
\def\@rotr#1{\@rotdimen=\ht#1\advance\@rotdimen by\dp#1%
   \hbox to\@rotdimen{\hskip\ht#1\vbox to\wd#1{\@rotstart{90 rotate}%
   \box#1\vss}\hss}\@rotfinish}
\def\@rotl#1{\@rotdimen=\ht#1\advance\@rotdimen by\dp#1%
   \hbox to\@rotdimen{\vbox to\wd#1{\vskip\wd#1\@rotstart{270 rotate}%
   \box#1\vss}\hss}\@rotfinish}%
\def\@rotu#1{\@rotdimen=\ht#1\advance\@rotdimen by\dp#1%
   \hbox to\wd#1{\hskip\wd#1\vbox to\@rotdimen{\vskip\@rotdimen
   \@rotstart{-1 dup scale}\box#1\vss}\hss}\@rotfinish}%
\def\@rotf#1{\hbox to\wd#1{\hskip\wd#1\@rotstart{-1 1 scale}%
   \box#1\hss}\@rotfinish}%
\def\rotate{\@ifnextchar[{\@rotate}{\@rotate[l]}}
\def\@rotate[#1]#2{\setbox\@rotbox=\hbox{#2}\@nameuse{@rot#1}\@rotbox}

\catcode`\@=12

\topmargin
-1.5cm
\textwidth
15.5cm
\textheight
23.5cm
\oddsidemargin
0.7cm
\evensidemargin
1.2cm

\begin{document}

\makeatletter
\@addtoreset{equation}{section}
\makeatother
\renewcommand{\theequation}{\thesection.\arabic{equation}}
\pagestyle{empty}
\rightline{ CPHT-RR0470605} \rightline{ CERN-PH-TH/2009-074}
\vspace{2cm}
\begin{center}
\LARGE{\bf Open string wavefunctions \\
 in flux compactifications  \\[12mm]}
\large{Pablo G. C\'amara$^{1}$ and Fernando Marchesano$^{2}$
\\[3mm]}

\bigskip

\footnotesize{

{}$^1$Centre de Physique Th\'eorique, UMR du CNRS 7644,\\
Ecole Polytechnique,
91128 Palaiseau, France\\

{}$^2$ PH-TH Division, CERN
CH-1211 Geneva 23, Switzerland}

\bigskip

\bigskip

\bigskip

\small{\bf Abstract} \\[5mm]
\end{center}
\begin{center}
\begin{minipage}[h]{16.0cm}

We consider compactifications of type I supergravity on manifolds
with $SU(3)$ structure, in the presence of RR fluxes and
magnetized D9-branes, and analyze the generalized Dirac and
Laplace-Beltrami operators associated to the D9-brane worldvolume fields.
These compactifications are T-dual to standard type IIB toroidal orientifolds
with NSNS and RR 3-form fluxes and D3/D7 branes. By using techniques of
representation theory and harmonic analysis, the spectrum of open string
wavefunctions can be computed for Lie groups and their quotients, as we
illustrate with explicit twisted tori examples. We find a correspondence between
irreducible unitary representations of the Kaloper-Myers  algebra
and families of Kaluza-Klein excitations. We perform the computation of 2-
and 3-point couplings for matter fields in the above flux compactifications,
and compare our results with those of 4d effective supergravity.

\end{minipage}
\end{center}
\newpage
\setcounter{page}{1}
\pagestyle{plain}
\renewcommand{\thefootnote}{\arabic{footnote}}
\setcounter{footnote}{0}

\vspace*{1cm}

\tableofcontents

\section{Introduction}

Realizing that background fluxes have a non-trivial effect on the spectrum of
a string compactification has been an important step towards constructing
realistic 4d string vacua. This is particularly manifest in those vacua that
admit a 10d supergravity description, where compactifications with fluxes
\cite{review1,review2,review3} have been shown to provide a powerful framework
to address moduli stabilization and supersymmetry breaking. Indeed, in
the regime of weak fluxes and constant warp factor, the effect of
fluxes on the light string modes can be summarized by adding a superpotential
to the 4d effective theory that arises in the fluxless limit \cite{gvw}. This
superpotential has then the effect of lifting a non-trivial set of moduli and producing
$\cn=0$ vacua at tree level \cite{sethi,gkp}.

While the above observation has mainly been exploited for the gravity sector of
the theory, it is easy to see that it also applies to the gauge sector. In particular,
in the context of type II compactifications with D-branes, it has
been shown that fluxes induce supersymmetric and soft term masses on the light
open string degrees of freedom of the theory. This can be seen both from a
microscopic \cite{ciu03,ggjl03,ciu04} and from a 4d effective field theory viewpoint
\cite{lrs04,geosoft}. In fact, in this particular case it turns out that the 4d effective sugra
approach is somehow more complete that the higher dimensional results, since it
allows to compute soft term masses for certain open strings modes that the analysis
in terms of D-brane actions has yet not been able to deal with. These modes
are nothing but open strings with twisted boundary conditions, and more precisely
those arising between two stacks of intersecting and/or magnetized D-branes.
Generically, these open string modes are the ones giving rise to the chiral content
of the 4d effective theory \cite{review3,reviews}. Hence, analyzing these modes is
crucial to describe the effect of fluxes on the visible sector of a realistic string
compactification.

Here we would like to improve the current situation by considering a string theory limit
where the coupling between open string modes and open and closed background fluxes
is well-defined. More precisely, we consider type I supergravity compactifications in the
presence of gauge bundles, torsion and non-trivial RR 3-form fluxes. Due to the closed
string fluxes and the torsion, the internal manifold is not Calabi-Yau, but possesses an
$SU(3)$-structure.
One can then analyze the effect of the closed string background fluxes on
open strings by directly looking at how their presence modifies the 10d equations
of motion for the fluctuations of the gauge sector of the theory. Such modification will
affect the spectrum of open string modes, which in this approach are described as
eigenfunctions of the flux-modified Laplace and Dirac operators. These new open
string wavefunctions, together with the new couplings induced by the background
fluxes, will dictate the effect of fluxes on the 4d effective action upon dimensional
reduction of the 10d supergravity background.

Note that this approach of computing explicit wavefunctions and using them in
the dimensional reduction is essentially the one used in \cite{yukawa}
to compute Yukawa couplings in toroidal models with magnetized D9-branes
(see also \cite{quevedo,ako08,diveccia,akp09}). In this sense, this work can be seen
as an extension of \cite{yukawa} to compactifications with non-vanishing closed
string fluxes. Moreover, here we will analyze the full spectrum of Kaluza-Klein modes,
which in fact can also be seen as open strings with twisted boundary conditions.\footnote{
Indeed, in our examples the
 wavefunctions are remarkably similar to the ones obtained in models with only open
 string fluxes, which can be interpreted as some sort of open/closed string duality.
 As we will see, this in turn leads to conjecture the existence of extra non-perturbative
 charged states.}
 Finally note that, unlike in the fluxless  case, the CFT techniques of
 \cite{cim03,cp03,ao03,tkahler2,lerda1,drs07} can no longer be used  and supergravity
 is the only available tool.

As pointed out in the literature, dimensional reduction in a fluxed closed string
background presents several subtleties that need to be addressed.
In fact, a concrete prescription for performing a consistent 4d truncation of the theory in
twisted tori (and more generical, in manifolds with SU(3) structure) is missing.\footnote{See
 however \cite{minasian,nearly,kashani2} for progress in this direction.}
The common practice is then to use instead the harmonic expansion of a standard
torsionless manifold. This indeed produces the right results for the light
modes in the 4d supergravity regime. Here we will follow an alternative,
more controlled strategy and use techniques of non-commutative
harmonic analysis to explicitly solve for the spectrum of eigenmodes of the flux-modified
Dirac and Laplace operators. In this way, we perform the computation of
wavefunctions for massless and massive Kaluza-Klein modes of vector bosons, scalars,
fermions and matter fields for magnetized D-branes in simple type I flux compactifications.
Interestingly, we find that the resulting spectrum can be classified in terms of irreducible
unitary representations of the Kaloper-Myers gauge algebra \cite{kaloper}.

The computation of the above wavefunctions carries a lot of information, that can be
used for several phenomenological applications.  First, by means of this formalism we
can show explicitly that some wavefunctions in flux compactifications are insensitive
to the flux background. Thus, if those are the lightest modes of the spectrum (as is indeed
the case for weak fluxes), it is justified to expand the fluctuations in fluxless harmonics.
 We can also compute physical observables in the 4d effective theory, such as Yukawa
 couplings, in terms of overlap integrals of the corresponding wavefunctions. As a last
 application, one may consider integrating the spectrum of massive charged excitations
in order to compute threshold corrections to the physical gauge couplings.
This will however be addressed in a separate publication \cite{wip}.

The above techniques are applied to three different classes of vacua: $\mathcal{N}=2$
vacua without flux-induced masses in the open string sector, $\mathcal{N}=1$ vacua
with flux-induced $\mu$-terms and $\mathcal{N}=0$ vacua, and more precisely to explicit
examples based on twisted tori. These examples are T-dual to type IIB flux compactifications
with D3/D7-branes \cite{sethi,gkp} and S-dual to heterotic compactifications with torsion
 \cite{intrinsic,becker}. It is then easy to see that our analysis can be easily extended to other
 families of flux compactifications.

The outline of the paper is as follows. In Section \ref{sec2} we identify the class of type I
flux vacua that we consider in the paper, and compute the modified
Dirac and Laplace operators for their open string modes. We also provide two explicit
 supersymmetric examples of such vacua, to which we will apply our techniques in the
 sections to follow.
 Indeed, in Section \ref{sec:wgauge} we address the computation of the wavefunctions
 for gauge bosons and introduce the necessary tools to solve for the spectrum of the
 Laplace-Beltrami operator in arbitrary twisted tori. Sections \ref{sec:scalars} and
 \ref{sec:fermions} are devoted respectively to the computation of wavefunctions for neutral
 scalars and fermions and, finally, matter field wavefunctions are considered in Section
 \ref{sec:wmatter}. In Section \ref{sec:app} we summarize the structure of massive excitations
 previously obtained, and then compare our results to those obtained from a 4d supergravity
 approach. We also translate our results to the more familiar context of type IIB flux
 compactifications. Section \ref{sec:conclu} contains our conclusions, while the most
 technical material has been left for the appendices. In particular, in Appendix \ref{ap:N=0}
 we show that our approach can also be applied to $\cn=0$ vacua.


\section{Dirac and Laplace equations in type I flux vacua}
\label{sec2}


\subsection{Type I Dirac and Laplace equations}\label{diraclap}

A simple way to construct a theory of gravity and non-Abelian
gauge interactions is to consider the low-energy limit of either
heterotic or type I superstring theories. Indeed, in such limit we
obtain a 10d $\cn=1$ supergravity whose bosonic and fermionic
degrees of freedom are contained in a gravity and a vector
multiplet as
\beq
\nonumber
\begin{array}{ccc}
 & \text{bosons} & \text{fermions} \\
\text{\bf gravity} \quad \quad &  g_{MN}, C_{MN}, \phi & \psi_M, \lam \\
\text{\bf vector} \quad \quad & A_M^\a & \chi^\a
\end{array}
\eeq
The gravitational content is then given by the 10d metric $g$, the
two-form $C_2$, the dilaton $\phi$ and the
Majorana-Weyl fermions $\psi$ and $\lam$, respectively dubbed
gravitino and dilatino. The vector multiplet is that of 10d $\cn
=1$ Yang-Mills theory, with both the gauge vector $A$ and
the gaugino $\chi$ transforming in the adjoint of the gauge group
$G_{gauge}$.

Both multiplets couple to each other via a relatively simple 10d
$\cn =1$ action which, in the Einstein frame, is given by
\cite{cham1,cham2,cham3}
\begin{multline}
S=-\int dx^{10}(\textrm{det }g)^{1/2}\textrm{Tr}\left[
\frac{e^{\phi/2}}{4}F^{\alpha}_{MN}F^{\alpha,MN} +
\bar\chi^\a\Gamma^{M}D_M\chi^\a+\frac{e^{\phi}}{24}F_{MNP}F^{MNP} \right. \\
\left.
+ \frac{1}{24}e^{\phi/2}F_{MNP}\bar\chi^\a\Gamma^{MNP}\chi^\a
-\frac12e^{\phi/4}F_{MN}^\a\bar \chi^\a\Gamma^Q\Gamma^{MN}(\psi_Q+\frac{\sqrt{2}}{12}\Gamma_Q\lambda)+\dots \right]
\label{accion}
\end{multline}
where all terms not involving $A$ or $\chi$ have been dropped. Here
$F_{MN}$ and $F_{MNP}$ are gauge-invariant field strengths
\begin{align}
&F^\alpha_{MN}\, =\, \partial_{M}A_N^\alpha-
\partial_NA_M^{\alpha}+g^\alpha_{\beta\gamma}A^\beta_MA^\gamma_N \\
&F_{MNP}\, =\, 3! \partial_{[M}C_{NP]}+3!\
A^\alpha_{[M}\partial_NA_{P]}^\alpha
+2g_{\alpha\beta\g}A^{\a}_MA^\b_NA^\g_P
\end{align}
that will be respectively written as $F_2$ and $F_3$ when expressed in
$p$-form language. Finally the gauge-covariant derivative $D_M$ acts
on the gaugino as
\beq
D_M \chi^\a\, =\, \nabla_M\chi^\a + g^\a_{\b\g} A^\b_M \chi^\g
\eeq
with $g^{\a}_{\b\g}$ the structure constant of $G_{gauge}$.

In bosonic backgrounds $\langle \psi \rangle = \langle \lam
\rangle \equiv 0$, and so the last piece of (\ref{accion}) does
not contribute to the equations of motion for the components of
the vector multiplet. Applying the Euler-Lagrange equations, it is
easy to see that those read
\begin{align}\label{gauginoeq}
&\left( \slashed{D}+\frac{1}{4}e^{\phi/2}\slashed{F}_{3}\right)\chi\, =\, 0 \\
&\nabla_KF^{KP}-i[A_K,F^{KP}]-\frac{e^{\phi/2}}{2}F_{MN}F^{MNP}=0
\label{gaugeeq}
\end{align}
where we have introduced the slashed notation $\slashed{A}_n\equiv
\frac{1}{n!}A_{i_1\ldots i_n}\Gamma^{i_1\ldots i_n}$, and we have made
use of the equation of motion for $F_3$ to discard terms proportional to
$\nabla_n F^{nkp}$ in (\ref{gaugeeq}).

In the spirit of \cite{intrinsic}, let us consider 4d
vacua with non-trivial $F_3$. In order to
preserve 4d Poincar\'e invariance one imposes an Einstein frame ansatz of the form
\beq
ds^2\, =\, Z^{-1/2} ds^2_{\IR^{1,3}} + ds^2_{\cam_6}
\label{metric10}
\eeq
where the warp factor $Z$ only depends on $\cam_6$, as well as
all $F_3$ indices lie along $\cam_6$. In general, vacua of this
kind are such that $\cam_6$ admits an SU(3) structure, specified in
terms of two globally well-defined SU(3) invariant forms $J$ and
$\Om$. In particular, we consider backgrounds where the following
relations are satisfied
\beqa
\label{rel1}
Z e^{\phi} & \equiv & g_s \, = \, \text{const.} \\
g_s^{1/2} e^{\phi/2} F_3 & = & *_{\cam_6}\, e^{-3\phi/2} d( e^{3\phi/2} J) \label{jj}\\
d\left( e^\phi J \wedge J \right)&= & 0 \label{rel2}\eeqa
Note that these equations are less restrictive than those
obtained in \cite{intrinsic}.\footnote{In order to compare to the
results in \cite{intrinsic} and related heterotic literature, one
has to replace $\phi \raw -\phi$, $H_3 \raw F_3$ and then convert all
quantities to the string frame.} As discussed in \cite{Schulz04,geosoft,dwsb},
these are necessary conditions to construct a 4d vacuum of no-scale type.
Sufficiency conditions also involve a constraint on $d\Omega$, which for
supersymmetric vacua reads $d(Z^{-5/4}\Om) = 0$ and implies that
$\cam_6$ is a complex manifold.

Due to the presence of $F_3$, the compactification manifold $\cam_6$
has intrinsic torsion and it is not Calabi-Yau. As a result, the usual Dirac
and Laplace equations of Calabi-Yau compactifications are also modified.
Let us then compute the new equations via a general dimensional reduction of
eqs.(\ref{gauginoeq}) and  (\ref{gaugeeq}) to 4d, closely following \cite{yukawa}.
For simplicity, we will consider a $U(N)$ gauge field $A$.\footnote{For $\cam_6$
a smooth manifold, one should in principle take $G_{gauge} = Spin(32)/ \IZ_2$. In this
sense, $A_M \in U(N)$ lies in a gauge subsector of the full theory.}
It can then be expanded as
\begin{equation}
A_M=B_M^\alpha U_\alpha+W^{\alpha\beta}_M e_{\alpha\beta}
\label{splita}
\end{equation}
with $B^\alpha_M$ real and $(W^{\alpha\beta}_M)^*=W^{\beta\alpha}_{M}$.
The $U(N)$ generators $U_\alpha$ and $e_{\alpha\beta}$ are given by
\begin{equation}
(U_\alpha)_{ij}=\delta_{\alpha i}\delta_{\alpha j}\quad \quad
(e_{\alpha\beta})_{ij}=\delta_{\alpha i}\delta_{\beta j}\quad \quad
\alpha\neq \beta \label{generators}
\end{equation}

In general, when performing a dimensional reduction on an SU(3)-structure
manifold several subtleties arise.\footnote{Familiar
examples are conformal CY manifolds, arising in the context of
warped compactifications. Dimensional reduction in those backgrounds has
been studied in detail in, e.g., \cite{giddings,warping,douglas,fershiu,luca09}.}
The first and most important one concerns the identification of a suitable basis to
expand the four dimensional fluctuations \cite{minasian}, since different choices should be
related by highly non-trivial field redefinitions in the 4d
effective theory. In our computations below, we find convenient to expand the vector
fields in terms of vielbein 1-forms $e^m$ of $\cam_6$\footnote{More precisely, $e^m$
stand for left-invariant 1-forms of a group manifold related to $\cam_6$, as in \cite{nearly}.}
\begin{align}
B(x^\mu,x^i) &\ =\ b_\mu(x^\mu)\ B(x^i)\ dx^\mu\ +\ \sum_{m}b^m(x^\mu)\ [\langle
B^m\rangle+ \xi^m](x^i)
\ e^m \label{splitboson}\\
W(x^\mu,x^i) &\ =\ w_\mu(x^\mu)\ W(x^i)\ dx^\mu\ +\ \sum_{m}w^m(x^\mu)\ \Phi^m(x^i)
\ e^m \label{splitboson2}
\end{align}
where $x^\mu$, $x^i$ denote respectively the 4d Minkowski and 6d internal coordinates. Here, as in \cite{yukawa}, we have set $\langle W\rangle = 0$ and
allowed for a non-trivial internal vev for $B$, which breaks the initial $U(N)$
gauge group into a subgroup $G_{unbr} = \prod_i U(n_i) \subset U(N)$.
The modes $b_\mu(x^\mu)$, $w_\mu(x^\mu)$, and $b^m(x^\mu)$, $w^m(x^\mu)$ transform
respectively as 4d Lorentz vector and scalar fields, while from the point
of view of $G_{unbr}$ the $b$'s transform in the adjoint and the $w$'s in
the bifundamental representation. Finally, these modes satisfy standard
equations of motion for 4d gauge bosons
\begin{equation}
\nabla_\mu F^{\mu\nu}-i[A_\mu,F^{\mu\nu}]=m^2_A A^\nu
\label{4dgaugeom}
\end{equation}
and Klein-Gordon fields
\begin{align}
\nabla^2_{\IR^{1,3}} b^{m}&=m^2_\xi\ b^{m} \\
\nabla^2_{\IR^{1,3}} w^{m}&=m^2_\Phi\ w^{m}
\end{align}
where in (\ref{4dgaugeom}) $A_\mu=b_\mu+w_\mu$ and $m^2_A=m^2_B+m^2_W$.

Similarly, the 10d Majorana-Weyl spinor $\chi$ can be decomposed as
\beq \chi\, =\, \zeta + \mathcal{B}^* \zeta^* \quad \quad \zeta\,
=\, \chi_4 \otimes \chi_6
\label{splitgaug}
\eeq
where  $\chi_6$ is a 6d Weyl spinor of negative chirality,
$\mathcal{B} = \mathcal{B}_4 \otimes \mathcal{B}_6$ a Majorana
matrix and $\chi_4$ is a 4d Weyl spinor of positive chirality satisfying
\begin{equation}
\gamma_{(4)}
\slashed{\p}_{\IR^{1,3}} \mathcal{B}_4^* \chi_4^* = - m_\chi\,
\chi_4\label{ferm4d}
\end{equation}
where the 4d fermionic modes will arise from. Just as in
eqs.(\ref{splitboson}), (\ref{splitboson2}), in the
decomposition (\ref{splitgaug}) there is a choice of basis for the
4d fluctuation modes, now implicit in the definition of $\chi_6$.
Such choice of basis is given in Appendix \ref{ap:ferm},
where the fermion conventions used in this
paper are specified. As one can check explicitly in the examples below,
 the choices performed in the bosonic and fermionic sectors are related to
each other via the 10d supersymmetry variation
\beq
\delta_\eps A_M\, =\, \frac{i}{2} \bar{\eps}\, \Gamma_M \chi
\label{10dSUSYt}
\eeq
where $\eps$ is the 10d Killing spinor of the background.\footnote{In
$\cn =0$ no-scale models, $\eps$ should be seen as an approximate
supersymmetry generator that nevertheless specifies an SU(3) structure
in $\cam_6$ \cite{geosoft,dwsb}.} As a result, the effective theory obtained
from the above dimensional reduction scheme will inherit a 4d SUSY
structure that can be obtained directly from reducing (\ref{10dSUSYt}).

In general, in order to fully specify the 4d couplings of the effective
action one first needs to compute internal wavefunctions of the fields
$B(y)$, $W(y)$, $\xi^m(y)$, $\Phi^m(y)$ and $\chi_6(y)$ that appear
in eqs.(\ref{splitboson}), (\ref{splitboson2}) and (\ref{splitgaug}). Such
wavefunctions can be obtained by solving the corresponding internal
6d Dirac and Laplace equations for a type I background with fluxes.
One can compute these equations by plugging (\ref{splitboson})-(\ref{ferm4d})
into (\ref{gauginoeq})-(\ref{gaugeeq}) and the ansatz (\ref{metric10}).
We obtain\footnote{In order to derive these equations we have neglected the 3- and
 4-point interactions and we have taken the gauge fixing conditions,
 $\nabla^{\mathcal{M}_6}_m\xi^{m,\alpha}=0$ and $\tilde{D}_m\Phi^{m,\alpha\beta}=0$
 as in \cite{quevedo}.}
\beqa
\label{Beq}
\nabla^{\mathcal{M}_6}{}^m\nabla^{\mathcal{M}_6}_m B - (\partial_m\textrm{log
}Z)\nabla^{\mathcal{M}_6}{}^{m} B\, = \,-Z^{1/2} m_B^2  B \\
\label{Weq}
\tilde{D}^m\tilde{D}_m W -2(\partial_m\textrm{log}Z)\tilde{D}^{m}W\, =\, -Z^{1/2} m_W^2  W
\eeqa
\vspace*{-.75cm}
\begin{multline}
\nabla^{\mathcal{M}_6}{}^m\nabla^{\mathcal{M}_6}_m\xi^{p,\alpha}-[\nabla_m^{\mathcal{M}_6},\nabla^{\mathcal{M}_6}{}^p]\xi^{m,\alpha}-2(\partial_k\textrm{log
}Z)\nabla^{\mathcal{M}_6}{}^{[k}\xi^{p],\alpha}+\\
+e^{\phi/2}(\nabla^{\mathcal{M}_6}_m\xi^{n,\alpha})F_n{}^{mp}=-Z^{1/2}m_\xi^2\xi^{p,\alpha}
\label{xieq}
\end{multline}
\vspace*{-.75cm}
\begin{multline}
\tilde{D}^m\tilde{D}_m\Phi^{p,\alpha\beta}-[\nabla_m^{\mathcal{M}_6},\nabla^{\mathcal{M}_6}{}^p]\Phi^{m,\alpha\beta}-2(\partial_k\textrm{log
}Z)\tilde{D}^{[k}\Phi^{p],\alpha\beta}+2i\Phi^{m,\alpha\beta}\langle G_m{}^{p,\alpha\beta}\rangle+\\
+e^{\phi/2}(\tilde{D}_m\Phi^{n,\alpha\beta})F_n{}^{mp}=-Z^{1/2}m_\Phi^2\Phi^{p,\alpha\beta}
\label{phieq}
\end{multline}
for the bosonic wavefunctions and
\begin{equation}
\G_{(4)}
\left( \slashed{D}^{\cam_6}  + \frac{1}{4} e^{\phi/2}
\slashed{F}_3 - \frac{1}{2} \slashed{\p} \ln Z \right) \chi_6 \,
=\, Z^{1/4}  m_\chi\, \mathcal{B}_6^* \chi_6^*
\label{dirac6d}
\end{equation}
for the fermionic wavefunctions, where $\nabla^{\mathcal{M}_6}_m$
and $\slashed{D}^{\cam_6} = \G^m D_m$ are the bosonic and
fermionic covariant derivatives in $\cam_6$ and we have introduced
the notation
\begin{align}
&\tilde
D_m\Phi^{\alpha\beta}_n=\nabla_m^{\mathcal{M}_6}\Phi^{\alpha\beta}_n-i(\langle
B^\alpha_m\rangle-\langle B^\beta_m\rangle)\Phi^{\alpha\beta}_n
\label{prot1}\\
&\langle
G^{\alpha\beta}_{mn}\rangle=2\nabla_{[m}^{\mathcal{M}_6}\langle
B^\alpha_{n]}\rangle-2\nabla_{[m}^{\mathcal{M}_6}\langle
B^\beta_{n]}\rangle\label{prot2}
\end{align}
Finally, note that if we expand the fermionic wavefunction as
$\chi_6 = \lam^\a U_\a + \Psi^{\a\b}e_{\a\b}$, we have that
$\slashed{D}^{\cam_6} \lam^\a =  \slashed{\nabla}^{\cam_6} \lam^\a$
and $\slashed{D}^{\cam_6} \Psi^{\a\b} =  \tilde{\slashed{D}} \Psi^{\a\b}$.


\subsection{Elliptic fibrations}\label{subsec:elliptic}

A simple way to find solutions to the equations
(\ref{rel1})-(\ref{rel2}) is to consider the particular case where
$\cam_6$ is an elliptic fibration of fiber $\Pi_2$ over a four dimensional
 base $B_4$ \cite{sethi,Schulz04,geosoft,dwsb}. In particular, we consider a
 metric ansatz of the form
\beq
ds^2_{\cam_6} \, =\, Z^{-1/2}  \sum_{a \in \Pi_2}
\left(e^a\right)^2 + Z^{3/2} ds^2_{B_4}
\label{mansatz}
\eeq
where neither the base metric $ds^2_{B_4}$ nor the vielbein 1-forms
of the fiber $e^a$ depend on the warp factor $Z$, which in turn only
depends on the $B_4$ coordinates. This will be indeed the case if $Z$ is sourced
by background fluxes and/or D5-branes/O5-planes wrapped on $\Pi_2$
(see e.g. \cite{Schulz04,dwsb} for explicit examples of this kind). The structure of
the (unwarped) fibration can then be parameterized as
\begin{equation}
de^a=\frac{1}{2}f^a_{mn}e^m\wedge e^n\in H^2(B_4,\IR)
\label{defviel}
\end{equation}
with $f^a_{mn}$ some structure constants.\footnote{Note that
these are not the usual integer-valued structure constants used
in, e.g., the twisted-tori literature, as they also include some
dependence on the compactification moduli. See below.}

In general, $\nabla^{\mathcal{M}_6}_m$, $\slashed{D}^{\cam_6}$ and
$e^{\phi/2} \slashed{F}$ will depend on the warp factor $Z$, that will
enter eqs.(\ref{Beq})-(\ref{dirac6d}) in a rather non-trivial way.
Even if as shown in Appendix \ref{ap:warp} the on-shell relations
(\ref{rel1})-(\ref{rel2}) simplify such dependence, we would like to simplify
the problem by taking a limit of constant warp factor. In practice, one can
achieve such limit via the non-isotropic fibration
$\textrm{Vol}_{B_4}^{1/2} \gg \textrm{Vol}_{\Pi_2}$, that in terms of mass scales
translates into the hierarchy
$m^{\text{KK}}_{\text{fib}} \gg m^{\text{KK}}_{\text{base}} \gg m_{\text{flux}}$
\cite{Schulz04}. Here $m_{\text{flux}}$ (denoted $\varepsilon$ in the following
sections) is the mass scale introduced by the presence of background fluxes,
and in particular the mass scale of closed and open string lifted moduli. As a result,
this hierarchy of scales is essential to understand the process of moduli
stabilization in terms of a 4d $\cn=1$ effective theory where all KK modes have
been integrated out. In addition, as discussed in section \ref{sugra}
the condition $m^{\text{KK}}_{\text{base}} \gg m_{\text{flux}}$ also ensures that
the warp factor can be taken to be constant, which is the approximation that we
would like to consider in the following.\footnote{In our analysis below we will not
be interested in closed strings dynamics and moduli stabilization,  and so the limit
$\textrm{Vol}_{B_4}^{1/2} \gg \textrm{Vol}_{\Pi_2}$ is in fact not essential for our
purposes. We will however take it for technical purposes, as it greatly simplifies
the open strings equations of motion.} Finally, imposing
$\textrm{Vol}_{B_4}^{1/2}, \textrm{Vol}_{\Pi_2} \gg \a'$ guarantees that
the supergravity approximation in which we are working remains valid.

Splitting the 2-form $J$ as $J = J_{\Pi_2} + J_{B_4}$ as in
\cite{geosoft}, introducing the projectors,
\beq
P_\pm^{\Pi_2}\, =\, \oh \left(1 \pm i\slashed{J}_{\Pi_2}
\g_{(6)} \right) \label{ex1proj}
\eeq
and taking $Z$ constant eq.(\ref{dirac6d}) becomes (see Appendix \ref{ap:warp})
\beq
\left(\slashed{D}^{\Pi_2} + \slashed{D}^{B_4} +
\frac{1}{2}\slashed{f}P_+^{\Pi_2} \right) \chi_6  \, =\, m_\chi\,
\mathcal{B}_6^* \chi_6^*
\label{dirac6duw}
\eeq
where we have absorbed the operator $\G_{(4)}$ in the definition
of slashed contraction. Indeed, in the expression above all slashed
quantities are constructed from the set of $\G$-matrices defined in
(\ref{commG}), a convention that we will take from now on. Finally,
we have defined the antisymmetrized geometric flux
\beq
f_{mnp}\, =\, 3\, \delta_{r[m}f^r_{np]}
\label{metricflux}
\eeq
The projector $P_+^{\Pi_2}$ corresponds to the chirality projector
of the 4d base $B_4$. One can then split the internal 6d
fermion as
\beq
\chi_6\, =\, \chi_{\Pi_2} + \chi_{B_4}
\label{split4+2}
\eeq
where $\chi_{\Pi_2,B_4}$ satisfy $P_+^{\Pi_2} \chi_{\Pi_2} = \chi_{\Pi_2}$ and
$P_+^{\Pi_2} \chi_{B_4} = 0$.\footnote{This splitting has a simple geometric
interpretation in the type IIB T-dual setup of Section \ref{D7dual}.} Since $\mathcal{B}_6$
changes the fiber chirality but not the base chirality, we can split the
Dirac equation as
\beqa \label{6dsplit1}
{\slashed{D}}^{\Pi_2}\chi_{B_4} + {\slashed{D}}^{B_4} \chi_{\Pi_2} & = & m_\chi \mathcal{B}_6^* \chi_{B_4}^*\\
\label{6dsplit2} {\slashed{D}}^{\Pi_2}\chi_{\Pi_2} + {\slashed{D}}^{B_4}
\chi_{B_4}  + \oh {\slashed{f}} \chi_{\Pi_2} & = & m_\chi \mathcal{B}_6^*
\chi_{\Pi_2}^* \eeqa
A similar analysis can be carried out for the scalar
wavefunctions, governed by eqs.(\ref{xieq}) and (\ref{phieq}).
Distinguishing between scalars corresponding to the base and
to the fiber,  the equations of motion (\ref{xieq}) and (\ref{phieq}) read
\begin{align}
\hat\partial_m\hat\partial^m\xi^{p}_{\Pi_2}
-(f^{p}_{mn}-e^{\phi/2}F_{nm}{}^{p})\hat\partial^m\xi^n_{B_4}
-\frac12f^{a}_{mn}(f^{p}_{mn}-e^{\phi/2}F_{nm}{}^{p})\xi^{a}_{\Pi_2}=-m_\xi^2\xi^p_{\Pi_2}
\label{xi+}\\
\hat\partial_m\hat\partial^m\xi^{p}_{B_4}-(f^{m}_{pn}-e^{\phi/2}F_{n{m}}{}^p)
\hat\partial^{m}\xi^n_{B_4}+(f_{mp}^{n}+e^{\phi/2}F_{{n}m}{}^p)\hat\partial^m\xi^{n}_{\Pi_2}
=-m_\xi^2\xi_{B_4}^p
\label{xi-}
\end{align}
and
\begin{align}
& \hat D_m\hat D^m\Phi^{p}_{\Pi_2}-(f^{p}_{mn}-e^{\phi/2}F_{nm}{}^{p})
\hat D^m\Phi^n_{B_4}-\frac12f^{a}_{mn}(f^{p}_{mn}-e^{\phi/2}F_{nm}{}^{p})
\Phi^{a}_{\Pi_2}=-m_\Phi^2\Phi^p_{\Pi_2}
\label{lapfin1}\\
& \hat D_m\hat D^m\Phi^{p}_{B_4}-(f^{m}_{pn}-e^{\phi/2}F_{n{m}}{}^p)
\hat D^{m}\Phi^n_{B_4}+(f_{mp}^{n}+e^{\phi/2}F_{{n}m}{}^p)
\hat D^m\Phi^{n}_{\Pi_2}+2i\Phi_{\Pi_2}^m\langle\hat G_m{}^p\rangle\nonumber\\
&\hspace{11cm}=-m_\Phi^2\Phi_{B_4}^p
\label{lapfin2}
\end{align}
where $\hat D_m\Phi_n^{\alpha\beta}$ and
$\langle \hat G^{\alpha\beta}_{mn}\rangle$ are respectively defined
as in (\ref{prot1}) and (\ref{prot2}), but replacing the covariant
derivative $\nabla_m^{\mathcal{M}_6}$ by twisted derivatives defined
in terms of the vielbein as
\beq
\hat\partial_a\equiv e_a{}^{\a}(x)\ \partial_{x^\a}
\label{hatted}
\eeq
Finally, we have assumed that $\langle B_m^\a\rangle$ is constant
along the fiber, as dictated by cancelation of Freed-Witten anomalies
\cite{fw1,kaloper}.


\subsection{Twisted tori examples}\label{subsec:twisted}

In order to provide explicit examples of the metric ansatz (\ref{mansatz})
one may consider the simple case where the base of the fibration
$B_4$ corresponds to a flat four-torus $T^4$. This basically implies that,
up to warp factors, $\cam_6$ lies within a particular class of twisted tori,
which are in fact the simplest non-trivial examples of SU(3) structure manifold.
A very interesting feature of twisted tori, and which will be crucial in the
discussion of next section, is that they can be defined in a group theoretic
way, and more precisely as a left quotient of groups
$\cam_6 = \G \backslash G$.

Indeed, let us consider a $d$-dimensional group manifold $G$ and its
Lie algebra $\mathfrak{g}=\textrm{Lie}(G)$. The latter is specified by a set of structure
constants $f^a_{bc}$ that satisfy the Jacobi identity $f^a_{[bc}f^g_{d]a}=0$.
In terms of a matrix representation of the Lie Group $g_G \in GL(n)$, one can
easily compute the vielbein left-invariant 1-forms as
$g_G^{-1}dg_G = e^a \mathfrak{t}_a$, with $\mathfrak{t}_a \in \mathfrak{g}$
the algebra generators, and hence the structure constants via
\begin{equation}
de^a=\frac{1}{2}f^a_{bc}e^b\wedge e^c
\quad  \quad \Leftrightarrow \quad \quad
[\hat \partial_b,\hat
\partial_c]=-f^a_{bc}\hat \partial_a
 \label{torsion}
\end{equation}
with $\hat{\p}_a$ defined as in (\ref{hatted}). These twisted
derivatives can then be identified with $\mathfrak{t}_a$.
While in general $G$ may not be a compact manifold, one can construct
such manifold by left-quotienting $G$ by a discrete, cocompact subgroup
$\G \subset G$.\footnote{One could actually be more general and quotient $G$
by a discrete subgroup of its affine group, $\pi\subset \textrm{Aff}(G)$, obtaining
a freely-acting orbifold of a twisted torus. Indeed, these kind of constructions are
well-known for $G\simeq \mathbb{R}^d$ and $\pi \in \text{Aff}(\IR^d)$ a torsion-free
crystallographic group (a.k.a. Bieberbach groups \cite{Bieber1,Bieber2}), that lead
to standard freely-acting orbifolds of $T^n$. Analogously, for $G$ a
nilpotent Lie group and $\pi$ an almost-Bieberbach group one obtains the so-called
infra-nilmanifolds \cite{dekimpe}.}
The resulting twisted torus $\cam_d = \G \backslash G$ is no longer
a group, but it is a parallelizable manifold since the left-invariant 1-forms
are still globally well-defined.

Given a set of structure constants
$f^a_{bc}$, constructing a compact manifold $\cam_d = \G \backslash G$
is usually a non-trivial problem. This is however greatly simplified
if we restrict ourselves to the case where $\mathfrak{g}$ is a nilpotent
Lie algebra.\footnote{See \cite{scan} for a discussion of this problem in
the more general context of solvable Lie algebras.} That is, we consider
the case where the series $\{\mathfrak{g}_s\equiv [\mathfrak{g}_{s-1},\mathfrak{g}_0]\}$,
with $\mathfrak{g}_0\equiv \mathfrak{g}=\textrm{Lie}(G)$, has $k$ non-vanishing
elements, in which case $\mathfrak{g}$ is said to be $k$-step nilpotent.
Then, in order for a cocompact $\G$ to exist, we only need to
require that $f^a_{ab} =0$ and that the structure constants are integers in
some particular basis \cite{Malcev}. The resulting nilmanifold is a non-flat,
compact  (usually iterated) fibration of tori. In particular, we will
obtain elliptic fibrations that fit into our metric ansatz (\ref{mansatz}).

If in particular we consider an elliptic fibration over $T^{d-2}$, then
 $\mathfrak{g}$ should be 2-step nilpotent. The associated Lie group has then
 the following faithful representation
\begin{equation}
g_G=
\begin{pmatrix}
\Id_{d} - \oh ad_{\vec X} & \vec X\\ 0& 1
\end{pmatrix}
\quad \quad \quad
[ad_{\vec X}]^i_j\, =\, X^k f^i_{kj}
\label{magicdisney}
\end{equation}
in terms of $GL(d+1,\mathbb{R})$ matrices. Here $\vec X$ is a $d$-dimensional
coordinate vector parameterizing $\mathfrak{g}$ and $ad$ is the adjoint
representation of the algebra, which due to 2-step nilpotency satisfies $ad^2 =0$.
Note that this implies that $e^i = dX^i + \oh f^i_{kj} X^kdX^j$.

A classical example of
this construction is given by the $(2p+1)$-dimensional
Heisenberg manifold $\mathcal{H}_{2p+1}$, the canonical
example of nilpotent Lie group. Here we can split
$\vec X^t = (z, \vec x ^{\, t}, \vec y^{\, t})$, $\vec x, \vec y \in \IR^p$
and express the algebra as
\begin{equation}
[\mathfrak{t}_{x^i}, \mathfrak{t}_{y^j}]\, =\, \d_{ij}\,\mathfrak{t}_{z}
\label{heisalg}
\end{equation}
so that (\ref{magicdisney}) reads
\begin{equation}
g_{\mathcal{H}_{2p+1}}\, =\,
\begin{pmatrix}1 & -\frac12\vec{y}^{\, t} &
\frac12\vec{x}^{\, t} & z\\
0 & 1 & 0 & \vec x \\
0 & 0 & 1 & \vec y\\
0 & 0 & 0 & 1\end{pmatrix}
\label{heis}
\end{equation}
In this case, a suitable choice for $\Gamma$ is the lattice
 $\G_{\ch_{2p+1}} = \{(\vec x,\vec y,z) = M (\vec{n}_x, \vec{n}_y, n_z)\}$ with
$M,  n_z \in \IZ$ and $\vec{n}_x, \vec{n}_y \in \IZ^n$.\footnote{In
fact, we need $M \in 2\IZ$ if we want $\G$ to be a subgroup.
Interestingly, the same condition is required by the presence of
orientifold planes \cite{fp03}.} One can then normalize the
generators as $\tilde{\mathfrak{t}}_{a} = M \mathfrak{t}_a$, so
that the algebra becomes
$[\tilde{\mathfrak{t}}_{x^i}, \tilde{\mathfrak{t}}_{y^j}]\, =\, \d_{ij}M\, \tilde{\mathfrak{t}}_{z}$
and the invariant 1-forms read
\beq
\tilde{e}^z\, =\, dz - \frac{M}{2}\left(\vec x^{\, t}d\vec y - \vec y^{\, t} d\vec x\right)
\quad \quad \tilde{e}^{x^i}\, =\, dx^i \quad \quad  \tilde{e}^{y^i}\, =\, dy^i
\label{1formheis}
\eeq
 The nilmanifold $\G \backslash G$ then corresponds to an $S^1$ fibration
 (whose fiber is parameterized by $z$) over a $T^{2p}$ (parameterized by
 $(\vec x,\vec y)$) and of Chern class $F_2 = M\sum_i dy^i \wedge dx^i$.
Such U(1)-bundle structure will become manifest below, when analyzing the
spectrum of the Laplace and Dirac operators in the (compactified)
Heisenberg manifold. Finally, a rescaling of the form
$\tilde{\mathfrak{t}}_a \raw (2 \pi R_a)^{-1} \tilde{\mathfrak{t}}_a$,
$X^a \raw 2\pi R_a\, X^a$ will take us to a moduli-dependent set of structure
constants, which are those that correspond to the set of vielbein left-invariant
1-forms in (\ref{defviel}) and (\ref{torsion}).

It follows from the above discussion that a good starting point to construct
explicit solutions to eqs.(\ref{rel1})-(\ref{rel2}) is to consider $\cam_6$  to
be either a nilmanifold or a product like $S^1 \times \G_{\ch_5}  \backslash \ch_5$.
In the following we will provide two different type I backgrounds based on
such strategy, for which we will later on explicitly solve the Laplace and
Dirac equations (see Sections \ref{sec:wgauge} to \ref{sec:wmatter}).

According to the open string spectrum, we can roughly classify nilmanifold type I flux
vacua in two different classes. The first one is that
where the spectrum of massless open string adjoint scalars $b^m$
(see (\ref{splitboson})) remains identical with respect to a toroidal
(or toroidal orientifold), fluxless compactification. The second class is that
where, because of the presence of the flux, some of these adjoint scalars
develop up a mass of the order of $m_{\text{flux}}$, just like the closed string moduli
of the compactification. We will dub such classes of vacua as vacua with vanishing
and non-vanishing flux-generated $\mu$-term, respectively, and present a
supersymmetric example for each of them below. A non-supersymmetric type I
flux vacua will be considered in Appendix \ref{ap:N=0}.

\subsubsection{Example with vanishing $\mu$-terms}\label{vmu}

Let us consider the following type I flux background, displayed in the
ten dimensional Einstein frame and $\alpha'$ units
\bes
\label{bg1}
\begin{align}
\label{bg11}
&ds^2=Z^{-1/2}(ds^2_{\IR^{1,3}}+ds^2_{\Pi_2})+Z^{3/2}ds^2_{T^4} \\
\label{bg12}
&ds^2_{T^4}=(2\pi)^2 \sum_{m=1,2,4,5}(R_mdx^m)^2 \\
\label{bg13}
&ds^2_{\Pi_2}= (2\pi)^2 \left[(R_3  dx^3)^2+(R_6\tilde e^{6})^2\right]  \\
\label{bg14}
&F_3=-(2\pi)^2  N ( dx^1\wedge dx^2+
dx^4\wedge dx^5)\wedge  \tilde e^6-g_s^{-1}*_{T^4}dZ^2 \\
&e^{\phi} Z=g_s = \text{const.}
\end{align}
\ees
where we have included the warp factor dependence, as well as
provisionally set $F_2=0$.

Let us first focus on the metric background (\ref{bg11})-(\ref{bg13}),
parameterized by the six compactification radii $R_i$. Here  $\tilde e^6$
stands for a left-invariant 1-form satisfying\footnote{
Recall that according to our definition (\ref{defviel}) the {\em vielbein}
left-invariant 1-form is not given by $\tilde e^6$, but rather by the
moduli-dependent 1-form $e^6 \equiv 2\pi R_6 \tilde e^6$.}
\begin{equation}
d \tilde e^6=M( dx^1\wedge dx^2+ dx^4 \wedge dx^5)\ .
\label{ejtwist}
\end{equation}
so from (\ref{heisalg}) it is easy to see that (up to warp factors) $\cam_6$
looks locally like $\IR \times \mathcal{H}_5$, and that $\tilde e^6$ is
associated to the center of the 5-dimensional Heisenberg group
$\mathcal{H}_5$. Following our general discussion above, we can easily
integrate eq.(\ref{ejtwist}) to obtain
\begin{equation}
\tilde e^6=dx^6+\frac{M}{2}(x^1dx^2-x^2dx^1+x^4dx^5-x^5dx^4)
\label{integrated1}
\end{equation}
as well as the vielbein 1-forms $e^a$. From the latter, we obtain the
twisted derivatives
\bes
\label{hatted1}
\begin{align}
\hat\partial_1&=(2\pi R_1)^{-1}\left(\partial_{x^1}+\frac{M}{2}x^2\partial_{x^6}\right)&
\hat\partial_4&=(2\pi R_4)^{-1}\left(\partial_{x^4}+\frac{M}{2}x^5\partial_{x^6}\right)\\
\hat\partial_2&=(2\pi R_2)^{-1}\left(\partial_{x^2}-\frac{M}{2}x^1\partial_{x^6}\right)&
\hat\partial_5&=(2\pi R_5)^{-1}\left(\partial_{x^5}-\frac{M}{2}x^4\partial_{x^6}\right)\\
\hat\partial_3&=(2\pi R_3)^{-1}\partial_{x^3}&\hat\partial_6&=(2\pi R_6)^{-1}\partial_{x^6}
\end{align}
\ees

Finally, the global structure of $\cam_6$ is not $\IR \times \mathcal H_5$
but rather the compact manifold $\cam_6 = \G \backslash (\IR \times \mathcal H_5)$,
where $\G$ is a cocompact subgroup of $\IR \times \mathcal H_5$, which we take
to be $\IZ \times \G_{\ch_5}$. Such quotient requires $M \in 2\IZ$ and produces the
identifications
\bes
\label{reli}
\begin{align}
&x^1\to x^1+1 \quad \quad x^6\to x^6-\frac{Mx^2}{2} \label{reli1}\\
&x^2\to x^2+1 \quad \quad x^6\to x^6+\frac{Mx^1}{2} \\
&x^3\to x^3+1 \\
&x^4\to x^4+1 \quad \quad x^6\to x^6-\frac{Mx^5}{2} \\
&x^5\to x^5+1 \quad \quad x^6\to x^6+\frac{Mx^4}{2} \\
&x^6\to x^6+1\label{reli2}
\end{align}
\ees
which by construction leave (\ref{integrated1}) and (\ref{hatted1}) invariant.

Taking now into account the RR flux (\ref{bg14}) it is easy to see that
eqs.(\ref{rel1})-(\ref{rel2}) are satisfied provided that the
on-shell relations $g_s N = M R_6^2$ and $R_1R_2=R_4R_5$ are imposed.
This implies that $d(Z^{-5/4}\Om) = 0$ for some suitable
choice of $\Omega$ (see below),  which in turn implies that our compactification
manifold $\cam_6$ is complex and our 4d theory supersymmetric. Finally, we
should also impose $N \in \IZ$ by standard Dirac quantization arguments.

Since $\cam_6$ is a compact manifold, we should check that both NSNS and
RR tadpoles are canceled globally. Before that, let us include in our background
an open string field strength of the form
\beq
F_2\, =\, F_{14}\, dx^1 \wedge dx^4 + F_{25}\, dx^2 \wedge dx^5
\eeq
as well as  D5-branes and O5-planes wrapping $\Pi_2$. The Bianchi
identity for $F_3$ then reads
\beqa\nonumber
dF_3 & =&-  \left( (2\pi )^2
 \frac{2NM + \tr (F_{14}F_{25})}{\text{Vol}(T^4)} + g_s^{-1}  \nabla^2_{T^4}Z^2 \right)
 e^1\wedge e^2\wedge e^3\wedge e^4\\
& =& (2\pi)^2 \sum_jq_j\delta_{T^4}(x - x_j)\,e^1\wedge
e^2\wedge e^3\wedge e^4
\label{BI1}
\eeqa
where in the second line we have made use of the warp factor
equation
\beq
-g_s^{-1} \nabla^2_{T^4}Z^2\, =\, (2\pi)^2 \left( \frac{2NM + \tr (F_{14}F_{25})}{\text{Vol}(T^4)}+\sum_jq_j\delta_{T^4}(x - x_j)\right)
\label{NSNSt1}
\eeq
where $q_j=1$ for D5-branes and $q_j = -2$ for O5-planes.
Note that (\ref{BI1}) does not imply that $\sum_j q_j = 0$, as it would
for $\cam_6 = T^6$, but rather that $\sum_j q_j = 0\, \text{mod}\, M$,
due to the torsional cohomology of $\cam_6$ \cite{Schulz04,torsion}.
On the other hand, the r.h.s. of (\ref{NSNSt1}) must vanish upon
integration on $B_4$. Since the background BPS conditions imply that
$NM > 0$ and that $\tr (F_{14}F_{25}) > 0$, this is only possible if
O5-planes are present on the compactification. We will thus implement
their presence via the additional orbifold quotient ${\cal R}: x^m \mapsto - x^m$,
where $x^m$ is a $B_4$ coordinate.

Finally, let us discuss the amount of supersymmetry preserved by this background.
The fact we are compactifying type I string theory sets the maximal amount
of supersymmetry to 4d $\cn=4$, which would be the case if we were
compactifying in $T^6$. Adding the orbifold quotient ${\cal R}$ above (or
equivalently adding the induced O5-planes) halves the amount of SUSY
to 4d $\cn =2$. These two generators of supersymmetry can be associated
with two different choices of complex structure, $(z^1, z^2, z^3)$ and
$(\bar{z}^1, \bar{z}^2, z^3)$, with $z^i = x^i + i \tau_i x^{i+3}$ and $\tau_i = R_{i+3}/R_i $,
that preserve the orientation of $T^6$ and of the 2-cycle $\Pi_2$ wrapped by
the O5-plane. If as a last ingredient we add the background flux (RR and geometric)
with the above choice of dilaton and compactification radii ($g_sN = MR_6^2$ and
$R_1R_2 = R_4R_5$) we see that no further supersymmetries are broken.
Indeed, taking for simplicity the $Z=1$ limit, this can be checked by noting that
$g_s F_3 - i dJ$ is a (2,1)-form for both
choices of complex structure, or by the fact that both choices of 3-form
$\Omega = e^{z^1} \wedge e^{z^2} \wedge e^{z^3}$ and
$\Omega' =  e^{\bar{z}^1} \wedge e^{\bar{z}^2} \wedge e^{z^3}$ are
closed, and so define a good complex structure even in the presence of the
geometric flux.

\subsubsection{Example with non-vanishing $\mu$-terms}\label{nvmu}

Let us now consider a slightly more involved solution to the
equations (\ref{rel1})-(\ref{rel2}), this time yielding supersymmetric
mass terms ($\mu$-terms) for some of the 4d adjoint multiplets.
Such background is given by
\bes
\label{bg2}
\begin{align}
&ds^2=Z^{-1/2}(ds^2_{\IR^{1,3}}+ds^2_{\Pi_2})+Z^{3/2}ds^2_{T^4} \\
&ds^2_{T^4}=(2\pi)^2 \sum_{m=1,2,4,5}(R_mdx^m)^2 \\
&ds^2_{\Pi_2}=(2\pi)^2\left[ (R_3\tilde e^3)^2+\left(R_6
\tilde e^{6}\right)^2\right] \\
&F_3=(2\pi)^2(N_6\, dx^2\wedge \tilde e^6
-N_3\, dx^5\wedge \tilde e^3)\wedge dx^4 -g_s^{-1}*_{T^4}dZ^2
\end{align}
\ees
and again $e^\phi Z = g_s$. This time the left-invariant 1-forms
satisfy
\begin{equation}
d \tilde e^3=M_3 dx^1\wedge dx^2 \quad \quad  \text{and} \quad \quad d\tilde e^6=M_6 dx^1
\wedge dx^5\label{nil2}
\end{equation}
which again corresponds to a nilpotent Lie algebra. The
twisted derivatives now read
\begin{align*}
\hat\partial_1&=(2\pi R_1)^{-1} \left(\partial_{x^1}+\frac{M_3}{2}x^2\partial_{x^3}+\frac{M_6}{2}x^5\partial_{x^6}\right)&\hat
\partial_4&=(2\pi R_4)^{-1}\partial_{x^4}\\
\hat\partial_2&=(2\pi R_2)^{-1}\left(\partial_{x^2}-\frac{M_3}{2}x^1\partial_{x^3}\right)&\hat\partial_5&=(2\pi R_5)^{-1}\left(\partial_{x^5}-\frac{M_6}{2}x^1\partial_{x^6}\right)\\
\hat\partial_3&=(2\pi R_3)^{-1}\partial_{x^3}&\hat\partial_6&=(2\pi R_6)^{-1}\partial_{x^6}
\end{align*}
and the quotient by $\G$ produces the identifications
\bes
\label{relii}
\begin{align}
&x^1 \to x^1 + 1 \quad \quad x^3 \to x^3 - \frac{M_3}{2}x^2 \quad \quad x^6
\to x^6 - \frac{M_6}{2}x^5 \label{relii1}\\
&x^2 \to x^2 + 1 \quad \quad x^3 \to x^3 + \frac{M_3}{2}x^1 \\
&x^3 \to x^3 + 1 \\
&x^4 \to x^4 + 1 \\
&x^5 \to x^5 + 1 \quad \quad x^6 \to x^6 + \frac{M_6}{2}x^1 \\
&x^6 \to x^6 + 1\label{relii2}
\end{align}
\ees
so that the resulting nilmanifold can be seen as the simultaneous fibration of two
$S^1$'s along a $T^4$ base.

The equations of motion for this background now require the on-shell
relations $M_3 R_3^2 R_4 R_5 = g_s N_3 R_1R_2$ and
$M_6 R_6^2 R_2R_4 = g_s N_6 R_1R_5$, with
$N_3, N_6, M_3, M_6 \in \IZ$. In addition, tadpole cancelation
will need of the presence of O5-planes wrapping the $\Pi_2$
fiber, that again will be introduced via the orbifold quotient
${\cal R}: x^m \mapsto - x^m$ on the base coordinates.

As before the presence of O5-planes will reduce the amount of supersymmetry as
$\cn =4 \raw \cn = 2$, while the background fluxes will further break the amount of
SUSY. More precisely, if we impose $M_3N_3 = M_6 N_6$, we will
satisfy the supersymmetry condition $d(Z^{-5/4} \Om) = 0$ for the choice
$\Omega = e^{z^1} \wedge e^{z^2} \wedge e^{z^3}$, this being the only choice of
 closed SU(3)-invariant 3-form. Hence, in general the fluxes will break the
4d supersymmetry as $\cn =2 \raw \cn =0$, while they will do as $\cn =2 \raw \cn =1$
if we impose that $M_3N_3 = M_6 N_6$. For simplicity, we will assume the latter
constraint to hold for the rest of the paper.


\section{Wavefunctions for gauge bosons}\label{sec:wgauge}

The simplest family of wavefunctions that one may analyze in type I
flux vacua correspond to the gauge bosons of the 4d gauge
group $G_{unbr}$ and their massive Kaluza-Klein excitations,
transforming in the adjoint representation of $G_{unbr}$.
Indeed, all these modes arise from the term $b_\mu (x^\mu) B(x^i)$ in the
expansion (\ref{splitboson}) and, as (\ref{Beq}) shows, their
internal Laplace equation for $B$ does not involve the flux $F_3$. In fact,
in the limit of constant warp factor (\ref{Beq}) reduces to the standard
Laplace-Beltrami equation in the manifold $\cam_6$. In the
notation of Sections \ref{subsec:elliptic} and \ref{subsec:twisted} such
equation can be written as
\begin{equation}
\Delta B\, =\,
\hat \partial_a \hat{\partial}^{a}B=-m_B^2B
\label{lap}
\end{equation}
where $B$ is a complex wavefunction describing two real d.o.f.
of the 4d gauge boson,\footnote{For massive gauge bosons there is
a third d.o.f. showing up as a scalar mode. See Section \ref{susyspect}.}
while $\hat{\p}_a$ are the twisted derivatives defined by (\ref{hatted}).

From (\ref{lap}) it is easy to see that, as expected, gauge boson zero modes
are given by constant internal wavefunctions $B= const$. Computing the
internal wavefunction of massive KK modes is however more involved, and
in general requires the explicit knowledge of the Laplace-Beltrami operator.
As shown in the previous section, twisted tori provide simple examples of
compactification manifolds where $\hat{\p}_a$
have a simple, globally well-defined expression, which allows to compute
analytically the full spectrum of KK masses and wavefunctions of $\Delta$.
Indeed, in this section we will compute such spectrum for the explicit twisted
tori examples described in Section \ref{subsec:twisted}. As we will see, in
simple twisted tori like that of subsection \ref{vmu} the spectrum of wavefunctions
is analogous to that of open strings in magnetized D-brane models, and so it
can be easily computed using the results of \cite{yukawa}. On the other hand,
for more involved nilmanifolds such analogy becomes less fruitful, and one is
led to apply group theoretic techniques as well as tools of non-commutative
harmonic analysis to compute the spectrum of $\Delta$. We will present below
a general description of the latter method, and apply it to the computation of
wavefunctions in the twisted torus background of subsection \ref{nvmu}.


\subsection{Vanishing $\mu$-terms}
\label{nili}

Let us then consider the Laplace-Beltrami
equation for the type I vacuum of subsection \ref{vmu}. As discussed above,
in the limit of constant warp factor this equation reduces to (\ref{lap}), where the
twisted derivatives are given by (\ref{hatted1}). Solving (\ref{lap}), however, does still
not guarantee that our wavefunction is well-defined globally, as the twisted derivatives
only see the local geometry $\IR \times \mathcal H_5$ of the twisted torus
$\cam_6 = \Gamma\backslash (\IR \times \mathcal H_5)$. Hence, proper wavefunctions
will also be invariant under the left action of the discrete subgroup $\G$, and more
precisely under the identifications (\ref{reli}).

Following a similar strategy to \cite{yukawa}, we will first impose $\G$-invariance via the ansatz
\begin{equation}
B_{k_3,k_6}(\vec
x)=\sum_{k_1,k_4}f_{k_1,k_3,k_4,k_6}(x^2,x^5)\, {e}^{2\pi i
(k_1 x^1+k_3 x^3+k_4 x^4+k_6\dot{x}^6)}
\quad \quad \quad k_i \in \IZ
\label{ansz}
\end{equation}
with
\begin{equation}
f_{k_1,k_3,k_4,k_6}(x^2+\ell_2,x^5+\ell_5)=f_{k_1+Mk_6\ell_2,k_3,k_4+Mk_6\ell_5,k_6}(x^2,x^5)
\quad \quad \quad \ell_2,\ell_5\in \mathbb{Z}
\label{period}
\end{equation}
and where we have performed the change of variables
\begin{equation}
\dot x^6\equiv x^6+\frac{M}{2}(x^1x^2+x^4x^5)
\label{change}
\end{equation}

Then, substituting into eq.(\ref{lap}) and proceeding by separation of variables
\begin{equation}
f_{k_1,k_3,k_4,k_6}(x^2,x^5)\equiv
f_{k_1,k_3,k_6}(x^2)f_{k_3,k_4,k_6}(x^5)
\label{sep}
\end{equation}
one can see that (\ref{lap}) is equivalent to a system of Weber differential
equations~\cite{whitaker}
\begin{align}
&\left[(\partial_{\dot x^2})^2-\frac{1}{4}(\dot
x^2)^2+\nu-\alpha\right]f_{k_1,k_3,k_6}(\dot x^2)=0 \label{web1}\\
&\left[(\partial_{\dot x^5})^2-\frac{1}{4}(\dot
x^5)^2+\alpha\right]f_{k_3,k_4,k_6}(\dot x^5)=0
\label{web2}
\end{align}
for some constant $\alpha$, where we have made the following
definitions:
\begin{align}
\dot x^2&\equiv\frac{2}{R_1}
a^{-1/2}(k_1+k_6Mx^2)
\label{x2}\\
\dot x^5&\equiv\frac{2}{R_4}
a^{-1/2}(k_4+k_6Mx^5)  \\
\nu&\equiv\frac{1}{a}\left(m_B^2 - \left[\left(\frac{k_6}{R_6}\right)^2+\left(\frac{k_3}{R_3}\right)^2\right]\right)
\label{x5}\\
a&\equiv \frac{|k_6 M|}{\pi R_1R_2}
\label{a}
\end{align}
The general solution is then given in terms of Hermite functions $\psi_n(x)$
as\footnote{There exist additional solutions given by general
parabolic cylinder functions. However it can be checked that these
do not lead to convergent sums when plugged into (\ref{sep}) and
(\ref{ansz}).}
\begin{align}
\label{wfunc1}
f_{k_1,k_3,k_6}(\dot x^2)&=\psi_{\nu-\alpha-\oh}\left(\frac{\dot
x^2}{\sqrt{2}}\right)\\
f_{k_3,k_4,k_6}(\dot
x^5)&=\psi_{\a -\oh}\left(\frac{\dot
x^5}{\sqrt{2}}\right)
\label{wfunc2}
\end{align}
where
\begin{equation}
\psi_n(x) \equiv
\frac{1}{\sqrt{n!2^n\pi^{1/2}}}e^{-x^2/2}H_n(x),\label{hermite}
\end{equation}
and $H_n(x)$ stands for the Hermite polynomial of degree $n$.
Note that this requires that the Hermite functions in (\ref{wfunc1})
and (\ref{wfunc2}) have subindices $\nu - \a -1/2,\ \a - 1/2 \in  \IN$
and, in particular, that $\nu -1 = n \in \IN$. This turns out to fix the
mass eigenvalues, obtaining the following KK mass spectrum
\begin{equation}
m_B^2=\frac{|k_6 M|}{\pi R_1R_2}(n+1)+
\left(\frac{k_6}{R_6}\right)^2+\left(\frac{k_3}{R_3}\right)^2
\label{eigen}
\end{equation}

Plugging back these solutions into (\ref{sep}) and (\ref{ansz}) and
defining $k_{1,4}=\delta_{1,4}+k_6Ms_{1,4}$ with $s_i \in \IZ$, we obtain the set
of eigenfunctions
\begin{multline}
B^{(k,\delta_1,\delta_4)}_{n,k_3,k_6}\, =\, \cn_B
\sum_{s_1,s_4}\psi_{n-k}\left(\frac{\dot
x^2}{\sqrt{2}}\right)\ \psi_{k}\left(\frac{\dot
x^5}{\sqrt{2}}\right) \, e^{2\pi i \left[(\delta_{1}+k_6Ms_{1}) x^1+k_3 x^3+(\delta_{4}+k_6Ms_{4})
x^4+k_6\dot{x}^6\right]} \\
\cn_B\, =\, \left(\frac{2\pi |k_6M|}{\textrm{Vol}_{\cam_6}}\frac{R_5}{R_1}\right)^{1/2}
\quad \quad \quad \textrm{Vol}_{\cam_6}=\prod_{i=1}^6 (2\pi R_i)
\quad \quad \quad \quad \quad \quad
\label{set1}
\end{multline}
where the indices run as $k=\a -1/2 = 0,\ldots, n$ and $\delta_{1,4}=0, \ldots, k_6M-1$.
As in \cite{yukawa}, the fact that different choices of $\d_1, \d_4$ give independent
wavefunctions is related to the recurrence relation (\ref{period}). Finally, the
normalization has been fixed so that
\begin{equation}
\langle
{B}^{(k,\delta_1,\delta_4)}_{n,k_3,k_6}\ , \
B^{(k',\delta_1',\delta_4')}_{n',k_3',k_6'}\rangle
=\prod_{i=n,k,k_3,k_6,\delta_1,\delta_4}\delta_{ii'}
\end{equation}
where $\langle\, , \rangle$ stands for the usual inner product of
complex functions.

Besides the set of wavefunctions (\ref{set1}) there is a different family of
solutions to (\ref{lap}). Indeed, simple inspection shows that these
are given by
\begin{align}
&B_{k_1,k_2,k_3,k_4,k_5}(\vec x)=\textrm{exp}
[2\pi i(k_1x^1+k_2x^2+k_3x^3+k_4x^4+k_5x^5)]
\label{set2}\\
&m_B^2=\sum_{i=1}^5\left(\frac{k_i}{R_i}\right)^2
\end{align}
so that, in terms of the ansatz (\ref{ansz}), correspond to the choice
$k_6 = 0$. We then find that there are two families of Kaluza-Klein
excitations for each 4d massless gauge boson, and that KK modes
enter in one family or the other depending on whether they have KK
momentum along the fiber coordinate $x^6$ or not.
The spectrum of KK modes which are not excited along $x^6$,
given by the wavefunctions (\ref{set2}), is the same than we would find in
an ordinary $T^5$.

\begin{figure}[!h]
\begin{center}
\includegraphics[width=9cm]{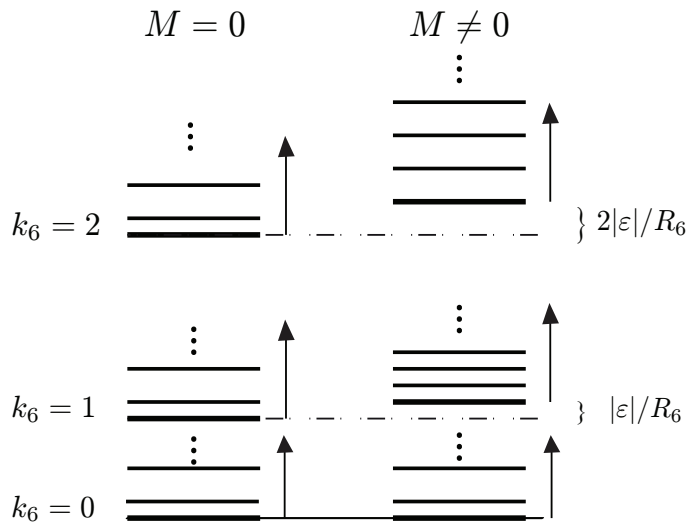}
\end{center}
\caption{\label{fig0} Spectra of massive gauge bosons in a fluxless
toroidal compactification (left) and in the fluxed example at hand (right), in the regime
$R_1R_2\gg M R_6$. The mass scale introduced by the fluxes is given by
$\varepsilon = M R_6/\pi R_1R_2$.}
\end{figure}

On the other hand, Kaluza-Klein modes excited along $x^6$, given by
the wavefunctions (\ref{set1}), present an interesting Landau degeneracy.
For each energy level there are exactly $(k_6M)^2(n+1)$ degenerate
modes, labeled by the triplet $(k,\delta_1,\delta_4)$. We have represented
in figure \ref{fig0} the resulting spectrum of particles associated to the gauge
boson, in the regime $R_1R_2\gg M R_6$, and we have compared with the
spectrum resulting in the fluxless case. As discussed in Section \ref{subsec:twisted},
in this regime the KK excitations along the base are much lighter than the
excitations along the fiber, and the mass scale induced by the fluxes is much
smaller than any KK scale. Hence, in analogy with standard type IIB flux
compactifications with large volumes and diluted fluxes, the effect of the flux
can be understood as a perturbation from the fluxless toroidal setup.

\begin{figure}[!h]
\begin{center}
\includegraphics[width=14cm]{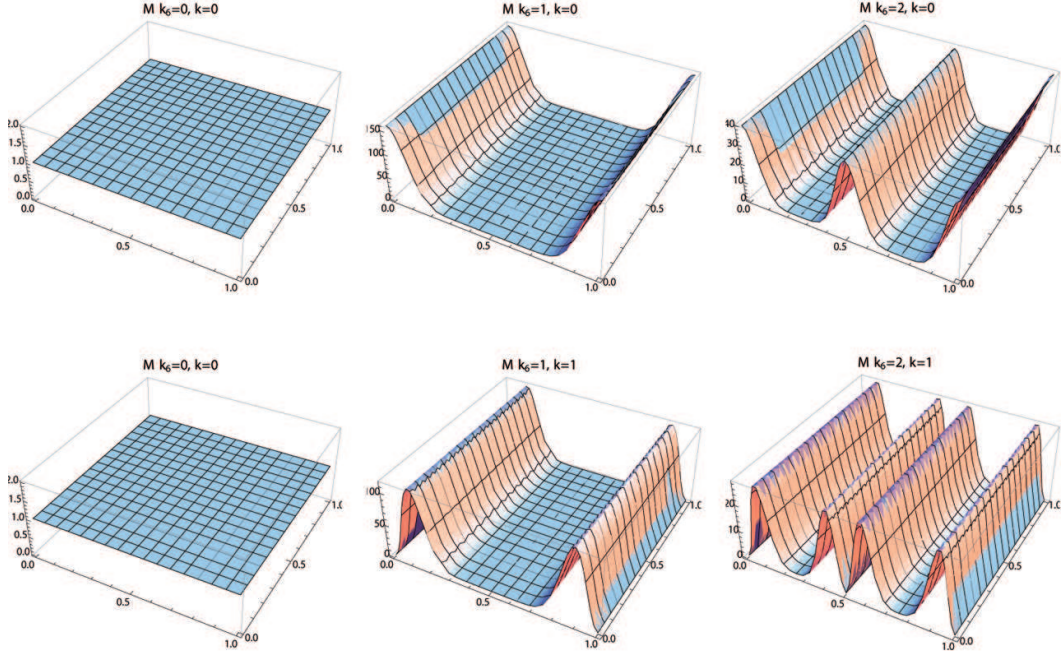}
\end{center}
\caption{\label{fig1} $|B^{(k,\delta_1,\delta_4)}_{n,k_3,k_6}|^2$
for $k=0,1$, $k_6M=0,1,2$, $n=k$ and arbitrary $\delta_1,
\delta_4,$ and $k_3$, in the plane $x^i=0, \ \ i=3\ldots 6$. The
normalization has been left unfixed.}
\end{figure}

Note that, even if we consider diluted fluxes, there are some qualitative differences
in the KK open string spectrum with respect to the fluxless case. In particular,
for $k_6\neq 0$ the masses of all the excitations along the base $B_4$ scale linearly
with respect to their KK quantum numbers, whereas in the toroidal case these scale
quadratically. In addition, the wavefunctions $|B^{(k,\delta_1,\delta_4)}_{n,k_3,k_6}|^2$
have a non-constant profile only along two directions, $x^1$ and $x^4$, as depicted in figure
\ref{fig1} for the first energy levels, reflecting the localization (independently of the warping)
of these Kaluza-Klein modes along those directions.  Note that the localization
of Kaluza-Klein excitations may affect in an interesting way the effective supergravity
description, leading to suppressions in the couplings of these modes to the low energy
effective theory.

Interestingly, the family of wavefunctions (\ref{set1}) can
be easily understood in terms of ordinary theta functions as follows.
First note that for $n=0$ and $k_6M >0$ we have
\begin{multline}
B^{(0,\delta_1,\delta_4)}_{0,k_3,k_6}= \left(\frac{2 |k_6M|R_5}
{R_1\ \textrm{Vol}_{\cam_6}}\right)^{1/2}
\vartheta\left[{-\frac{\delta_1}{k_6M} \atop 0}\right]\left(
k_6M \tilde z_1;\ k_6M\tilde \tau_1\right)\
\vartheta\left[{-\frac{\delta_4}{k_6M} \atop 0}\right]\left(k_6M \tilde z_2;\ k_6M \tilde \tau_2\right) \\
\times \textrm{exp}\left[i\pi k_6M\left(\frac{\tilde z_1\textrm{Im }\tilde z_1}{\textrm{Im }\tilde \tau_1}+
\frac{\tilde z_2\textrm{Im }\tilde z_2}{\textrm{Im }\tilde \tau_2}\right)\right]
\textrm{exp}\left[2\pi i(k_6x^6+k_3x^3)\right]
\label{thetawave}
\end{multline}
where we have defined a non-standard complex structure
\beq
\begin{array}{ccc}
\tilde z_1=x^1+\tilde \tau_1x^2& \qquad  & \tilde \tau_1=iR_2/R_1\\
\tilde z_2=x^4+\tilde \tau_2x^5  & \qquad & \tilde \tau_2=iR_5/R_4
\end{array}
\label{zeff}
\eeq
The higher energy levels corresponding to $n>0$ can then be built by acting
with the following raising operators
\begin{equation}
a^\dagger_1\equiv \hat\partial_1-i\hat\partial_2 \qquad a^\dagger_2\equiv \hat\partial_4-i\hat\partial_5
\label{raising}
\end{equation}
which act on the wavefunctions (\ref{set1}) as
\begin{align}
&a^\dagger_1B^{(k,\delta_1,\delta_4)}_{n,k_3,k_6}=
i\sqrt{\frac{k_6 M(n-k+1)}{\pi R_1R_2}}B^{(k,\delta_1,\delta_4)}_{n+1,k_3,k_6} \\
&a^\dagger_2B^{(k,\delta_1,\delta_4)}_{n,k_3,k_6}=
i\sqrt{\frac{k_6 M(k+1)}{\pi R_1R_2}}B^{(k+1,\delta_1,\delta_4)}_{n+1,k_3,k_6}
\end{align}
Similarly, for $k_6M < 0$ we should complex conjugate $\tilde z_k, \tilde \tau_k$ in
(\ref{thetawave}) and $a^\dagger_k$ in (\ref{raising}).

Note that the kind of wavefunctions (\ref{thetawave}) are precisely those arising
from open string zero modes charged under a constant $U(1)$ field strength
$F_2$ in toroidal magnetized compactifications \cite{yukawa}. This was indeed
expected, as nilmanifolds $\G\backslash\ch_{2p+1}$ based on the
Heisenberg manifold are standard examples of $S^1\simeq U(1)$ bundles,
and so both kind of wavefunctions can be understood mathematically
in terms of sections of the same vector bundle. It is amusing, however, to note
that the physical origin of the bundle geometry is quite different for these two
cases. Indeed, while in \cite{yukawa} the bundle arises from an open string flux
$F_2$ and the $U(1)$ fiber is not a physical dimension, in the present case
the bundle geometry is sourced entirely form closed string fluxes,
and all the coordinates of the fibration correspond to the background geometry.
This multiple interpretation of the wavefunctions (\ref{thetawave}) could presumably
be understood as a particular case of open/closed string duality, where the closed
string background (\ref{bg1}) is dual to a background of magnetized D9-branes.
More precisely, one can build a dictionary between both classes of backgrounds
as
\begin{center}
\begin{tabular}{ccc}
\underline{closed string} & & \underline{open string}\\
$e^6$ & $\leftrightarrow$ & $A$\\
$x^6$ & $\leftrightarrow$ & $\Lambda$\\
$F_3^{\text{cl}}$ & $\leftrightarrow$ & $\omega_3$
\end{tabular}
\end{center}
where $F_3 = F_3^{\text{cl}} + \om_3$, $\omega_3$ is the Chern-Simons 3-form
for the open string gauge bundle and $\Lambda$ the gauge transformation
parameter.

To finish our discussion let us comment on the uniqueness
of the above solutions. Note in particular that the change of variables in
(\ref{change}) is not unique, and one can check that taking different choices
for $\dot{x}^6$ leads to wavefunctions that are localized along different directions.
Again, this fact is not totally unexpected, since similar effects occur in the context
of magnetized D-branes in toroidal compactifications \cite{yukawa}.
Let us then consider the following change of coordinates
\begin{equation}
{\dot x}^6\equiv x^6 +
\frac{M}{2}(\epsilon_ax^1x^2+\epsilon_bx^4x^5)
\label{xgen}
\end{equation}
with $\epsilon_a, \ \epsilon_b=\pm 1$. From a group theoretical point
of view, this choice of signs are nothing but the four possible
{manifold polarizations}\footnote{Not to be confused
with the gauge boson polarization to be discussed below.} of the
5-dimensional Heisenberg group $\ch_5$. Proceeding as we did in
the previous sections, we obtain the following set of wavefunctions
\begin{multline}
B^{(k,\delta_a,\delta_b)_{\epsilon_a,\epsilon_b}}_{n,k_3,k_6}= \cn_{\eps_a\eps_b}\sum_{s_a, s_b\in\mathbb{Z}}\psi_{n-k}\left(\frac{\dot{x}^a}{\sqrt{2}}\right)\ \psi_{k}\left(\frac{\dot{x}^b}{\sqrt{2}}\right) \, e^{2\pi i \left((\delta_a+k_6Ms_a) x^a+k_3 x^3+(\delta_b+k_6Ms_b) x^b+k_6\dot
x^6\right)} \\
\cn_{\eps_a\eps_b} \, =\,  \left(\frac{2\pi |k_6M|}{\textrm{Vol}_{\cam_6}}\frac{R_1R_2}{R_aR_b}\right)^{1/2}
\quad \quad \quad \quad \quad \quad \quad \quad \quad \quad
\label{generic}
\end{multline}
with
\begin{align}
\dot{x}^a&\equiv\begin{cases}\frac{2}{R_a}
a^{-1/2}(\delta_a+k_6M(x^2+s_a)) & \textrm{for } \epsilon_a=+1\\
\frac{2}{R_a} a^{-1/2}(\delta_a-k_6M(x^1-s_a)) & \textrm{for }
\epsilon_a=-1
\end{cases}\\
\dot{x}^b&\equiv\begin{cases}\frac{2}{R_b}
a^{-1/2}(\delta_b+k_6M(x^5+s_b))  & \textrm{for } \epsilon_b=+1 \\
\frac{2}{R_b} a^{-1/2}(\delta_b-k_6M(x^4-s_b)) & \textrm{for }
\epsilon_b=-1
\end{cases}
\end{align}
\begin{equation}
x^a\equiv \begin{cases}x^1 &\textrm{for } \epsilon_a=+1\\
x^2 &\textrm{for } \epsilon_a=-1\end{cases} \qquad x^b\equiv \begin{cases}x^4 &\textrm{for } \epsilon_b=+1\\
x^5 &\textrm{for } \epsilon_b=-1\end{cases}
\label{xab}
\end{equation}
and an analogous definition to (\ref{xab}) for $R_{a,b}$. Note that
$\epsilon_a=+1\ (-1)$ leads to wavefunctions localized in $x^1\ (x^2)$,
whereas $\epsilon_b=+1\ (-1)$ leads to wavefunctions localized in
$x^4\ (x^5)$. Each choice of polarization, however, leads to a complete set
of wavefunctions. Therefore any wavefunction within a given polarization
can be expressed as a linear combination of wavefunctions in a different
polarization through a discrete Fourier transform \cite{yukawa}. See Appendix
\ref{kirillov}  for a more general, formal presentation of manifold polarizations
for the case of nilmanifolds.


\subsection{Laplace-Beltrami operators for group manifolds}
\label{gener}

When finding solutions to the equation (\ref{lap}) in our
previous example, a key ingredient was to impose
$\G$-invariance via the ansatz (\ref{ansz}). While such ansatz is
easy to guess either from the identifications (\ref{reli}) or from the
magnetized D-brane literature, it is a priori not obvious how to
formulate such an ansatz for arbitrary twisted tori.

In the following
we would like to systematize the procedure above and generalize it
to solve the Laplace-Beltrami equation in arbitrary manifolds of the form
$\cam_6 = \G\backslash G$. As we will see, the method described
below not only leads automatically to the two families of KK towers
(\ref{set1}) and (\ref{set2}) that we found for $\cam_6 = \G\backslash(\IR \times \ch_5)$,
but also gives a simple group theoretical understanding of their
existence in terms of the irreducible representations of $\IR \times \ch_5$.

In fact, the relation between families of KK modes on $\cam_6 = \G\backslash G$
and irreducible representations of a group $G$ can be traced back to the
mathematical literature that analyzes the spectrum of Laplace-Beltrami
operators in group manifolds.
Particularly useful for our purposes will be the tools developed in the context
of non-commutative harmonic analysis (see e.g. \cite{taylor,heis}), a field
aiming to extend the results of Fourier analysis to non-commutative
topological groups.

In order to motivate this approach let us first consider the Laplace eigenvalue
problem in the Abelian case $\cam_n = \IZ^n \backslash \IR^n = T^n$.
Here the twisted derivative operators $\hat\partial_m$ are nothing but
ordinary derivatives, so (\ref{lap}) reduces to
\begin{equation}
\partial_{x^i}\partial^{x^i}B=-m_B^2B
\label{laptorito}
\end{equation}
and the underlying algebra of isometries is Abelian.
A standard approach to solve this Laplace equation is to apply Fourier analysis.
More precisely, we can apply the Fourier transform
\begin{equation}
\hat{f}_{\vec \omega}=\int_{\mathbb{R}^n}B(\vec
x)e^{i\vec\omega\cdot\vec x}d\vec x
\label{fourierTn}
\end{equation}
to rewrite (\ref{laptorito}) in the dual space of momenta. We then obtain
\begin{equation}
\int_{\IR^n} |\vec \omega|^2\hat{f}_{\vec \omega}\,=\, \int_{\IR^n}m_B^2\hat{f}_{\vec \omega}
\end{equation}
which easily gives $\hat{f}_{\vec \omega} = \d(\vec \omega - \vec \omega_0)$
and $|\vec \omega_0|^2 = m_B^2$.  Hence, the eigenfunctions of the Laplace
operator correspond to Kaluza-Klein excitations with constant momentum of
norm $m_B$. Applying the inverse Fourier transform we find that these are
given by $B_{\vec \omega}(\vec x)=e^{-i\vec \omega_0\cdot\vec x}$. The
eigenfunctions of $\Delta$ are then nothing but the irreducible unitary representations
$e^{i\vec\omega\cdot\vec x}$ of the group $\IR^n$, which are also the ``coefficients"
entering the Fourier transform (\ref{fourierTn}). Finally, imposing invariance
under the compactification lattice $\G = \IZ^n$ restricts $\vec \omega$ to the
dual sublattice $2\pi \IZ$.

So one interesting observation that we can extract from this example is that
the irreducible unitary representations  $\pi_{\vec \omega}(\vec x) =
e^{i\vec\omega\cdot\vec x}$ of the Abelian group $G= \IR^n$ correspond
to the eigenfunctions of the Laplace operator. In particular, those which are
invariant under the subgroup $\G = \IZ^n$ are well-defined in the
compact quotient $\G\backslash G$, and so describe the KK wavefunctions
of $T^n$.

Naively, we would expect that some sort of analogous statement can be made
for $G$ a non-Abelian group. Again, a good starting point is to consider the
non-commutative version of (\ref{fourierTn}),\footnote{For a recent application
of this Fourier transform in a different physical context see \cite{pioline}.}
which reads \cite{taylor,heis}
\begin{equation}
\hat f_{\vec \omega}\, \varphi(\vec s)=\int_{G}B(g)\pi_{\vec
\omega}(g)\varphi(\vec s)dg
\label{fourier}
\end{equation}
with $\pi_{\vec \omega}(g)$ a complete set of
inequivalent irreducible unitary representations of $G$.
An important difference with respect to the Abelian case is
that the irreducible representations $\pi_{\vec \omega}(g)$
are no longer simple functions, but rather operators acting on
a Hilbert space of functions, $\varphi(\vec s)\in L^2(\mathbb{R}^{p(\pi)})$
with $p(\pi)\in \mathbb{N}$, and so is $\hat f_{\vec \omega}$. Remarkably,
the set $\pi_{\vec \omega}$ can be computed systematically by means of
the so-called orbit method, mainly developed by A.~Kirillov \cite{kirillovv},
and which we briefly summarize in Appendix \ref{kirillov}.

In principle, one could follow the standard strategy of the Abelian case and
make use of (\ref{fourier}) to write down eq.(\ref{lap}) in the space of momenta,
and then apply the inverse Fourier transform to obtain our wavefunction $B$.
An alternative approach, which we will adopt here, is to start with an educated
ansatz for $\G$-invariant wavefunctions, based on the close relation between
Laplace-Beltrami eigenfunctions and unitary irreps of $G$.

Indeed, consider a complex valued function $B_{\vec \omega} :G \raw \IC$,
defined as
\beq
B^{\vphi, \psi}_{\vec \omega} (g) \, =\, \left( \pi_{\vec \omega}(g) \vphi , \psi\right)\,
\equiv\, \int_{\mathbb{R}^{p(\pi)}} \bar \psi(\vec s) \cdot [\pi_{\vec \omega}(g)\varphi(\vec s)]\,
d\vec s
\eeq
where $(\, , \, )$ is the usual $L^2(\IC^{p(\pi)})$ norm.
If $\cl$ is a differential operator acting on the space of
wavefunctions $L^2(G)$ that can be expressed as a polynomial
$P(\{\mathfrak{t}_a\})$ of the algebra generators, then it is easy to see that
\beq
\cl \left( \pi_{\vec \omega}(g) \vphi , \psi\right)\, = \,
\left( \pi_{\vec \omega}(g) \pi_{\vec \omega}(\cl) \vphi , \psi\right)
\eeq
where $\pi_{\vec \omega}(\cl)$ is defined in the obvious way \cite{taylor,heis}.
Hence, finding eigenfunctions of $\cl$ reduces to finding eigenfunctions of
$\pi_{\vec \omega}(\cl)$ in the auxiliary space $L^2(\mathbb{R}^{p(\pi)})$,
since $\pi_{\vec \omega}(\cl) \vphi = \lam \vphi \Raw
\cl B^{\vphi, \psi}_{\vec \omega} = \lam B^{\vphi, \psi}_{\vec \omega}$.
Note that this is independent of our choice of $\psi$, which we can take
to be, e.g., a delta function $\delta(\vec s - \vec s_0)$. A suitable set of
eigenfunctions of $\cl$ is then given by
\beq
B^{\vphi_\a}_{\vec \omega} (g) \, =\, \pi_{\vec \omega}(g) \vphi_\a (\vec s_0)
\label{ansatzLB1}
\eeq
where $\vphi_\a$ is an eigenfunction of $\pi_{\vec \omega}(\cl)$.
In particular, this result applies to the Laplace-Beltrami operator $\Delta$, which
can be written as a quadratic form on $\{\mathfrak{t}_a\}$. Hence, (\ref{ansatzLB1})
provides a clear correspondence between unitary irreps of $G$ and families of
eigenfunctions of its Laplace-Beltrami operator.

As stressed before, we also need to impose that our wavefunctions are well-defined
in the quotient space $\cam = \G \backslash G$. A simple way to proceed is to
consider the sum
\beq
B_{\vec \omega} (g) \, =\, \sum_{\g \in \G} \pi_{\vec \omega}(\g g) \vphi (\vec s_0)
\, \equiv\, \pi_{\vec \omega}^\G(g) \vphi (\vec s_0)
\label{ansatzLB2}
\eeq
keeping only the wavefunctions $B_{\vec \omega}$ belonging to $L^2(\cam)$.\footnote{
This procedure may present some subtleties. For instance, if
$\pi_{\vec \om}(\vec x) = e^{i \vec \om \cdot \vec x}$ and $\vec \om \in 2\pi \IZ$,
then the sum over $\G = \IZ^n$ does not converge. In those cases, one should rather
think of (\ref{ansatzLB2}) as a way of replacing $\pi_{\vec \om}$ with $\G$-invariant irreps
$\pi_{\vec \omega}^\G$ in (\ref{ansatzLB1}). We have followed this philosophy
in eqs.(\ref{rep1}) and (\ref{rep2}) below.} Again, if $\vphi$ is an eigenfunction of
$\pi_{\vec \omega}(\cl)$ then (\ref{ansatzLB2}) is automatically an eigenfunction of
$\Delta$. Alternatively, one may consider $\vphi$ an unknown function and the
expression (\ref{ansatzLB2}) an educated ansatz to be plugged into the
Laplace-Beltrami equation (\ref{lap}).

In order to illustrate how this ansatz works, let us again consider the $(2p+1)$
dimensional Heisenberg manifold $\mathcal{H}_{2p+1}$, discussed in Section
\ref{subsec:twisted}. The Stone-von Neumann theorem~\cite{taylor,heis} states that the
irreducible unitary representations for $\mathcal{H}_{2p+1}$ are given by two
inequivalent sets\footnote{See Appendix \ref{kirillov} for an
alternative derivation of this result.}
\begin{align}
&\pi_{k_z^\prime} (\vec X)\, u(\vec s)\,=\, e^{2\pi ik_z^\prime[z + \vec x\cdot\vec y/2 +
\vec y\cdot \vec s ]}\, u(\vec s+\vec x)\ & u(\vec s)&\in
L^2(\mathbb{R}^p)
\label{rep1}\\
&\pi_{\vec k_x^\prime,\vec k_y^\prime} (\vec X) \,=\,
e^{2\pi i(\vec k_x^\prime\cdot \vec x\ +\ \vec k_y^\prime\cdot\vec y)}& &
\label{rep2}
\end{align}
where we are taking the same parameterization $\vec X^t = (z, \vec x ^{\, t}, \vec y^{\, t})$
 of $\mathcal{H}_{2p+1}$ as in (\ref{heis}). Considering the cocompact subgroup
 $\G_{\ch_{2p+1}} = \{(\vec x,\vec y,z) = M (\vec{n}_x, \vec{n}_y, n_z) \in M\IZ^{2p+1}\}$,
 $M \in 2\IZ$, and the $\G_{\ch_{2p+1}}$-invariant representations $\pi^\G$ we obtain
 \begin{align}
&\pi_{k_z}^\G (\vec X)\, u(\vec s)= \hspace*{-.35cm}\sum_{\vec s_x, \vec s_y \in \IZ^p}
\hspace*{-.35cm} e^{2\pi i k_z
[z +  \frac{M}{2} \vec x \cdot \vec y + (\vec y + \vec s_y)\cdot (\vec s  + M\vec s_x)]}
\, u(\vec s+ M(\vec s_x + \vec x)) \quad \quad  \quad   k_z\in \mathbb{Z}
\label{rep1inv}\\
&\pi_{\vec k_x,\vec k_y}^\G (\vec X) = e^{2\pi i(\vec k_x\cdot \vec x\ +\ \vec k_y\cdot\vec y)}
\hspace*{7.25cm}
 \vec k_x, \vec k_y   \in \IZ^p
\label{rep2inv}
\end{align}
where as before we have normalized the generators of the algebra as
$\tilde{\mathfrak{t}}_\a = M\mathfrak{t}_\a$, and in addition we have relabeled
the unirreps as $\vec k_a = M \vec k_a^\prime$, $a = x, y, z$. An interesting
effect of considering the invariant unirreps $\pi^\G$ is that the allowed choices for
$\vec s \in \IR^p$ become discrete. Indeed, note that (\ref{rep1inv}) vanishes
unless $k_z \vec s \in \IZ^p$, and that if we impose the latter condition we no longer
need to sum over $\vec s_y$ to produce an invariant unirrep. Hence, we can identify
our set of $\G$-invariant unirreps producing our ansatz (\ref{ansatzLB2}) as
 \begin{align}
&\pi_{k_z}^\G (\vec X)\, \vphi_{\vec \d}= \hspace*{-.2cm} \sum_{\vec s_x \in \IZ^p}
 e^{2\pi i k_z \left(z +  \frac{M}{2} \vec x \cdot \vec y\right)}
e^{2\pi i \left(\vec y \cdot (\vec \d  + k_zM\vec s_x)\right)}
\vphi (\vec \d + k_zM(\vec s_x + \vec x)) \quad \quad   k_z\in \mathbb{Z}
\label{rep1invb}\\
&\pi_{\vec k_x,\vec k_y}^\G (\vec X) = e^{2\pi i(\vec k_x\cdot \vec x\ +\ \vec k_y\cdot\vec y)}
\hspace*{7.25cm}
 \vec k_x, \vec k_y   \in \IZ^p
\label{rep2invb}
\end{align}
where $\vphi(\vec s) = u(k_z^{-1} \vec s)$ and $\vec \d \in \IZ^p$. Note that because of the
sum over $\vec s_x$, for fixed $k_z$ there are only $|k_z M|^p$ independent choices of
$\vec \d$ that we can take. Moreover, all these choices can be related via a redefinition of
$\vec x$, so if we find a solution to the Laplace equation via the ansatz (\ref{rep1invb}) in
general we will have $|k_z M|^p$ independent solutions.

To be more concrete, let us go back to the twisted torus example of subsection \ref{vmu}.
Recall that there the internal geometry is given by $\cam_6 = \G \backslash (\IR \times \ch_5)
 = S^1 \times \G_{\ch_5}\backslash \ch_5$, and that  in (\ref{rep1}) and (\ref{rep2}) we
 should take $p=2$ and identify  $z\equiv x^6$, $\vec x\equiv(x^2,x^5)$ and
 $\vec y\equiv(x^1,x^4)$. The ansatz (\ref{ansatzLB2}) then amounts to take the invariant
 unirreps (\ref{rep1invb}) and (\ref{rep2invb}) with the same identifications, and tensored
  with the unitary irreps of $S^1\simeq U(1)$, given by $e^{2\pi i k_3x^3}$. More precisely
  we obtain
\begin{align}
&B^{(\delta_1,\delta_4)}_{k_3,k_6}(\vec x)\, =\,
\sum_{k_1,k_4} \vphi \left(k_1+k_6Mx^2, k_4+k_6Mx^5\right)
e^{2\pi i \left( k_1 x^1 + k_3 x^3 + k_4 x^4 + k_6\dot x^6\right)}
\label{nili1} \\
& \hspace*{5cm} k_i = \d_i + k_6 M s_i \quad \quad \quad n_i \in \IZ \nonumber\\
&B_{k_1,k_2,k_3,k_4,k_5}(\vec x)=\textrm{exp}[2\pi
i(k_1x^1+k_2x^2+k_3x^3+k_4x^4+k_5x^5)] \label{nili2}
\end{align}
with $\vphi(x,y)$ a function to be determined. Eq.(\ref{nili1}) is indeed
the ansatz considered in eq.(\ref{ansz}), while (\ref{nili2}) gives the set
of wavefunctions (\ref{set2}) obtained by inspection. Finally, plugging
(\ref{nili1}) into (\ref{lap}), directly leads to
$\vphi(x,y)=\psi_k(\mu_1x)\psi_{n-k}(\mu_2y)$, with
$\mu_{1}^2=4\pi k_6 R_{2}/R_{1}$ and $\mu_{2}^2=4\pi k_6 R_{5}/R_{4}$
reproducing the results of the previous section.

As promised, the ansatz (\ref{ansatzLB2}) gives a direct relation between
families of KK modes on $\cam_6 = \G\backslash G$ and invariant unirreps of $G$.
In this respect, note that the inequivalent unirreps of $G = {\rm exp\, } \mathfrak{g}$
can be extracted from its Lie algebra $\mathfrak{g}$, given by (\ref{torsion}).
Now, from the 4d effective theory point of view $\mathfrak{g}$ is nothing but
the  4d gauge algebra resulting from dimensional reduction of the
10d metric \cite{kaloper}.  Hence, we can establish a correspondence between
inequivalent unirreps of the 4d gauged isometry algebra and families of
eigenfunctions of the internal Laplace-Beltrami operator. Note also that $\mathfrak{g}$
is only part of the full 4d $\mathcal{N}=4$ gauged supergravity algebra, as there are
further gauge symmetries arising from dimensional reduction of the 10d
$p$-forms. As we will argue below, by making use of the global $SL(2)\times SO(6,6+n)$
 symmetry one should be able to extend such correspondence to the full 4d gauged
 algebra and the full set of massive modes of the untwisted D9-brane sector.


\subsection{Non-vanishing $\mu$-terms}
\label{conmu}

Let us now apply the ansatz (\ref{ansatzLB2}) to a more involved
background, namely the twisted torus compactification with  flux-generated $\mu$-terms
of subsection \ref{nvmu}. Again, the
wavefunctions for the 4d gauge boson are given by the eigenfunctions
of the Laplace-Beltrami operator $\Delta$, and more precisely by the
solutions to eq.(\ref{lap}), with the twisted derivatives given by
(\ref{hatted1}). As before, the first step of the ansatz is to find the set of
inequivalent unirreps of the Lie group $G$. This can be done via
 the orbit method, as shown in Appendix \ref{kirillov}. We then find
 four families of irreducible unitary representations
 associated to the Lie algebra defined by eq.(\ref{nil2}), given
 in eqs.(\ref{pi1})-(\ref{pi6}).

As a second step, we need to impose $\G$-invariance on these
unirreps. For this purpose it is useful to introduce the variables
\begin{equation}
\dot x^3=x^3-\frac{M_3}{2}x^1x^2 \quad \text{and} \quad
\dot x^6=x^6-\frac{M_6}{2}x^5x^1
\end{equation}
so that the action of $\G$, given by (\ref{reli}), now reads
\begin{equation}
x^1\to x^1+1 \quad \quad \dot x^3\to \dot x^3-M_3x^2
\quad \quad \dot x^6 \to \dot x^6 - M_6x^5
\label{lat}
\end{equation}
with all the other coordinates being periodic, $x^i\to x^i+1$ for
$i=2,4,5$, and $\dot x^i \to \dot x^i+1$ for $i=3,6$. Imposing
invariance of (\ref{pi1})-(\ref{pi6}) under (\ref{lat}) and plugging
the result into (\ref{lap}), leads to the following $6\times 4=24$ towers of
KK gauge boson wavefunctions:\\

\noindent \underline{\emph{Modes not excited along the fiber $\{\dot x^3, \dot x^6\}$}}\\

These are given by standard toroidal wavefunctions in the base
\begin{align}
B_{k_1,k_2,k_4,k_5}=e^{2\pi i(k_1 x^1+k_2x^2+k_4x^4+k_5x^5)}
\label{mod1}
\end{align}
with mass eigenvalue
\begin{equation}
m_B^2=\sum_{a=1,2,4,5}\left(\frac{k_a}{R_a}\right)^2\label{mod1m}
\end{equation}
In particular, this includes the massless gauge boson.\\

\noindent \underline{\emph{Modes excited along $\dot x^r$, with $r=3$ or $6$}}\\

Their wavefunction is given by
\beq
B^{(\delta)}_{k_r,k_4,k_{8-r},n}\, =\,
\cn \, \sum_{s\in\mathbb{Z}}\psi_n\left(\frac{\dot
x^1}{\sqrt{2}}\right) e^{2\pi i\left(k_r\dot
x^r+k_4x^4+k_{8-r}x^{8-r}+(\delta+sk_rM_r)x^{r-1}\right)}
\label{rr}
\eeq
with $\delta=0\ldots k_rM_r-1$ and where
$\varepsilon_\mu  \equiv {M_3R_3}/{2\pi R_1R_2}$ stands for the mass scale of the flux
\begin{equation}
\cn = \left(\frac{2\pi R_1}{\textrm{Vol}_{\cam_6}}\sqrt{\frac{|k_r\varepsilon_\mu|}{R_r}}\right)^{1/2}
\quad \quad
\dot x^1\equiv 2\pi R_1 \left(\frac{2|\varepsilon_\mu
k_r|}{R_r}\right)^{1/2}\left(x^1-s-\frac{\delta}{k_rM_r}\right)
\end{equation}
The corresponding mass eigenvalues are
\begin{equation}
m_B^2=\frac{|\varepsilon_\mu
k_r|}{R_r}(2n+1)\, +\sum_{a=r,4,8-r}\left(\frac{k_a}{R_a}\right)^2\label{rrm}
\end{equation}

\noindent \underline{\emph{Modes excited along both $\dot x^3$ and $\dot x^6$}}\\

The wavefunctions for these modes are
\beq
B^{(\delta_2,\delta_5)}_{k_3,k_4,k_6,n}=\cn\, \sum_{s\in\mathbb{Z}}\psi_n\left(\frac{\dot x^1}{\sqrt{2}}\right)
\, e^{2\pi i\left[k_4x^4+k_3\dot x^3+k_6\dot
x^6+(\delta_5+sk_6M_6)x^5+(\delta_2+sk_3M_3)x^2\right]}
\label{mod2}
\eeq
with
\begin{equation}
\cn = \left(\frac{2\pi R_1\sqrt{\Delta_{k_3,k_6}|\varepsilon_\mu|}}{\textrm{Vol}_{\cam_6}}\right)^{1/2}
\quad \quad
\dot x^1\equiv
2\pi(2\Delta_{k_3,k_6}|\varepsilon_\mu|)^{1/2}R_1\left(x^1-s-\frac{\delta_2}{k_3M_3}\right)
\end{equation}
\beq
\Delta_{k_3,k_6}^2\equiv\left(\frac{k_3}{R_3}\right)^2+\left(\frac{k_6}{R_6}\right)^2
\label{delta}
\eeq
and where $\delta_2,\delta_5\in \mathbb{Z}$
are related through the constraint $k_6\delta_2M_6=\delta_5k_3M_3$.
Finally, the mass eigenvalues are
\begin{equation}
m_B^2=\Delta_{k_3,k_6}^2+\left(\frac{k_4}{R_4}\right)^2+|\varepsilon_\mu|\Delta_{k_3,k_6}(2n+1)
\label{mod2m}
\end{equation}


\section{Scalar wavefunctions}\label{sec:scalars}

In this section we proceed with the computation of the wavefunctions for
the 4d scalar modes transforming in the adjoint representation of the gauge
group $G_{unbr}$. These modes arise from the term $b^m (x^\mu) \xi^m(x^i)$ in the
dimensional reduction (\ref{splitboson}) of the 10d gauge boson, so they can
be thought as Wilson line moduli of
the compactification plus their KK replicas. Note that the choice of expansion
(\ref{splitboson}) in terms of the left-invariant 1-forms $e^m$ indeed
simplifies the computation of the wavefunctions $\xi^m(x^i)$ which, just as the
previous gauge boson wavefunction $B(x^i)$, are invariant under the action of the
 subgroup $\G$ in $\cam_6 = \G \backslash G$.

 In fact, we will see that having
 computed the spectrum of $B(x^i)$'s, the computation of  $\xi^m(x^i)$'s reduces to a
 purely algebraic problem. This problem is easily solved in the case of our
 compactification with vanishing $\mu$-terms, since it basically amounts to
 diagonalize a $3 \times 3$ matrix with commuting entries. The case with
 non-vanishing $\mu$-term, on the other hand, turns out to be more
  involved, as the entries of this $3 \times 3$ matrix become
  non-commutative.\footnote{The fact  that we have to diagonalize a $3 \times 3$
  matrix instead of a general $6 \times 6$ mass matrix is due
  to the fact that the 4d vacua considered in this section are supersymmetric, and to the
  exact pairing between bosonic and fermionic wavefunction that this implies.
 See Appendix \ref{ap:N=0} for a non-supersymmetric example where bosonic and
 fermionic wavefunctions are no longer the same.}


\subsection{Vanishing $\mu$-terms}
\label{vanish}

As discussed in Section \ref{subsec:elliptic}, for elliptic fibrations of
the form (\ref{mansatz}) the internal profiles $\xi^p_{\Pi_2,B_4}$ of the 4d
scalars in the adjoint of $G_{unbr}$ are real functions satisfying eqs.(\ref{xi+}) and
(\ref{xi-}). These equations of motion can be summarized in matrix
notation as
\begin{equation}
\left[\mathbb{M}+m_b^2\ \mathbb{I}_{6}\right]\mathbb{V}=0
\label{eigenval}
\end{equation}
where
\beq
\mathbb{V}=
\begin{pmatrix}\xi^1\\
\xi^2\\ \xi^3\\ \xi^{*\, 1}\\ \xi^{*\, 2}\\ \xi^{*\, 3}
\end{pmatrix}
\quad \quad \quad
\begin{array}{c}
\xi^1\, \equiv\,  \xi_{B_4}^1+i\xi_{B_4}^4\\
\xi^2\, \equiv\,  \xi_{B_4}^2+i\xi_{B_4}^5\\
\xi^3\, \equiv\,  \xi_{\Pi_2}^3+i\xi_{\Pi_2}^6
\end{array}
\quad \quad \quad
\begin{array}{c}
\xi^{*\, 1}\, \equiv\,  \xi_{B_4}^1-i\xi_{B_4}^4\\
\xi^{*\, 2}\, \equiv\,  \xi_{B_4}^2-i\xi_{B_4}^5\\
\xi^{*\, 3}\, \equiv\,  \xi_{\Pi_2}^3-i\xi_{\Pi_2}^6
\end{array}
\label{standcom}
\eeq
and the matrix $\mathbb{M}$ has as entries differential operators whose general expression is given in Appendix \ref{ap:matrix}. For the type I vacuum of subsection \ref{vmu}, one
can check that $\mathbb{M}$ reduces to
\begin{equation}
\mathbb{M}\,=\,
\begin{pmatrix}
\hat\partial_m\hat\partial^m& -\varepsilon\hat\partial_6&0&0&0&0\\
\varepsilon\hat\partial_6&\hat\partial_m\hat\partial^m&0&0&0&0\\
0&0&\hat\partial_m\hat\partial^m&0&0&0\\
0&0&0&\hat\partial_m\hat\partial^m&-\varepsilon\hat\partial_6&0\\
0&0&0&\varepsilon\hat\partial_6&\hat\partial_m\hat\partial^m&0\\
0&0&0&0&0&\hat\partial_m\hat\partial^m
\end{pmatrix}
\label{system1}
\end{equation}
where as before $\varepsilon = M R_6/\pi R_1R_2$ is the flux scale.
Note that all the entries of the matrix $\mathbb{M}$ commute, and so
(\ref{eigenval}) can be treated as an ordinary eigenvalue problem. Moreover,
$\mathbb{M}$ is block diagonal, with no entries mixing holomorphic and
antiholomorphic states. This can be traced back to the fact that our compactification
manifold $\cam_6$ is complex, as required by $\mathcal{N}=1$ supersymmetry.
 Therefore, it is enough to solve for one of the $3\times 3$ blocks in (\ref{system1}).

In order to do so let us distinguish again between states which are excited along
the fiber coordinate $x^6$ and states which are not excited along it. For the latter
the wavefunction should not depend on $x^6$, and so they are annihilated by
$\hat \partial_6$. Therefore, for those states $\mathbb{M}$ is proportional to the
Laplace-Beltrami operator, whose eigenvalues were solved for in Section \ref{nili}.
It is then straightforward to verify that the wavefunctions associated with these modes
are given by the same functions $B_{k_1,k_2,k_3,k_4,k_5}$ defined in equation
(\ref{set2}). Similarly, the mass eigenvalues are
\begin{equation}
m_\xi^2=\sum_{i=a}^5\left(\frac{k_a}{R_a}\right)^2\nonumber
\end{equation}
so the wavefunction $B_{0,0,0,0,0} = const.$ corresponds to the six
real Wilson line moduli.

\begin{figure}[!h]
\begin{center}
\includegraphics[width=14cm]{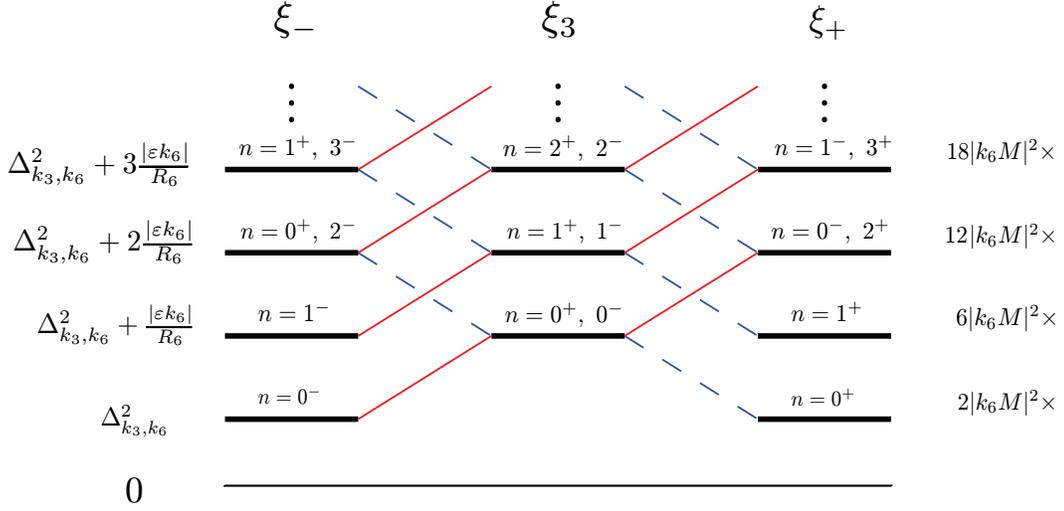}
\end{center}
\caption{\label{splitlevel} Mass spectra for the
complex scalar modes $\xi_3$ and $\xi_{\pm}$ excited along
the fiber with same momentum $|k_6|$ in the example with vanishing $\mu$-terms. Continuous red lines relate states
with same $n$ and $k_6<0$, whereas dashed blue lines relate states
with same $n$ and $k_6>0$. We have labeled the energy levels by
$n^{s_{k_6M}}$. The spectrum of gauge boson excitations
coincide with the one of $\xi_3$. The flux mass scale is given by
$\varepsilon=\frac{MR_6}{\pi R_1R_2}$, whereas $\Delta_{k_3,k_6}^2$
is defined in (\ref{delta}). We have also indicated the number of real scalars
at each energy level, for fixed $s_{k_6}$, $s_{k_3}$.}
\end{figure}

On the other hand, $\hat \partial_6$ does not act trivially on modes excited along
the fiber, as they depend on $x^6$. Note however that $\hat\partial_6$ belongs to
the center of the Lie algebra $\mathfrak{g}$ of our twisted torus
$\cam_6 = \G \backslash G$. Hence,
 $\hat \partial_6$ commutes with the Laplace-Beltrami operator
$\hat\partial_m\hat\partial^m$, and so they can be simultaneously diagonalized.
In fact, it turns out that the family of wavefunctions (\ref{set1}) obtained above are
not only eigenfunctions of $\hat\partial_m\hat\partial^m$ but also of $\hat \partial_6$, their eigenvalue for
the latter being $i k_6/R_6$. This allows to diagonalize the upper $3 \times 3$
block of (\ref{system1}) for the fiber KK modes as
\begin{equation}
(\xi_\pm)_{n,k_3,k_6}^{(k,\delta_1,\delta_4)}\equiv \begin{pmatrix}1\\ \pm i\\ 0\end{pmatrix}B_{n,k_3,k_6}^{(k,\delta_1,\delta_4)}
\label{neutpm}
\end{equation}
with mass eigenvalue
\begin{equation}
m^2_{\xi_{\pm}}\, = \, \frac{|\varepsilon k_6|}{R_6}
\left(n + 1 \mp s_{k_6M} \right) +\Delta^2_{k_3,k_6}
\end{equation}
where $s_{k_6M} = \text{sign}(k_6M)$ and $\Delta_{k_3,k_6}$ is given by (\ref{delta}).
The effect of the off-diagonal entries in (\ref{system1}) is then to shift up or down the mass eigenvalues with respect to the ones computed in Section \ref{nili} for the gauge bosons.
In figure \ref{splitlevel} we have represented the splitting of the Laplace-Beltrami energy
levels due to this mass shift effect.

The remaining eigenvector is
\begin{equation}
(\xi_3)_{n,k_3,k_6}^{(k,\delta_1,\delta_4)}\equiv\begin{pmatrix}0\\ 0\\ 1\end{pmatrix}B_{n,k_3,k_6}^{(k,\delta_1,\delta_4)}
\label{extrapol}
\end{equation}
with mass eigenvalue
\begin{equation}
m^2_{\xi_{3}}\, =\, \frac{|\varepsilon k_6|}{R_6} (n+1) + \Delta^2_{k_3,k_6}
\end{equation}
identical to the KK masses of the corresponding massive gauge boson.
In fact, the degrees of freedom coming from (\ref{extrapol}) should be seen
as the extra polarizations that massive gauge bosons have with respect to
massless ones.

Putting these results together with the spectrum of gauge bosons computed in Section \ref{nili}, and the fermionic spectrum (to be computed in next section), one can observe that the content of massive Kaluza-Klein replicas can be arranged into $\mathcal{N}=4$ vector multiplets, except for the levels $k_6\neq 0$, $n=0$, which only fit into ultrashort $\mathcal{N}=2$ hypermultiplets. See Section \ref{susyspect} for a more detailed discussion.


\subsection{Non-vanishing $\mu$-terms}
\label{nonvanish}

Let us now turn to the type I vacuum of subsection \ref{nvmu}, where
the background induces a non-vanishing mass
term for one of the chiral multiplets. The internal profiles of the
adjoint scalars must again satisfy the eigenvalue problem
(\ref{eigenval}), now with $\mathbb{M}$ given by
\begin{equation}
\mathbb{M}=
\begin{pmatrix}
\hat\partial_m\hat\partial^m&-\varepsilon_\mu\hat\partial_{z^3}&-\varepsilon_\mu\hat\partial_{z^2}&0&0&0\\
\varepsilon_\mu\hat\partial_{\bar z^3}&\hat\partial_m\hat\partial^m&\varepsilon_\mu\hat\partial_{z^1}&0&0&0\\
\varepsilon_\mu\hat\partial_{\bar z^2}&-\varepsilon_\mu\hat\partial_{\bar z^1}&\hat\partial_m\hat\partial^m-\varepsilon_\mu^2&0&0&0\\
0&0&0&\hat\partial_m\hat\partial^m&-\varepsilon_\mu\hat\partial_{\bar z^3}&-\varepsilon_\mu\hat\partial_{\bar z^2}\\
0&0&0&\varepsilon_\mu\hat\partial_{z^3}&\hat\partial_m\hat\partial^m&\varepsilon_\mu\hat\partial_{\bar z^1}\\
0&0&0&\varepsilon_\mu\hat\partial_{z^2}&-\varepsilon_\mu\hat\partial_{z^1}&\hat\partial_m\hat\partial^m-\varepsilon_\mu^2
\end{pmatrix}
\label{system2}
\end{equation}
with $\varepsilon_\mu\equiv M_3R_3/2\pi R_1R_2$ and where the complexification
\begin{equation}
\hat\partial_{z^k}\equiv \hat\partial_k-i\hat\partial_{k+3}
\label{comphat}
\end{equation}
is related to the standard choice of complex structure $z^k = x^k + i(R_{k+3}/R_k)x^{k+3}$.
Note that again the mass matrix is
block diagonal, as expected for a 4d SUSY vacuum. We will thus solve (\ref{eigenval})
for the upper block and obtain the other eigenfunctions by
complex conjugation.

An important qualitative difference with the case of vanishing
$\mu$-term (\ref{system1}), is that the entries of the matrix $\mathbb{M}$
are operators that no longer commute. However, using the following
commutation relations
\begin{align}
&[\hat{\p}_{z^1}, \hat{\p}_{z^2}] \, =\, [\hat{\p}_{\bar{z}^1}, \hat{\p}_{z^2}]\, =\, - \varepsilon_\mu \hat \p_{z^3}  \qquad \quad [\hat{\p}_{z^1}, \hat{\p}_{\bar{z}^2}] \, =\, [\hat{\p}_{\bar{z}^1}, \hat{\p}_{\bar{z}^2}]\,
 =\, - \varepsilon_\mu \hat \p_{\bar{z}^3}  \label{com}\\
&[\hat{\p}_m\hat{\p}^m, \hat{\p}_{z^2}]\, =\,-\varepsilon_\mu\hat{\p}_{z^3}(\hat{\p}_{z^1}+\hat{\p}_{\bar z^1})\qquad [\hat{\p}_m\hat{\p}^m, \hat{\p}_{\bar z^2}]\, =\,-\varepsilon_\mu\hat{\p}_{\bar z^3}(\hat{\p}_{z^1}+\hat{\p}_{\bar z^1}) \nonumber\\
&[\hat{\p}_m\hat{\p}^m, \hat{\p}_{z^1}]\, =\, [\hat{\p}_m\hat{\p}^m, \hat{\p}_{\bar z^1}]\, = \, \varepsilon_\mu\left(\hat{\p}_{\bar z^2}\hat{\p}_{z^3}+\hat{\p}_{z^2}\hat{\p}_{\bar z^3}\right) \nonumber
\end{align}
one can still diagonalize this matrix. Indeed, after some little effort one can
check that the above system have a complex eigenvector
\begin{equation}
\xi_3 \equiv
\begin{pmatrix}
\hat\partial_{\bar z^1}\\ \hat\partial_{\bar z^2}\\\hat\partial_{\bar z^3}
\end{pmatrix}
B(\vec x)
\label{muxi3}
\end{equation}
with mass eigenvalue $m^2_{\xi_3}=m_B^2$, and two complex eigenvectors
\begin{equation}
\xi_{\pm}\equiv
\begin{pmatrix}
\hat\partial_{z^3}\hat\partial_{\bar z^1}+m_{\xi_\pm}\hat\partial_{z^2}\\ \hat\partial_{z^3}\hat\partial_{\bar z^2}-m_{\xi_\pm}\hat\partial_{z^1}\\\hat\partial_{z^3}\hat\partial_{\bar z^3}+m_{\xi_\pm}^2
\end{pmatrix}B(\vec x)
 \label{muxipm}
\end{equation}
with mass eigenvalues
\begin{equation}
m_{\xi_\pm}^2-\varepsilon_\mu m_{\xi_\pm}-m_B^2=0\quad \Longrightarrow \quad m_{\xi_\pm}^2=\frac14\left(\varepsilon_\mu\pm\sqrt{\varepsilon_\mu^2+4m_B^2}\right)^2
\label{cuadr}
\end{equation}
Here $B(\vec x)$ is any of the gauge boson wavefunctions
(\ref{mod1}), (\ref{rr}) or (\ref{mod2}) with mass $m_B^2$ given
respectively by eqs.(\ref{mod1m}), (\ref{rrm}) and (\ref{mod2m}).
Hence, for each Kaluza-Klein boson with mass $m_B^2$, there is
one complex scalar with the same mass (eaten by the massive gauge
boson via a Higgs mechanism) and two complex scalars
whose masses are solutions to the quadratic equation in (\ref{cuadr}).

Note that for the lowest modes of the neutral gauge boson, $B=\textrm{const.}$,
the eigenvector parametrization (\ref{muxi3}) and (\ref{muxipm}) breaks down,
and does not constitute a good representation of the lightest modes for the
scalar fields. Instead, these states correspond to the constant eigenvectors
\begin{equation}
(\xi_\pm)_{0}\equiv
\begin{pmatrix}1\\ \pm i\\ 0\end{pmatrix} \times \text{const.}
 \qquad (\xi_3)_{0}\equiv\begin{pmatrix}0\\ 0\\ 1
\end{pmatrix}\times \text{const.}
\end{equation}
with masses $m_{\xi_\pm}^2=0$ and $m_{\xi_3}^2=\varepsilon_\mu^2$, respectively,
recovering in this way the low energy effective supergravity result \cite{geosoft}.
We will come back to this point in Section \ref{sugra}.


\section{Fermionic wavefunctions}\label{sec:fermions}

Let us now turn to the equation (\ref{dirac6duw}) describing the wavefunctions
of fermionic eigenmodes. As in the two previous sections, we will consider those
modes transforming in the adjoint representation of the unbroken gauge group
$G_{unbr}$, computing them explicitly for the two examples of
Section \ref{subsec:twisted}. In general, for compactifications preserving 4d $\cn=1$
supersymmetry, one expects all those modes belonging to the same supermultiplet
to share the same internal wavefunction. This should in particular apply to the two type
I vacua examples analyzed above, and so the eigenvalue problem for fermionic modes
should reduce to the one already solved in Sections \ref{sec:wgauge} and \ref{sec:scalars}.
We will see that this is indeed the case. Let us however stress that, as our approach treats
bosons and fermions independently, the method below could also be applied to
type I backgrounds where the flux breaks 4d supersymmetry and so wavefunctions
no longer match. An example of such $\cn=0$ flux vacuum is discussed in Appendix
\ref{ap:N=0}, where both classes of open string wavefunctions are computed.

Following the conventions of Appendix \ref{ap:ferm}, we can take our wavefunction as a
linear combination of the fermionic basis (\ref{basisMW}). Defining the vector
\beq
\Psi\, =\,
\left(
\begin{array}{c}
\psi^0 \\ \psi^ 1\\ \psi^2 \\ \psi^3
\end{array}
\right)
\label{fvector}
\eeq
it is then easy to see that (\ref{dirac6duw}) can be expressed as
\beq
i ({\bf D} + {\bf F}) \Psi\, =\,  m_\chi \Psi^*
\label{6db}
\eeq
where
\beq
{\bf D} \, =\,
\left(
\begin{array}{cccc}
0 & \hat{\p}_{{z}^1} & \hat{\p}_{{z}^2} & \hat{\p}_{{z}^3} \\
- \hat{\p}_{{z}^1} & 0 & - \hat{\p}_{\bar{z}^3} & \hat{\p}_{\bar{z}^2} \\
 - \hat{\p}_{{z}^2} & \hat{\p}_{\bar{z}^3} & 0 & - \hat{\p}_{\bar{z}^1} \\
 - \hat{\p}_{{z}^3} & - \hat{\p}_{\bar{z}^2} &  \hat{\p}_{\bar{z}^1} & 0
\end{array}
\right)
\eeq
and ${\bf F}$ contains the contribution of the term proportional to $\slashed{f}$
in eq.(\ref{dirac6duw}). In particular, we have that ${\bf F} = 0$ for vanishing
$\mu$-terms.

Eq.(\ref{6db}) implies that
\beq
({\bf D} + {\bf F})^* ({\bf D} + {\bf F})\Psi \, =\, |m_\chi|^2 \Psi
\label{6dsq}
\eeq
which is the fermionic equivalent to (\ref{eigenval}).


\subsection{Vanishing $\mu$-terms}

Let us then consider the internal Dirac equation in the vanishing $\mu$-term
background of subsection \ref{vmu}. First, given the choice of fibration and the
conventions of Appendix \ref{ap:ferm}, the splitting (\ref{split4+2}) reads
\beqa
\chi_{\Pi_2}\, =\, \psi^0 \, \chi_{---} + \psi^3\, \chi_{++-}\\
\chi_{B_4}\, =\, \psi^1 \, \chi_{-++} + \psi^2\, \chi_{+-+}
\eeqa
Second, recall that the contraction of indices in (\ref{dirac6d}) and (\ref{dirac6duw}) is performed with the internal gamma matrices in (\ref{commG}), which are essentially the 6d
matrices in (\ref{tilgamma}). Then, the contribution of the geometric flux to the Dirac equation
 (\ref{dirac6duw}) reads
\beq
\slashed{f}\, =\, \frac{MR_6}{2\pi} \left( \frac{ \tilde{\g}^{126}}{R_1R_2} + \frac{ \tilde{\g}^{456}}{R_4R_5}\right)\, =\, (2\pi)^{-1} \frac{M R_6}{R_1R_2} \left( \tilde{\g}^{126} + \tilde{\g}^{456}\right)
\eeq
where we have used the condition $R_1R_2 = R_4R_5$. In addition we have that
\beq
\label{SUSYf1}
\tilde{\g}^{126} + \tilde{\g}^{456}\, =\, -i\left( \sig_1 \otimes \sig_2  - \sig_2 \otimes \sig_1\right) \otimes \sig_2\, =\, \oh \left(\sig_{z} \otimes \sig_{\bar{z}} - \sig_{\bar{z}} \otimes \sig_{z}\right) \otimes \sig_2
\eeq
where
\beq
\sig_z\, =\,
\left(
\begin{array}{cc}
0 & 2 \\ 0 & 0
\end{array}
\right)\quad \quad \quad
\sig_{\bar{z}}\, =\,
\left(
\begin{array}{cc}
0 & 0 \\ 2 & 0
\end{array}
\right)
\eeq
Hence, we see that $\slashed{f} \chi_{\Pi_2} \equiv \slashed{f} P_+^{\Pi_2}\chi = 0$, and
so ${\bf F} = 0$, as expected from the fact that in this background no $\mu$-term is
generated for D9-brane moduli.

In order to solve the squared Dirac equation (\ref{6dsq}) we just need to compute the action of ${\bf D}^*{\bf D}$, which in general reads
{\footnotesize \beq
\begin{array}{c}
-{\bf D}^* {\bf D} \, =\, \hat{\p}_m\hat{\p}^m\, \mathbb{I}_4 \, + \\
\sum_a
\left(
\begin{array}{cccc}
\oh(f_{1\bar{1}}^a + f_{2\bar{2}}^a + f_{3\bar{3}}^a)
& f_{\bar{2}\bar{3}}^a & f_{\bar{3}\bar{1}}^a & f_{\bar{1}\bar{2}}^a \\
f_{{3}{2}}^a & \oh(f_{1\bar{1}}^a - f_{2\bar{2}}^a - f_{3\bar{3}}^a)
& f_{{2}\bar{1}}^a & f_{{3}\bar{1}}^a \\
f_{{1}{3}}^a & f_{{1}\bar{2}}^a
& \oh(-f_{1\bar{1}}^a + f_{2\bar{2}}^a - f_{3\bar{3}}^a) & f_{{3}\bar{2}}^a \\
f_{{2}{1}}^a & f_{{1}\bar{3}}^a & f_{{2}\bar{3}}^a
& \oh( - f_{1\bar{1}}^a - f_{2\bar{2}}^a +f_{2\bar{3}}^a) \\
\end{array}
\right) \hat{\p}_a
\end{array}\nonumber
\label{6dsqgen}
\eeq}
and that for the case at hand reduces to
\beq
-{\bf D}^* {\bf D} \, =\,
\left(
\begin{array}{cccc}
\hat{\p}_m\hat{\p}^m & 0 & 0 & 0 \\
0 & \hat{\p}_m\hat{\p}^m & -\varepsilon\hat{\p}_6 & 0 \\
0 & \varepsilon\hat{\p}_6 & \hat{\p}_m\hat{\p}^m & 0 \\
0 & 0 & 0 & \hat{\p}_m\hat{\p}^m
\end{array}
\right)
\label{6dsq1}
\eeq
This operator matrix is block diagonal, and it is easy to see that the upper
$1\times 1$ box, containing the Laplace-Beltrami operator, corresponds to the
eigenvalue problem for the 4d gaugino and its KK replicas, arising from $\psi_0$
in (\ref{fvector}). The lower $3\times 3$ block, on the other hand, corresponds to
the squared Dirac operator for the fermionic superpartners of the 4d scalars,
since it exactly matches the $3\times 3$ blocks of (\ref{system1}).
As the diagonalization of (\ref{6dsq1}) proceeds exactly as in Section \ref{vanish},
we will not repeat it here.


\subsection{Non-vanishing $\mu$-terms}
\label{scalarmu}

Let us now turn to the type I vacuum with $\mu$-term of subsection \ref{nvmu}.
An obvious difference with respect to the case without $\mu$-term is the contribution
of the background fluxes to the internal Dirac equation, which now reads
\beq
\slashed{f}\, =\, (2\pi)^{-1} \left( \frac{M_3R_3}{R_1R_2} \tilde{\g}^{123} + \frac{M_6R_6}{R_1R_5} \tilde{\g}^{156}\right)\, =\,\varepsilon_\mu \left( \tilde{\g}^{123} + \tilde{\g}^{156}\right)
\eeq
where we have again used the condition $M_3R_3/R_2 = M_6R_6/R_5$.
 Hence now we have that
\beq
\label{SUSYf2}
\tilde{\g}^{123} + \tilde{\g}^{156}\, =\, \sig_1 \otimes \sig_1 \otimes i\sig_2 - \sig_1 \otimes i\sig_2 \otimes \sig_1\, =\, \oh \sig_1 \otimes \left( \sig_{\bar{z}} \otimes \sig_{z} - \sig_{z} \otimes \sig_{\bar{z}}\right)
\eeq
and so $\slashed{f}$ does not kill $\chi_{\Pi_2}$, as expected from a compactification with
non-trivial $\mu$-terms. As a result, ${\bf F}$ does not vanish, and we have that
\beq
{\bf D} + {\bf F}\, =\,
\left(
\begin{array}{cccc}
0 & \hat{\p}_{{z}^1} & \hat{\p}_{{z}^2} & \hat{\p}_{{z}^3} \\
- \hat{\p}_{{z}^1} & 0 & - \hat{\p}_{\bar{z}^3} & \hat{\p}_{\bar{z}^2} \\
 - \hat{\p}_{{z}^2} & \hat{\p}_{\bar{z}^3} & 0 & - \hat{\p}_{\bar{z}^1} \\
 - \hat{\p}_{{z}^3} & -\hat{\p}_{\bar{z}^2} &  \hat{\p}_{\bar{z}^1} &  \varepsilon_\mu
\end{array}
\right)
\eeq
where $\varepsilon_\mu$ is now defined as in (\ref{system2}).
The r.h.s. of eq.(\ref{6dsq}) then reads
\beq
-({\bf D} + {\bf F})^* ({\bf D} + {\bf F}) \, =\,
\left(
\begin{array}{cccc}
\hat{\p}_m\hat{\p}^m & 0 & 0 & 0 \\
0 & \hat{\p}_m\hat{\p}^m & -\varepsilon_\mu \hat{\p}_{{z}^3} &
- \varepsilon_\mu \hat{\p}_{{z}^2} \\
0 & \varepsilon_\mu \hat{\p}_{\bar{z}^3} & \hat{\p}_m\hat{\p}^m
& \varepsilon_\mu\hat{\p}_{{z}^1} \\
0 & \varepsilon_\mu \hat{\p}_{\bar{z}^2} & -\varepsilon_\mu \hat{\p}_{\bar{z}^1} &
\hat{\p}_m\hat{\p}^m - \varepsilon_\mu^2
\end{array}
\right)
\label{6dsq2}
\eeq

Note that even in this more involved case, where ${\bf F} \neq 0$, the operator
matrix (\ref{6dsq2}) is block diagonal, as expected from 4d supersymmetry. Again, we can identify the upper block with the gaugino + KK modes
eigenvalue equation and the lower one with that for the 4d holomorphic scalars of
Section \ref{nonvanish}.\footnote{Had we chosen to write eq.(\ref{6dsq}) in terms of
$\Psi^*$, we would have obtained the lower $3 \times 3$ block of (\ref{system2}) instead
of the upper one.} Hence, the diagonalization of (\ref{6dsq2}) proceeds exactly
as for the bosonic sector of the theory.


\section{Matter field wavefunctions}\label{sec:wmatter}

Recall that in our general discussion of Section \ref{sec2}, we considered a gauge
subsector $U(N) \subset G_{gauge}$ and a $U(N)$ gauge field (\ref{splita}) whose
vev broke this gauge symmetry as $U(N) \raw \prod_i U(n_i) \equiv G_{unbr}$.
Just like in the more familiar fluxless case \cite{yukawa}, from this gauge breaking
pattern we obtain 4d fields transforming in the adjoint representation of each $U(n_i)$
factor, arising from the fluctuations contained in (\ref{splitboson}) and their fermionic
partners, as well as 4d fields in the bifundamental representations $({n}_i, \bar{n}_j)$,
arising from those  in (\ref{splitboson2}).\footnote{In a more complete discussion of
these type I flux vacua one should consider {\it i)} The full gauge sector
$G_{gauge} = Spin(32)/\IZ_2$, that could in
principle give rise to $(n_i, n_j)$, symmetric and antisymmetric representations of
$G_{unbr}$. {\it ii)} The spectrum arising from the inclusion of D5-branes. {\it iii)} The action
of the orbifold on the open string sector of the theory. None of these points will be essential
for the computations of this section, so in order to simplify our discussion we will not consider
them for the time being. A more detailed analysis will be carried on \cite{wip}.}
Up to now we have focused on those open string modes that
correspond to $U(n_i)$ adjoint representations or, otherwise said, on those wavefunctions
arising from (\ref{splitboson}). As we have seen, both the mass spectrum and the internal
wavefunctions of these modes are directly modified by the closed string background
flux $F_3$ and by the torsional metric of the compactification manifold $\cam_6$.

While this sector of adjoint representation modes already gives us a lot of information
on the interplay between open strings and background fluxes, for phenomenological
purposes it is clearly not the most interesting one. Indeed, from our recent experience
with D-brane model building (see, e.g., \cite{review3,reviews,denef08}) we know that
the bifundamental
modes arising from (\ref{splitboson2}) and their fermionic partners can in principle
reproduce the matter content of the MSSM from their lightest modes. Since these
light matter fields wavefunctions are crucial to compute effective theory quantities
like Yukawa couplings and soft terms, an essential question to be answered is how
they are affected in the presence of background fluxes. We will devote this section to
obtain the spectrum of bifundamental eigenmodes and eigenfunctions arising from the
expansion (\ref{splitboson2}), leaving the discussion in terms of 4d effective theory for
the next section.

As can be guessed from the magnetized D-brane literature, matter field
wavefunctions will not only be affected by closed string fluxes like $F_3$, but also
by the open string magnetic flux $F_2 = dA$ under which they are charged. As we
will see, the resulting wavefunction can be understood as an open string mode
charged under an effective closed + open string magnetic flux, with the relative
densities of both kind of fluxes entering the wavefunction in a rather interesting way.

Let us be more precise and let us consider the gauge symmetry breaking
$U(N)\to U(p_\alpha)\times U(p_\beta)$. In the twisted tori examples of
Section \ref{subsec:twisted}, this breaking will be induced by an open string flux
$F_2$ with indices on the $T^2\times T^2$ base of $\cam_6$ and of the
form\footnote{Note that the Bianchi identity, $dF_2=0$, does not allow to turn on
a magnetic flux along the fiber of $\cam_6$. As a result, for the examples at hand
$\int _{\cam_6} F_2^3 = 0$ and the resulting 4d spectrum will be non-chiral.
We nevertheless expect that the general results for matter field wavefunctions obtained
below remain valid for more involved, chiral flux vacua.}
\begin{equation}
F_2=2\pi \sum_{k=1,2}\begin{pmatrix}m^k_\alpha\mathbb{I}_{n_\alpha}& 0\\ 0&m^k_\beta\mathbb{I}_{n_\beta}\end{pmatrix}\ dx^k\wedge dx^{k+3}
\label{f2}
\end{equation}
with $n_\a + n_\b = N$ and
$p_\Lambda \equiv g.c.d.(n_\Lambda, n_\Lambda m_\Lambda^1, n_\Lambda
m_\Lambda^2, n_\Lambda m_\Lambda^1 m_\Lambda^2)$, $\Lambda = \a,\b$.
For simplicity, we will assume that $n_\Lambda, m_\Lambda^k \in \IZ$,
 which in the language of \cite{yukawa} corresponds to
a compactification with Abelian Wilson lines.

Given this particular choice of open string flux, we can proceed with our
dimensional reduction scheme of eqs.(\ref{splitboson}) and (\ref{splitboson2}).
Here $\langle B \rangle = \langle B^\Lambda \rangle U_\Lambda$, where
 $U_\Lambda$ is defined by (\ref{generators}) and $ \langle B^\Lambda \rangle$
 can be chosen to be
\begin{equation}
\langle B^\Lambda\rangle = \pi \sum_{k=1,2}m_\Lambda^k[x^kdx^{k+3}-x^{k+3}dx^{k}] \qquad \qquad \Lambda=\alpha,\beta
\end{equation}
where, again for simplicity, we have set all Wilson lines to zero. The gauge
transformation of a $U(N)$ adjoint field
 along a non-trivial closed path $\gamma$ is then
\begin{equation}
W \ \to \ \textrm{exp}\left[i\oint_\gamma \langle B^\Lambda\rangle U_\Lambda\right]\cdot W \cdot \textrm{exp}\left[- i\oint_\gamma \langle B^\Lambda\rangle U_\Lambda\right]
\label{wilson}
\end{equation}
so for a  $(\bar{n}_\a, {n}_\b)$ representation we have
\begin{align}
x^k\to x^k+1\ , \ \ldots \ : \qquad &W^{\alpha\beta}\to
e^{i\pi I^k_{\alpha\beta} x^{k+3}}W^{\alpha\beta}
\label{boundary}\\
x^{k+3}\to x^{k+3}+1\ , \ \ldots \ : \qquad &W^{\alpha\beta}\to
e^{-i\pi I^k_{\alpha\beta}x^k}W^{\alpha\beta} \nonumber
\end{align}
where the dots in the l.h.s. indicate a possible accompanying action on the fiber,
as dictated by the structure of our twisted torus $\cam_6$ (see e.g. (\ref{reli}) or (\ref{relii})), and $k=1,2$.
We have also defined $I^k_{\alpha\beta}=m^k_\alpha-m^k_\beta$, following
the conventions in \cite{yukawa}.

Finally, consistency with the equations of motion for $F_2$ requires that
$I^1_{\alpha\beta}I^2_{\alpha\beta}<0$. Let us in particular assume that
$I^2_{\alpha\beta}>0>I^1_{\alpha\beta}$, and introduce the quantities
\begin{equation}
\sigma_{\pm}=\frac{1}{2\pi}\left(\frac{I^2_{\alpha\beta}}{R_2R_5}\pm\frac{I^1_{\alpha\beta}}{R_1R_4}\right)
\end{equation}
so that $\sigma_-$ is the total density of flux $F_2$, whereas $\sigma_+$ is
proportional to the D-term induced by $F_2$ \cite{cim02}. One can check
that the SUSY conditions for $F_2$ amount to \cite{mmms99,raul01}
\begin{equation}
J^2 \wedge F_2  - \frac{1}{3}F_2^3  = J^2 \wedge F_2 = 0 \quad \Leftrightarrow \quad \sigma_+=0
\label{susy}
\end{equation}



\subsection{W bosons}

Let us start considering the 4d vector bosons $w_\mu$ in (\ref{splitboson2}),
transforming in the bifundamental representation of $G_{unbr} = U(n_\a) \times U(n_\b)$.
The internal profile of such open string mode is given by the scalar wavefunction
$W = W^{\a\b}e_{\a\b}$, with components $W^{\a\b}$ satisfying the equation of motion
\begin{equation}
\hat{D}^m\hat{D}_m W^{\alpha\beta} = -m_{W}^2 W^{\alpha\beta}
\label{gaugebos}
\end{equation}
with
\begin{equation}
\hat D_m W^{\alpha\beta}=\hat\partial_m W^{\alpha\beta}
-i(\langle B_m^\alpha\rangle-\langle B_m^\beta\rangle)W^{\alpha\beta}
\label{covariant}
\end{equation}
in agreement with our notation in eqs.(\ref{lapfin1}) and (\ref{lapfin2}).
Note that (\ref{gaugebos}) reduces to (\ref{lap}) if we set $\langle B^\Lambda \rangle = 0$,
so it is reasonable to expect a structure of KK modes similar to the one found in
Section \ref{sec:wgauge}.

In particular, for bosons in the adjoint representation we have seen that KK modes not
excited along the fiber do not feel the closed string fluxes at all, and so they present the
spectrum of a standard, fluxless toroidal compactification. The same result applies to W
bosons, in the sense that if $W^{\alpha\beta}$ does not depend on the coordinates of the
fiber $\hat{\p}$ becomes the standard partial derivative. As a result, (\ref{gaugebos}) becomes in this case the equation of motion for a $W$ boson in a magnetized $T^2\times T^2$, and
their spectrum follows from the results in \cite{yukawa}. Indeed, the lightest mode,
of mass $m^2_W=|\sigma_-|$, is given by
\begin{multline}
W^{\alpha\beta, \ (0,j_1,j_2)}_{0}(\tilde z_1', \tilde z_2') =\cn \prod_{k=1,2}
 e^{i\pi |I^k_{\alpha\beta}| \tilde z_k' \textrm{Im }\tilde z_k' / \textrm{Im }\tilde \tau_k'}\ \vartheta\left[{\frac{j_k}{|I^k_{\alpha\beta}|} \atop 0}\right](|I^k_{\alpha\beta}|\tilde z_k'\, {\Large ;} \, |I^k_{\alpha\beta}|\tilde \tau_k')\\
\cn \, =\,   \left(\frac{2}{\textrm{Vol}_{\cam_6}}\right)^{1/2}
\prod_{k=1,2}(|I^k_{\alpha\beta}|\textrm{Im }\tilde \tau_k')^{1/4} \qquad
\begin{array}{ccc}
\tilde{z}_1' =x^4+\tilde \tau_1' x^1 & \quad   & \tilde \tau_1'=iR_1/R_4 \\
\tilde z_2' =x^2+\tilde \tau_2' x^5  & \quad & \tilde \tau_2' =iR_5/R_2
\end{array}
\label{magtheta}
\end{multline}
where again we have defined a non-standard choice of complex structure.
As in \cite{yukawa}, a full KK tower can be constructed from (\ref{magtheta})
by applying appropriate raising operators.

On the other hand, for modes with non-vanishing Kaluza-Klein momentum along
the fiber, some subtleties arise. Let us for concreteness focus on the
example without flux-generated $\mu$-terms of subsection \ref{vmu}, whose gauge boson
spectrum was analyzed in Section \ref{nili}. There, we saw that in practice one can
trade the effect of a closed string flux on $\cam_6$ by an appropriate
magnetic flux $F_2^{\text{cl}} = 2\pi k_6M (dx^1 \wedge dx^2 + dx^4 \wedge dx^5)$
on the $T^2\times T^2$ base of the fibration. Since now our W boson also feels the genuine
open string flux
$(F_2^{\a\b})^\text{op} = 2\pi (I^1_{\a\b} dx^1 \wedge dx^4 + I^1_{\a\b} dx^2 \wedge dx^5)$,
it is natural to consider a total, effective magnetic flux defined as
$(F_2)_{\text{eff}} = F_2^\text{op} + F_2^\text{cl}$, which in this case reads
\begin{equation}
(F^{\alpha\beta}_2)_{\rm eff}\,=\,2\pi\, dx^1\wedge (k_6M dx^2+I^1_{\alpha\beta} dx^4)+2\pi\, (k_6Mdx^4+ I^2_{\alpha\beta}dx^2)\wedge dx^5
\label{totalflux}
\end{equation}
and to expect that our open string modes behave as particles charged under
$(F_2)_{\rm eff}$.\footnote{In the next section we will see that, in the T-dual setup
of type IIB flux compactifications, (\ref{totalflux}) translates into the gauge invariant
field strength $\cf = F_2 + B$ in the worldvolume of a D7-brane.}

In our example, the choice of $T^2\times T^2$ metric (\ref{bg12}), guarantees that both fluxes
$F_2^{\text{cl}}$ and $F_2^\text{op}$ are factorizable, in the sense that they can be
decomposed as  $F_2 = F_2|_{(T^2)_i} + F_2|_{(T^2)_j}$, with $(T^2)_i$ and
$(T^2)_j$ two orthogonal two-tori. In turn, this property implies that their associated lowest
KK mode can be written as a product of two theta functions, which in the case at hand
are given by (\ref{thetawave}) for $F_2^\text{cl}$ and (\ref{magtheta}) for $F_2^\text{op}$.
Note, however, that $(F_2)_{\text{eff}} = F_2^\text{op} + F_2^\text{cl}$ will in general
not be factorizable, and so we cannot expect the associated lowest KK mode to be
again a product of two Jacobi theta functions, but rather a Riemann $\vartheta$-function
\cite{yukawa}. Hence, for matter modes excited along the fiber we would expect
a lowest KK mode wavefunction of the form
\begin{multline}
W^{\alpha\beta,\ (j_1,j_2)}_{0,0,k_3,k_6}=
\mathcal{N}\ e^{i\pi(\mathbf{N}\cdot\vec z)\cdot (\textrm{Im }\mathbf{\Omega}_{\bf U})^{-1}\cdot \textrm{Im }\vec z}\ \vartheta\left[{\vec j \atop 0}\right](\mathbf{N}\cdot \vec z\, ;\, \mathbf{N}\cdot \mathbf{\Omega}_{\bf U})\  e^{2\pi i(k_3x^3+k_6x^6)}
\label{wavematter}
\end{multline}
where $\vec z\in \mathbb{C}^2$, and $\mathbf{N}$ and $\mathbf{\Omega}_{\bf U}$
are $2\times 2$ real and complex matrices, respectively. The definition of the
Riemann $\vartheta$-function and its properties can be found in Appendix \ref{riem}.

Indeed, one can check that the ansatz (\ref{wavematter}) is a solution of
(\ref{gaugebos}), (\ref{reli}), (\ref{boundary}), with mass eigenvalue $m^2_W=\Delta^2_{k_3,k_6}+\rho$, if we set\footnote{See \cite{akp09}
 for a similar set of wavefunctions recently derived in the context of magnetized
 D9-brane models without closed string background fluxes.}
\begin{equation}
\vec{z}=\begin{pmatrix}x^4\\ x^2\end{pmatrix}
+\mathbf{\Omega}_{\bf U}\cdot\begin{pmatrix}x^1\\ x^5\end{pmatrix} \quad \qquad
\mathbf{\Omega}_{\bf U}=\bar{\bf B}^{-1}\cdot \bar{\bf U}\cdot\bar{\bf B}\cdot\bf{\Omega}
\label{rotcpxst}
\end{equation}
and
\begin{align}
\mathbf{N}&=\begin{pmatrix}-I^1_{\alpha\beta}&-k_6M\\ k_6M&I^2_{\alpha\beta}\end{pmatrix}
& \mathbf{\Omega}&=i\begin{pmatrix}\frac{R_1}{R_4}& 0\\
0& \frac{R_5}{R_2}\end{pmatrix}\label{nomega}\\
\mathbf{B}&= \sqrt{2} \pi \begin{pmatrix}R_4& 0\\ 0&R_2\end{pmatrix}
& \mathbf{U}&=\begin{pmatrix}\textrm{cos }\phi &\textrm{sin }\phi\\
-\textrm{sin }\phi&\textrm{cos }\phi\end{pmatrix} \\
\mathcal{N}&=\left(\frac{2R_5\, {\rm det\, } {\bf N}}{R_2 \textrm{Vol}_{\cam_6}}
\right)^{1/2}
& & \vec j^{\, t} {\bf N} \in \IZ^2
\end{align}
where we have defined the effective flux density $\rho$ and the interpolation angle $\phi$
as\footnote{In terms of the open/closed string correspondence of Section \ref{nili},
we have the relation $\rho_{\rm cl} = g\, q\, m_\text{flux}$, with $g = R_6^{-1}$ a coupling
constant, $q = k_6$ an integer charge and $m_{\text{flux}} = \varepsilon$ the flux mass scale.}
\begin{equation}
\rho=\sqrt{\rho_{\text{op}}^2 + \rho_{\text{cl}}^2} = \sqrt{\sigma_-^2+
\left(\frac{k_6 \varepsilon}{R_6}\right)^2}\quad
 \qquad   \textrm{tan }\phi= \frac{\rho_{\text{cl}}}{\rho_{\text{op}}}=
 \frac{k_6\varepsilon}{R_6 \sigma_-}
\end{equation}
Note that ${\bf N}$ and ${\bf \Omega}_{\bf U}$ satisfy the convergence conditions
(\ref{conv}) that allow (\ref{wavematter}) to be well-defined, and that the degeneracy
of each level is given by $\textrm{det }\mathbf{N}$. Moreover, under the lattice transformations
(\ref{reli1})-(\ref{reli2}), $W^{\alpha\beta}$ transforms as dictated by (\ref{boundary}).
Finally, $\phi$ interpolates between the two choices of complex structures
(\ref{zeff}) and (\ref{magtheta}). In the limit $\phi \raw \pi/2$ we recover from
(\ref{rotcpxst}) $z_k = \tilde z_k$ and the factorized wavefunctions (\ref{thetawave}) for
neutral bosons, while in the limit $\phi \raw 0$ we obtain $z_k = \tilde z_k'$ and the
wavefunctions for charged bosons without KK momentum along the fiber, given by
(\ref{magtheta}).

In order to build the full tower of Kaluza-Klein excitations for the charged bosons, we
can systematically act on (\ref{wavematter}) with the holomorphic covariant derivatives
defined in Appendix \ref{riem}, which for the case at hand read
\begin{align}
a^\dagger_1& \equiv \hat D_2+i\ \textrm{sin }\phi\, \hat D_1-i\ \textrm{cos }\phi\, \hat D_5 \\
a^\dagger_2& \equiv \hat D_4-i\ \textrm{cos }\phi\, \hat D_1-i\ \textrm{sin }\phi\, \hat D_5
\end{align}
Indeed, note that the deformation angle $\phi$ is such that
\begin{equation}
\textrm{Im }{\bf \Omega}_{\bf U}^{-1}\cdot {\bf N}^t=
(\textrm{Im }{\bf \Omega}^{-1}_{\bf U})^t\cdot {\bf N}
\end{equation}
and as a result $[a_1^\dagger,a_2^\dagger]=0$.
This allows us to write a number operator
\begin{equation}
N=\hat{D}_m\hat{D}^m+\rho\
\end{equation}
and to build the full KK tower of states by applying
$(a_1^\dagger)^{n-k} (a_2^\dagger)^{k}$ to (\ref{wavematter}).
The resulting spectrum of masses is given by
\begin{equation}
m_W^2=\left(\frac{k_3}{R_3}\right)^2+\left(\frac{k_6}{R_6}\right)^2+(n+1)\rho
+ (2k-n)\frac{\sigma_+\sigma_-}{\rho}
\quad \qquad k, n \in \IZ
\label{masquasi}
\end{equation}
where $k=0, \ldots, n$.

Interestingly, for a vanishing D-term $\sigma_+=0$
the effective flux (\ref{totalflux}) factorizes, and so the Riemann $\vartheta$-function
in (\ref{wavematter}) becomes the product of two ordinary $\vartheta$-functions.
 In that case, the complete tower of wavefunctions is given by
\beqa
\label{setmagnetico}
& & W^{\alpha\beta, (k,\delta_1,\delta_2)}_{n,k_3,k_6} \, = \, \left(\frac{2\pi \rho R_1R_5}{\textrm{Vol}_{\cam_6}}\right)^{1/2}\sum_{s_1,s_2\in\mathbb{Z}}\psi_{n-k}\left(\frac{\dot
x^a}{\sqrt{2}}\right)\ \psi_{k}\left(\frac{\dot
x^b}{\sqrt{2}}\right) \times \\ \nonumber
& & \times \, \textrm{exp}\left[2\pi i \left((-\delta_2-k_6Ms_1+I^2_{\alpha\beta}s_2) x^2+k_3 x^3+(\delta_1-I^1_{\alpha\beta}s_1+k_6Ms_2)
x^4+k_6\dot{x}^6\right)\right]
\eeqa
with $\delta_{k}=0\ldots \textrm{g.c.d}(k_6M,I^k_{\alpha\beta})-1$.
Note that these wavefunctions have the same structure as in (\ref{set1}),
but now they localize along the tilted coordinates
\begin{align*}
&\dot x^a\equiv \frac{1}{R_2}\sqrt{\frac{4\pi}{\rho}}[\delta_2+k_6M(x^1+s_1)-I^2_{\alpha\beta}(x^5+s_2)] \\
&\dot x^b\equiv \frac{1}{R_4}\sqrt{\frac{4\pi}{\rho}}[\delta_1-I^1_{\alpha\beta}(x^1+s_1)+k_6M(x^5+s_2)]
\end{align*}

Alternatively, we could have derived all the above results by considering an extended
version of the algebra (\ref{torsion}), accounting for the D9-brane gauge generators,
and then making use of the representation theory methods described in Section
\ref{gener}. More precisely, we know that the algebra (\ref{torsion}) is part of the four
dimensional gauge algebra, corresponding to the gauge symmetries which arise from
dimensional reduction of the metric tensor. This, however, is not the full 4d gauge
algebra. In particular, in the presence of D9-branes, we should also include the generators
of the $U(1)$ gauge symmetries arising from such open string sector \cite{kaloper,algebras}
\begin{align}
[\hat D_m,\hat D_n] &=-f^p_{mn}\hat D_p+i F^\alpha_{mn}U_\alpha
 \label{extalgebra}\\
[\hat D_m,U_\alpha] &=[U_\alpha,U_\beta]=0
\nonumber
\end{align}
where the covariant twisted derivatives $\hat D_m$ are defined as in (\ref{covariant}) and
the Abelian gauge generators $U_\alpha$ by (\ref{generators}).

Given such extended algebra, it is straightforward to apply the methods of Section
\ref{gener} and Appendix \ref{kirillov} to compute its irreducible unitary
representations. For the case at hand, we find the following two sets of irreducible
unitary representations\footnote{For completeness, let us present the coadjoint action
of the algebra:
\begin{multline*}
K(G)(g_1,g_2,g_3,g_4,g_5,g_6,g_\Lambda)=(g_1,\ g_2,\ g_3+
\frac12(Mg_2x^4+I^2_{\alpha\beta}g_1x^2),\ \\
g_4-\frac12(g_2Mx^5-g_1I^1_{\alpha\beta}x^1),\ g_5,\ g_6
+\frac12(g_2Mx^1-g_1I^2_{\alpha\beta}x^5),\ g_\Lambda
-\frac12(g_2Mx^2+g_1I^1_{\alpha\beta}x^4))
\end{multline*}}
\begin{align}
&\pi_{k_1,k_2,k_3,k_4,k_5}=\prod_{r=1}^5\textrm{exp}[2\pi ik_rx^r]
 \label{generirr}\\
&\pi_{k_3,k_6,k_q}=\textrm{exp}\left[2\pi i\left(k_3x^3+k^{\alpha\beta}
\left(\textrm{Tr }\Lambda_{\alpha\beta}+I^2_{\alpha\beta}x^2\left(s_5+\frac{x^5}{2}\right)-
I^1_{\alpha\beta}x^4\left(s_1+\frac{x^1}{2}\right)\right)\right)\right.\nonumber\\
&\hspace*{.5cm}  \left. + k_6\left(x^6-Mx^2\left(s_1+\frac{x^1}{2}\right)+Mx^4\left(s_5+\frac{x^5}{2}\right)\right)\right]
 u(s_1+x^1,s_5+x^5)
\label{generirr2}
\end{align}
where $u(\vec s)\in L^2(\mathbb{R}^2)$ and $\textrm{Tr }\Lambda_{\alpha\beta}$ is the
trace of the gauge parameter (i.e. the unphysical coordinate in the $U(1)\simeq S^1$
D9-brane gauge fibers). Note that there is a new natural quantum number
$k^{\alpha\beta}$ which we did not find in our previous analysis.

Plugging now (\ref{generirr}) and (\ref{generirr2}) into (\ref{gaugebos}) we find that, indeed,
the unirreps (\ref{generirr2}) with $k^{\alpha\beta}=1$ lead to the matter wavefunction solutions
(\ref{wavematter}), as well as the more massive replicas produced by acting with
$a_1^\dag$, $a_2^\dag$. It would then seem that those unirreps in (\ref{generirr})
with $k^{\alpha\beta} \neq 1$ would not correspond to any physical modes, somehow
against the general philosophy of Section \ref{gener}. Let us try to argue that such modes
do exist.

First, let us consider the meaning of $k^{\a\b} \in \IZ$. If we set $k^{\a\b} = 0$, then from
(\ref{generirr}) and (\ref{generirr2}) we recover the vector boson adjoint modes (\ref{set1})
and (\ref{set2}) of Section \ref{sec:wgauge}. Indeed, (\ref{generirr}) directly correspond
to adjoint bosons without Kaluza-Klein momentum along the fiber, given by (\ref{set2}),
while (\ref{generirr2}) with $k^{\alpha\beta}=0$ correspond to the adjoint bosons with
Kaluza-Klein momentum along the fiber, given by (\ref{set1}). This is not a surprise
since, after all, the KK modes of Section \ref{sec:wgauge} arose from the irreducible
unitary representations of a subalgebra of (\ref{extalgebra}). What is perhaps more
illuminating is the fact that neither of the above subset of modes satisfy eq.(\ref{gaugebos}),
but rather the Laplace-Beltrami equation (\ref{lap}) for a neutral boson. This clearly
suggest that the internal differential equation that should be satisfied by an arbitrary
wavefunction arising from (\ref{generirr2}) is given by (\ref{gaugebos}), but with the
gauge covariant derivative defined as
\begin{equation}
\hat D_m W^{\alpha\beta}=\hat\partial_m W^{\alpha\beta}
-ik^{\a\b} (\langle B_m^\alpha\rangle-\langle B_m^\beta\rangle)W^{\alpha\beta}
\quad \quad k^{\a\b} \in \IZ
\label{covariantk}
\end{equation}
instead of (\ref{covariant}). In this sense, the massive modes corresponding to
$k^{\a\b} \neq 1$ should be understood as states with $U(1)$ charges
$k^{\a\b}(-n_\a, n_\b)$, which hence undergo the gauge transformations
\begin{align}
x^k\to x^k+1\ , \ \ldots \ : \qquad &W^{\alpha\beta}\to
e^{i\pi k^{\a\b} I^k_{\alpha\beta} x^{k+3}}W^{\alpha\beta}
\label{boundaryk}\\
x^{k+3}\to x^{k+3}+1\ , \ \ldots \ : \qquad &W^{\alpha\beta}\to
e^{-i\pi k^{\a\b} I^k_{\alpha\beta}x^k}W^{\alpha\beta} \nonumber
\end{align}
In particular, those states with $k^{\a\b} = -1$ correspond to the bifundamental
representation $(n_\a, \bar{n}_\b)$, whose wavefunction can be obtained by
complex conjugation of (\ref{wavematter}). Finally, those modes with
$|k^{\a\b}| > 1$ should be non-perturbative in nature, as they cannot arise from
the perturbative open string spectrum.

Note that the existence of these exotic non-perturbative charged vector states is
not only suggested by the spectrum of unirreps (\ref{generirr2}), but also required
by global symmetry arguments. Indeed, the 4d effective action of the
untwisted D9-brane sector is given by a $\mathcal{N}=4$ gauged supergravity, whose
global symmetry group is $SL(2)\times SO(6,6+N)$, and where $N=n_\alpha+n_\beta$ is the
number of extra vector multiplets coming from $D9$-brane gauge symmetries.
The spectrum of 4d particles is therefore naturally arranged in multiplets of this global
symmetry. In the particular example at hand, the global symmetry group includes a
 $\mathbb{Z}_2$ generator corresponding to the open/closed string correspondence
  discussed in Section \ref{nili}. This generator maps neutral bosons with Kaluza-Klein
  momentum $|k_6| > 1$ along the fiber to non-perturbative charged bosons
  with $U(1)$ charge $|k^{\a\b}| > 1$.
Making use of this global symmetry, we would expect the following masses for the
non-perturbative modes
\begin{equation}
m_W^2=\left(\frac{k_3}{R_3}\right)^2+\left(\frac{k_6}{R_6}\right)^2+(n+1)\rho_{\rm n.p.}
+ (2k -n) (k^{\a\b})^2 \frac{\sigma_+\sigma_-}{\rho_{\rm n.p.}}
\label{masquasi2}
\end{equation}
where
\begin{equation}
\rho_{\rm n.p.}=\sqrt{(k^{\alpha\beta}\sigma_-)^2+\left(\frac{k_6\varepsilon}{R_6}\right)^2}
\end{equation}

Finally, note that the algebra (\ref{extalgebra}) is still not the full four dimensional gauge
algebra. There are further gauge symmetries which arise from dimensional reduction of e.g.
the RR 2-form. In particular, the RR 3-form fluxes enter as structure constants of the
complete 4d algebra \cite{algebras}. We expect the irreducible unitary representations of
the complete four dimensional algebra to encode further untwisted states of the higher
dimensional string theory. We leave the exploration of these issues for future work.

Similarly, we can work out the wavefunctions for the charged bosons in the example
with non-vanishing $\mu$-term of subsection \ref{nvmu}. In that case, the total effective
magnetic flux is given by
\begin{equation}
(F_2^{\alpha\beta})_{\rm eff}=2\pi\left(I^1_{\alpha\beta}dx^1\wedge dx^4+I^2_{\alpha\beta}dx^2\wedge dx^5+k_3M_3dx^1\wedge dx^2+k_6M_6dx^1\wedge dx^5\right)
\label{totalflux2}
\end{equation}
Recall that for this vacuum we should distinguish between bosons with no Kaluza-Klein
momentum along the $S^1$ fibers, bosons with Kaluza-Klein momentum along only one
 of the fibers, and bosons with momentum along both of the fibers. We can easily adapt
 our previous discussion in this section to describe the wavefunctions for the first two types
 of bosons. Indeed, it is not difficult to see that the wavefunctions for charged bosons
 without momentum along any fiber are given again by eq.(\ref{magtheta}), whereas the
  wavefunctions for charged bosons with Kaluza-Klein momentum along only one of the
  fibers, e.g. $k_3\neq 0, \ k_6=0$, are given by eq.(\ref{wavematter}) with the same
  parameters $\vec z$, ${\bf \Omega}$ and ${\bf B}$, but with charge matrix, deformation
  angle and effective flux density given by
\begin{equation}
{\bf N}=\begin{pmatrix}-I^1_{\alpha\beta}& -k_3M_3\\ 0&I^2_{\alpha\beta}\end{pmatrix}
 \quad \quad \quad \textrm{tan }\phi=\frac{k_3\varepsilon_\mu}{R_3\sigma_-}
 \quad \quad \quad \rho=\sqrt{\sigma_-^2+\left(\frac{k_3\varepsilon_\mu}{R_3}\right)^2}
\end{equation}
The set of charged bosons excited along both fibers, with arbitrary $k_3$ and $k_6$,
is however a more involved sector, and in particular does not fall into the class of functions
 (\ref{wavematter}). This basically comes from the fact that $F_2$ has then all the
 possible components of the form $dx^1 \wedge dx^\alpha$.
 We refer to the reader to Appendix \ref{riem} for a more precise statement
 as well as a more detailed discussion of this point.


\subsection{Bifundamental scalars and fermions}
\label{bifund}

Just like for adjoint KK modes, the wavefunctions for bifundamental scalars and fermions
are easily worked out once that the 4d vector boson wavefunctions are known. Note
that these bifundamental KK modes are particularly interesting in semi-realistic
flux compactification vacua, since they correspond to the MSSM matter fields and
their KK replicas.

As before, let us start our analysis by considering the scalars in the bifundamental.
From eqs.(\ref{lapfin1})-(\ref{lapfin2}), we see that the corresponding mass matrix can
be obtained from the one for adjoint scalars analyzed in Appendix \ref{ap:matrix}, by
simply replacing twisted derivatives $\hat \p_m$ by covariant twisted derivatives
$\hat D_m$, and adding a term proportional to $\langle G_{mp}^{\alpha\beta}\rangle$.
In particular, for the example without $\mu$-terms discussed in subsection \ref{vmu} we
obtain the mass matrix
\beq
\mathbb{M}= \hat D_m\hat D^m \mathbb{I}_6 +
\begin{pmatrix}
-\frac{I^1_{\alpha\beta}}{\pi R_1R_4}& -\varepsilon \hat D_6&0&0&0&0\\
\varepsilon\hat D_6&-\frac{I^2_{\alpha\beta}}{\pi R_2R_5}&0&0&0&0\\
0&0&0&0&0&0\\
0&0&0&\frac{I^1_{\alpha\beta}}{\pi R_1R_4}&-\varepsilon\hat D_6&0\\
0&0&0&\varepsilon\hat D_6&\frac{I^2_{\alpha\beta}}{\pi R_2R_5}&0\\
0&0&0&0&0&0\end{pmatrix}
\label{chargscalar}
\eeq
where again $\varepsilon = MR_6/\pi R_1R_2$ and we are using the conventions of
(\ref{standcom}), with $\xi^p_{\Pi_2,B_4}$ now complex functions. Like in the case of
adjoint scalars, this matrix is block diagonal, and
so it will be enough to diagonalize the upper $3\times 3$ block. Note that both blocks
are related by an $\mathcal{N}=2$ R-symmetry transformation. However,
it is important to notice that, since we are dealing with charged modes, this transformation
takes $\sigma_+\to-\sigma_+$.

For the upper $3\times 3$ block we find the eigenvectors
\begin{equation}
\Phi_3^{\alpha\beta}\equiv\begin{pmatrix}0\\ 0\\ 1\end{pmatrix}W^{\alpha\beta}(\vec x)
\label{phi3}
\end{equation}
with mass eigenvalue $m^2_{\Phi_3}=m^2_W$, and with $W^{\alpha\beta}(\vec x)$
 the wavefunction of a charged boson. In addition, we find
\begin{equation}
\Phi_{\pm}^{\alpha\beta}\equiv\begin{pmatrix}\sigma_-\mp\rho\\ i\varepsilon \frac{k_6}{R_6}\\ 0\end{pmatrix}W^{\alpha\beta}(\vec x) \label{phipm}
\end{equation}
with mass eigenvalues $m^2_{\Phi_\pm}=m^2_W+\sigma_+\pm\rho$. The R-symmetry
conjugates $\bar \Phi_\pm$ has then mass eigenvalues
$m^2_{\bar{\Phi}_\pm}=m^2_W-\sigma_+\pm\rho$. Thus, $\Phi_\pm$ and $\bar \Phi_\pm$
lead to scalars with masses
\begin{align}
m^2_{\Phi_+}&=\left(\frac{k_3}{R_3}\right)^2+\left(\frac{k_6}{R_6}\right)^2+
(n+2)\rho+(2k-n)\sigma_+\sigma_-\rho^{-1} \\
m^2_{\Phi_-}&=\left(\frac{k_3}{R_3}\right)^2+\left(\frac{k_6}{R_6}\right)^2+n\rho+(2k-n)
\sigma_+\sigma_-\rho^{-1} \\
m^2_{\bar{\Phi}_+}&=\left(\frac{k_3}{R_3}\right)^2+\left(\frac{k_6}{R_6}\right)^2+(n+2)\rho
-(2k-n)\sigma_+\sigma_-\rho^{-1} \\
m^2_{\bar{\Phi}_-}&=\left(\frac{k_3}{R_3}\right)^2+\left(\frac{k_6}{R_6}\right)^2+n\rho
-(2k-n)\sigma_+\sigma_-\rho^{-1}
\end{align}
Then, as expected, for supersymmetry preserving open string fluxes we observe two
massless modes, whereas for generic  fluxes there is always a single tachyonic mode.

Similarly, if we analyze the charged scalars in the example with non-vanishing $\mu$-term
of subsection \ref{nvmu} we have to diagonalize the following mass matrix
\beq
\mathbb{M}= \hat D_m\hat D^m \mathbb{I}_6 +
\begin{pmatrix}
-\frac{I^1_{\alpha\beta}}{\pi R_1R_4}&-\varepsilon_\mu\hat D_{z^3}&
-\varepsilon_\mu\hat D_{z^2}&0&0&0\\
\varepsilon_\mu\hat D_{\bar z^3}&-\frac{I^2_{\alpha\beta}}{\pi R_2R_5}&
\varepsilon_\mu\hat D_{z^1}&0&0&0\\
\varepsilon_\mu\hat D_{\bar z^2}&-\varepsilon_\mu\hat D_{\bar z^1}&-|\varepsilon_\mu|^2&0&0&0\\
0&0&0&\frac{I^1_{\alpha\beta}}{\pi R_1R_4}&-\varepsilon_\mu\hat D_{\bar z^3}&
-\varepsilon_\mu\hat D_{\bar z^2}\\
0&0&0&\varepsilon_\mu\hat D_{z^3}&\frac{I^2_{\alpha\beta}}{\pi R_2R_5}&
\varepsilon_\mu\hat D_{\bar z^1}\\
0&0&0&\varepsilon_\mu\hat D_{z^2}&-\varepsilon_\mu\hat D_{z^1}&-|\varepsilon_\mu|^2
\end{pmatrix}
\label{chargscalarm}
\eeq
with $\varepsilon_\mu = M_3R_3/2\pi R_1R_2$. This is again a non-commutative
eigenvalue problem, that  can be solved with the aid of the commutation relations
\begin{align}
&[\hat{D}_{z^1}, \hat{D}_{z^2}] \, =\, [\hat{D}_{\bar{z}^1}, \hat{D}_{z^2}]
\, =\, - \varepsilon_\mu \hat D_{z^3} \label{comuta1} \\
& [\hat{D}_{z^1}, \hat{D}_{\bar{z}^2}] \, =\, [\hat{D}_{\bar{z}^1}, \hat{D}_{\bar{z}^2}]\, =
\, - \varepsilon_\mu \hat D_{\bar{z}^3}  \nonumber \\
&[\hat{D}_m\hat{D}^m, \hat{D}_{z^2}]\, =\,-\varepsilon_\mu\hat{D}_{z^3}(\hat{D}_{z^1}+
\hat{D}_{\bar z^1}) -\frac{I^2_{\alpha\beta}}{\pi R_2R_5}\hat{D}_{z^2} \nonumber\\
& [\hat{D}_m\hat{D}^m, \hat{D}_{\bar z^2}]\, =\,-\varepsilon_\mu\hat{D}_{\bar z^3}(\hat{D}_{z^1}
+\hat{D}_{\bar z^1}) + \frac{I^2_{\alpha\beta}}{\pi R_2R_5}\hat{D}_{\bar z^2}\nonumber\\
&[\hat{D}_m\hat{D}^m, \hat{D}_{z^1}]\, =\, \varepsilon_\mu\left(\hat{D}_{\bar z^2}\hat{D}_{z^3}
+\hat{D}_{z^2}\hat{D}_{\bar z^3}\right) -\frac{I^1_{\alpha\beta}}{\pi R_1R_4} \hat{D}_{z^1}
 \nonumber\\
& [\hat{D}_m\hat{D}^m, \hat{D}_{\bar z^1}]\, = \, \varepsilon_\mu\left(\hat{D}_{\bar z^2}\hat{D}_{z^3}
+\hat{D}_{z^2}\hat{D}_{\bar z^3}\right) +\frac{I^1_{\alpha\beta}}{\pi R_1R_4}
\hat{D}_{\bar z^1}\nonumber
\end{align}
in close analogy with what we did for the neutral scalars in Section \ref{scalarmu}.
For the upper $3\times 3$ block in $\mathbb{M}$, we obtain the eigenvectors
\begin{equation}
\Phi_3=\begin{pmatrix}\hat D_{\bar z^1}\\ \hat D_{\bar z^2}\\ \hat D_{\bar z^3}
\end{pmatrix}W^{\alpha\beta}(\vec x)
\label{eig1}
\end{equation}
with mass eigenvalue $m^2_{\Phi_3}=m_W^2$ and
\begin{equation}
\Phi_\pm=\begin{pmatrix}\hat D_{z^3}\hat D_{\bar z^1}+\tilde m_\pm\hat D_{z^2}\\ \hat D_{z^3}\hat
D_{\bar z^2}-\tilde m_\pm \hat D_{z^1}\\ \hat D_{z^3}\hat D_{\bar z^3}+\tilde m^2_
\pm-2\varepsilon_\mu^{-1}\tilde m_\pm\sigma_+\end{pmatrix}W^{\alpha\beta}(\vec x)
\label{eig2}
\end{equation}
with mass eigenvalues $m_{\Phi_{\pm}}^2=m_W^2+\varepsilon_\mu\tilde m_\pm$, and
$\tilde m_\pm$ given by the quadratic equation
\begin{equation}
-m_{W}^2\varepsilon_\mu+\varepsilon_\mu \tilde m^2_\pm-
\tilde m_\pm(\varepsilon_\mu^2\pm 2\sigma_+)
\pm \varepsilon_\mu\sigma_+=0
\end{equation}
so that
\begin{equation}
m^2_{\Phi_\pm}=\frac14\left(\varepsilon_\mu\pm\sqrt{\varepsilon_\mu^2+
4m^2_W+(\varepsilon_\mu^{-1}\sigma_+)^2}\right)^2
-(\varepsilon_\mu^{-1}\sigma_+)^2+\sigma_+
\end{equation}

Analogously, the lower $3\times 3$ block in $\mathbb{M}$ leads to the conjugate scalars
\begin{equation}
\bar \Phi_3=
\begin{pmatrix}\hat D_{z^1}\\ \hat D_{z^2}\\ \hat D_{z^3}\end{pmatrix}
W^{\alpha\beta}(\vec x) \qquad \bar \Phi_\pm=
\begin{pmatrix}
\hat D_{\bar z^3}\hat D_{z^1}+\tilde m_{\mp}\hat D_{\bar z^2}\\
\hat D_{\bar z^3}\hat D_{z^2}-\tilde m_{\mp} \hat D_{\bar z^1}\\
\hat D_{\bar z^3}\hat D_{z^3}+\tilde m_{\mp}^2+2\varepsilon_\mu^{-1}\tilde m_{\mp}\sigma_+
\end{pmatrix}W^{\alpha\beta}(\vec x)
\label{eig3}
\end{equation}
with mass eigenvalues
\begin{equation}
m^2_{\bar \Phi_3}=m_W^2\quad \text{and} \quad
m^2_{\bar \Phi_\pm}=\frac14\left(\varepsilon_\mu\pm\sqrt{\varepsilon_\mu^2+4m^2_W+(\varepsilon_\mu^{-1}\sigma_+)^2}\right)^2
-(\varepsilon_\mu^{-1}\sigma_+)^2-\sigma_+
\end{equation}

As in Section \ref{nonvanish}, special care has to be taken with the zero modes.
The vectors (\ref{eig1})-(\ref{eig3}) break down for the lightest modes, and the latter
have to be taken apart. After some thinking, it is not difficult to see that the vectors
(\ref{eig1})-(\ref{eig3}) have to be supplemented with the lightest modes
\begin{equation}
(\Phi^{\alpha\beta})_0=\begin{pmatrix}
1\\
0 \\
0
\end{pmatrix}W^{\alpha\beta}_{0}
\quad \quad
({\bar \Phi}^{\alpha\beta})_0=\begin{pmatrix}
0 \\
1 \\
0
\end{pmatrix}W^{\alpha\beta}_{0}
\end{equation}
where $W^{\alpha\beta}_{0} \equiv W^{\alpha\beta, \ (0,j_1,j_2)}_{0}(\tilde z_1', \tilde z_2')$
is given by eq.(\ref{magtheta}). The mass eigenvalues are respectively
$m^2_{\Phi_0}=\sigma_+$ and $m^2_{\bar \Phi_0}=-\sigma_+$.

Finally, let us compute the wavefunctions for the bifundamental fermions.
These again satisfy an equation of the form (\ref{6db}), where now
the covariant derivative must be incorporated
\beq
{\bf D}\, \raw\, {\bf D_A}\, =\,
\left(
\begin{array}{cccc}
0 & \hat{D}_{{z}^1} & \hat{D}_{{z}^2}  & \hat{D}_{{z}^3}  \\
-\hat{D}_{{z}^1} & 0 & - \hat{D}_{\bar{z}^3} & \hat{D}_{\bar{z}^2}  \\
 - \hat{D}_{{z}^2} & \hat{D}_{\bar{z}^3} & 0 & - \hat{D}_{\bar{z}^1} \\
 -\hat{D}_{{z}^3} & -\hat{D}_{\bar{z}^2} &  \hat{D}_{\bar{z}^1} & 0
\end{array}
\right)
\label{repcov}
\eeq

Taking into account that the commutation relation for these operators is given
by (\ref{extalgebra}), that $f^i_{k\bar{k}} = 0$ and that the only non-vanishing
components of the open string magnetic flux are $F_{1\bar{1}}$ and $F_{2\bar{2}}$,
we have that
{\begin{equation}
\begin{array}{c}
-{\bf D_A}^* {\bf D_A} \, =\, \hat{D}_m\hat{D}^m\, \mathbb{I}_4 \, +
\sum_a
\left(
\begin{array}{cccc}
- \sigma_+ & f_{\bar{2}\bar{3}}^a\, \hat{D}_a & f_{\bar{3}\bar{1}}^a \,\hat{D}_a
& f_{\bar{1}\bar{2}}^a\, \hat{D}_a \\
f_{{3}{2}}^a\, \hat{D}_a& \sigma_- & f_{{2}\bar{1}}^a\,\hat{D}_a & f_{{3}\bar{1}}^a\,\hat{D}_a \\
f_{{1}{3}}^a\,\hat{D}_a & f_{{1}\bar{2}}^a\,\hat{D}_a & - \sigma_- & f_{{3}\bar{2}}^a\,\hat{D}_a \\
f_{{2}{1}}^a\,\hat{D}_a & f_{{1}\bar{3}}^a\, \hat{D}_a& f_{{2}\bar{3}}^a\,\hat{D}_a  & \sigma_+ \\
\end{array}
\right)
\end{array}
\label{6dsqgenA}
\end{equation}}
Hence, in our example without $\mu$-term
{\beq
-{\bf D_A}^* {\bf D_A} \, =\,  \hat{D}_m\hat{D}^m\, \mathbb{I}_4 \, +\,
\left(
\begin{array}{cccc}
- \sigma_+ & 0 & 0 & 0 \\
0 &  \sigma_- & -\varepsilon\hat{D}_6 & 0 \\
0 & \varepsilon\hat{D}_6 &  - \sigma_- & 0 \\
0 & 0 & 0 & \sigma_+
\end{array}
\right)
\label{6dsq1A}
\eeq}
which again contains the upper $3\times 3$ block of the scalar mass matrix
(\ref{chargscalar}), with the diagonal shifted by $\sigma_+$. Therefore we obtain
the same eigenvectors (\ref{phi3}) and (\ref{phipm}), but now with masses
\begin{equation}
\Psi_\pm\, \raw\, m^2_{\Phi_\pm} - \sigma_+\qquad  \bar \Psi_\pm \, \raw \, m^2_{\bar \Phi_\pm} + \sigma_+\label{dterm}
\end{equation}
and similarly for $\Psi_W$ and $\Psi_3$. This indeed reflects the D-term breaking
of the charged $\mathcal{N}=2$ supermultiplets
caused by an open string flux with $\sigma_+\neq 0$.

Similar considerations apply also for the charged fermions in the example with
non-vanishing $\mu$-term. Indeed, in that case we have that
\begin{equation*}
\begin{array}{c}
-({\bf D_A} + {\bf F})^* ({\bf D_A} + {\bf F}) \, =\,  \hat{D}_m\hat{D}^m\, \mathbb{I}_4 \, +
\left(
\begin{array}{cccc}
- \sigma_+ & 0 & 0 & 0 \\
0 & \sigma_- & -\varepsilon_\mu \hat{D}_{{z}^3} & - \varepsilon_\mu \hat{D}_{{z}^2} \\
0 & \varepsilon_\mu \hat{D}_{\bar{z}^3} & - \sigma_- &  \varepsilon_\mu\hat{D}_{{z}^1} \\
0 & \varepsilon_\mu \hat{D}_{\bar{z}^2} & -\varepsilon_\mu \hat{D}_{\bar{z}^1}
& \sigma_+ - \varepsilon_\mu^2
\end{array}
\right)
\end{array}
\label{6dsq2A}
\end{equation*}
so again the eigenvalue problem is already solved by the knowledge of the bosonic sector.
Indeed, comparing with (\ref{chargscalarm}) we see that these states have the same
eigenvectors than their scalar superpartners (\ref{eig1}) and (\ref{eig2}), with their
masses given again by eq.(\ref{dterm}).


\section{Applications}\label{sec:app}

Having computed the open string spectrum in several type I flux vacua,\footnote{
It should be noted that our discussion misses those open string modes which are
genuine stringy oscillations and therefore cannot be captured by a supergravity analysis.}
we now would like to apply these results to understand better the effect of fluxes
on open strings. First we will consider the effect of fluxes on the open string
massive spectrum, and in particular how they may break the degeneracies present in
fluxless compactifications. Second, we will focus on the light spectrum of the theory,
and compare our results with those derived from a 4d effective supergravity analysis.
Finally, we will consider a type IIB T-dual setup, where the open strings arise from a
stack of D7-branes in the presence of $G_3$ fluxes, and translate the effect of fluxes
on open strings to this more familiar picture. Further applications of the above results
will be explored in \cite{wip}.


\subsection{Supersymmetric spectrum}
\label{susyspect}

As emphasized in the literature, flux vacua based on twisted tori are special in the
sense that they are directly related to 4d $\cn=4$ gauged supergravity. Moreover,
in the vanishing flux limit ($\varepsilon \raw 0$ for the vacua of
Section \ref{subsec:twisted}) one should recover the $\cn =4$ spectrum of a toroidal
compactification. Hence, in general one would expect that the flux lifts
the mass degeneracies of the $\cn = 4$ spectrum by an amount directly related to
$\varepsilon$, so that the previous 4d $\cn=4$ supermultiplets split into smaller ones.

In particular, for the type I flux vacua of subsections \ref{vmu} and \ref{nvmu} the flux
breaks the bulk $\cn=4$ supersymmetry down to $\cn=2$ and $\cn=1$, respectively,
so the neutral open string modes of Sections \ref{sec:wgauge}, \ref{sec:scalars} and
\ref{sec:fermions} should feel such kind of splitting.\footnote{In fact, recall that for
consistency we need to add a $\IZ_{2}$ (or $\IZ_{2n}$) orbifold that induces O5-planes
wrapping the twisted torus fiber, and that this already breaks $\cn=4 \raw \cn=2$ at the
string scale. However, for those open string sectors that are untwisted (i.e., not fixed
by the orbifold action) neutral, and not related to D5-branes, the tree-level fluxless
spectrum arranges indeed into $\cn=4$ multiplets, and the present discussion applies.
For the twisted open string spectrum one just needs to take into account the effect of the
orbifold on the wavefunctions, along the lines of \cite{ako08}.} On the other hand, the open
string flux $F_2$ already breaks $\cn=4 \raw \cn=2$,\footnote{For simplicity, we will assume
 a supersymmetric ($\sigma_+ = 0$) open string flux $F_2$.} and so the charged,
 bifundamental modes of Section \ref{sec:wmatter} could feel the effects of fluxes in a rather
 different way. Finally, let us recall that for the vacua of subsection \ref{vmu}, no multiplet
 splitting occurs at the massless level, while for the vacua of subsection \ref{nvmu} this is
 clearly the case. It is then natural to wonder how these facts will translate in terms of the
 full massive spectrum of the theory.

In order to classify our spectrum let us recall the content of massless and
massive 4d $\cn=1$ vector and chiral multiplets. Following a notation similar
to that of Sections \ref{sec:wgauge} to \ref{sec:wmatter}, we have for the
massless $\mathcal{N}=1$ multiplets

\begin{center}
\begin{tabular}{ccc}
&{neutral}&  {charged}\\
{\bf vector}
&$({\mathcal{A}}^\alpha)_0=(B^\alpha, \Psi_B^\alpha)$
&$(\mathcal{A}^{\alpha\beta})_0=(W^{\alpha\beta}, \Psi_W^{\alpha\beta})$
\quad $(\mathcal{\bar{A}}^{\alpha\beta})_0=(\bar{W}^{\alpha\beta}, \bar{\Psi}_W^{\alpha\beta})$
\\
{\bf chiral}
&$(\mathcal{C}^\alpha_p)_0=(\xi^\alpha_p, \Psi^\alpha_p)$
 &$(\mathcal{C}^{\alpha\beta}_p)_0=(\Phi^{\alpha\beta}_p, \Psi^{\alpha\beta}_p)$
 \quad $(\mathcal{\bar{C}}^{\alpha\beta}_p)_0=(\bar{\Phi}^{\alpha\beta}_p, \bar{\Psi}^{\alpha\beta}_p)$
 \\
\end{tabular}
\end{center}

\noindent where neutral multiplets contain particles in a real (in our case adjoint)
representation of the gauge group $G_{unbr}$, while charged multiplets transform
in complex representations (in our case the bifundamental rep. of Section
\ref{sec:wmatter}). The index $\a$ runs over the factors of $G_{unbr} = \prod_\a U(n_\a)$,
and the same applies for $\b$. The index $p$ labels instead different chiral multiplets
inside the same representation, and in our case takes the three different values
$p = \pm, 3$, as in (\ref{neutpm}) and (\ref{extrapol}). Finally, $\mathcal{A}^{\alpha\beta}$
and $\mathcal{C}^\alpha_p$ contain 4d spinors of positive chirality and
$\mathcal{\bar{A}}^{\alpha\beta}$ and $\mathcal{\bar{C}}^\alpha_p$ of negative chirality,
and the above degrees of freedom should be completed with their CPT conjugates.

For massive $\cn=1$ multiplets the above picture has to be slightly modified. In particular,
gauge bosons eat extra degrees of freedom in order to become massive through
 the standard Higgs mechanism, whereas chiral fields group into vector-like
 combinations. We can thus express their field content as

\begin{center}
\begin{tabular}{ccc}
&{neutral}&  {charged}\\
{\bf vector}
&${\mathcal{A}}^\alpha=({\mathcal{A}}^\alpha)_0+(\mathcal{C}_3^\alpha)_0$
&${\mathcal{A}}^{\alpha\beta}=(\mathcal{A}^{\alpha\beta})_0+(\mathcal{\bar{A}}^{\alpha\beta})_0+(\mathcal{C}_3^{\alpha\beta})_0+(\mathcal{\bar{C}}_3^{\alpha\beta})_0$\\
{\bf chiral}
&${\mathcal{C}}^\alpha_p=(\mathcal{C}^\alpha_p)_0$
 &${\mathcal{C}}^{\alpha\beta}_\pm=(\mathcal{C}^{\alpha\beta}_\pm)_0+(\mathcal{\bar{C}}^{\alpha\beta}_\pm)_0$
\end{tabular}
\end{center}

\noindent where we have taken $\mathcal{C}_3$ to contain the
degrees of freedom eaten by the gauge bosons, in agreement with the notation in
Sections \ref{nonvanish}, \ref{scalarmu} and \ref{bifund}.

On the other hand, massless $\mathcal{N}=2$ vector and hyper multiplets are given by

\begin{center}
\begin{tabular}{ccc}
&{neutral}& {charged}\\
{\bf vector}&$\mathcal{B}^\alpha={\mathcal{A}}^\alpha$&$\mathcal{B}^{\alpha\beta}
= {\mathcal{A}}^{\alpha\beta}$\\
{\bf hyper}&$\mathcal{H}^\alpha_\pm={\mathcal{C}}_\pm^\alpha$
&$\mathcal{H}^{\alpha\beta}_\pm={\mathcal{C}}_\pm^{\alpha\beta}$
\end{tabular}
\end{center}

\noindent where $\mathcal{H}^\alpha_p$ are in fact half-hypermultiplets.
For $\cn=2$ massive multiplets we have

\begin{center}
\begin{tabular}{c}
$\mathcal{V}^\alpha=\mathcal{B}^\alpha+\mathcal{H}^\alpha_+ + \mathcal{H}^\alpha_-$
\end{tabular}
\end{center}


\noindent and similarly for $\cv^{\a\b}$, looking like $\mathcal{N}=4$
vector multiplets. Finally, we may also have ultrashort $\cn=2$ massive multiplets,
containing the same  particle content as massless $\cn=2$ multiplets $\cb$ and $\ch$
and corresponding to $\oh$-BPS objects of the theory.

Let us now go back to the two main families of flux vacua analyzed in the previous sections.
In Tables \ref{table0} and \ref{table1} we summarize, respectively, the resulting neutral and
charged spectrum for the class of $\mathcal{N}=2$ compactifications with vanishing
$\mu$-term introduced in subsection \ref{vmu}. We have taken a supersymmetric configuration
of the open string flux (i.e.,  $\sigma_+=0$) and we have introduced the
shorthand notation
\begin{equation}
\Delta_{k_{i_1},k_{i_2},\ldots}^2\equiv \sum_{r=i_1,i_2,\ldots}\left(\frac{k_r}{R_r}\right)^2
\end{equation}
for the squared mass of a fluxless, toroidal KK mode.
The open string field content in this class of compactifications can be arranged into
different 4d $\mathcal{N}=2$ multiplets. More precisely, for the neutral sector
of the open string spectrum there is a tower of standard $\mathcal{N}=2$ massive
multiplets $\cv^\a$ associated to each irreducible unitary representation of the closed
string algebra (\ref{torsion}), plus an extra tower of ultrashort $\mathcal{N}=2$ hypers
$\ch^\a$. Since in principle the multiplets $\cv^\a$ can be identified with vector $\cn=4$
multiplets and $\ch^\a$ cannot, the latter can be seen as a clear effect of the $\cn=4 \raw \cn=2$
supersymmetry breaking induced by the closed string fluxes into the open string sector.

\begin{table}[!h]
\begin{center}
\begin{tabular}{|c||c|c|}
\hline
Multiplets& $(\textrm{Mass})^2$ & Degeneracy\\
\hline \hline
$(\mathcal{V}^\alpha)_{k_1,k_2,k_3,k_4,k_5}$&$\Delta^2_{k_1,k_2,k_3,k_4,k_5}$&$1$\\
\hline
$(\mathcal{V}^\alpha)_{n,k_3,k_6}^{(k,\delta_1,\delta_4)}$&
$\Delta^2_{k_3,k_6} + |\varepsilon|\Delta_{k_6} (n+1)$&$(k_6M)^2(n+1)$\\
\hline
$(\mathcal{H}^\alpha_{s_{k_6M}})_{k_3,k_6}^{(\delta_1,\delta_4)}$&$\Delta^2_{k_3,k_6}$&$(k_6M)^2$\\
\hline
\end{tabular}
\end{center}
\caption{Spectrum of neutral $\mathcal{N}=2$ multiplets for D9-brane fields
in the model with vanishing $\mu$-terms of subsection \ref{vmu}.}
\label{table0}
\end{table}

At the massless level the theory contains of a single neutral $\mathcal{N}=4$
vector multiplet $(\mathcal{V}^\alpha)_{0,0,0,0,0}$ for each adjoint representation of
$G_{unbr} = \prod_\a U(n_\a)$, and $|I^1_{\alpha\beta}I^2_{\alpha\beta}|$ charged
$\mathcal{N}=2$ hypermultiplets $(\mathcal{H}^{\alpha\beta})_{0}^{(j_1,j_2)}$ in the
 bifundamental representation of $U(n_\alpha)\times U(n_\beta)$. Therefore the massless
 open string spectrum is the same than in flat space, and the same applies to the open string
 wavefunctions. In fact, in the limit of diluted closed string fluxes, on which the size of
 the fiber is much smaller than any other size ($R_{6}\ll R_k$ with $k\neq 6$)
 the lightest Kaluza-Klein modes (which correspond to the modes
 $(\mathcal{V}^\alpha)_{k_1,k_2,k_3,k_4,k_5}$ in Table \ref{table0}) also match with
 the ones in the fluxless case.

\begin{table}[!h]
\begin{center}
\begin{tabular}{|c||c|c|}
\hline
Multiplets& $(\textrm{Mass})^2$ & Degeneracy\\
\hline \hline
$(\mathcal{V}^{\alpha\beta})_{n,k_3,k_6}^{(k,\delta_1,\delta_4)}$&$\rho(n+1)
+\Delta^2_{k_3,k_6}$&$[(k_6M)^2-I^1_{\alpha\beta}I^2_{\alpha\beta}](n+1)$\\
\hline
$(\mathcal{H}^{\alpha\beta}_-)_{k_3,k_6}^{(j_1,j_2)}$&$\Delta^2_{k_3,k_6}$&
$(k_6M)^2-I^1_{\alpha\beta}I^2_{\alpha\beta}$\\
\hline
$(\mathcal{H}^{\alpha\beta}_-)_{0}^{(j_1,j_2)}$&$0$&$|I^1_{\alpha\beta}I^2_{\alpha\beta}|$\\
\hline
\end{tabular}
\end{center}
\caption{Spectrum of charged $\mathcal{N}=2$ multiplets for D9-brane
fields in the vanishing $\mu$-term model of subsection \ref{vmu},  for supersymmetric
open string fluxes.}
\label{table1}
\end{table}

For the class of $\mathcal{N}=1$ compactifications with non-vanishing $\mu$-term
introduced in subsection \ref{nvmu}, the further breaking of the supersymmetry to
$\mathcal{N}=1$ and the presence of the $\mu$-term makes the spectrum slightly
more complicated. We have summarized in Tables \ref{table2} and \ref{table3} the
resulting neutral and charged spectra.\footnote{Actually, we present only that part of
the charged spectrum computed in Section \ref{sec:wmatter}.} The field content again
corresponds to a tower of $\mathcal{N}=4$ vector multiplets $\cv$ for each set
of unirreps of the closed string algebra, but now with a mass
mass splitting on their $\mathcal{N}=1$ constituents induced by the fluxes. Indeed, in
terms of $\mathcal{N}=1$ representations, each multiplet $\cv$ leads to one massive vector multiplet and two chiral multiplets. For compactifications
 with vanishing $\mu$-terms all these multiplets have degenerate Dirac
 mass $m_{\mathcal{B}}$, thus assembling into a $\mathcal{N}=4$ vector
 representation. For compactifications with non-vanishing $\mu$-term, however, the closed
 string background induces a Majorana mass $\varepsilon_\mu$ for one of the two
 chiral multiplets, leading to a mass-matrix which is of the
 form\footnote{We thank E. Dudas for pointing out this structure to us.}
\begin{equation}
\begin{pmatrix}{\mathcal{C}}_1& {\mathcal{C}}_2\end{pmatrix}\begin{pmatrix}\varepsilon_\mu & m_{\mathcal{B}}\\ m_{\mathcal{B}}& 0\end{pmatrix}\begin{pmatrix}{\mathcal{C}}_1\\ {\mathcal{C}}_2\end{pmatrix}
\end{equation}
The mass eigenvalues for this matrix are then given by
\begin{equation}
m_{{\mathcal{C}}_\pm}^2-\varepsilon_\mu m_{{\mathcal{C}}_\pm}
-m_{\mathcal{B}}^2=0\quad \Longrightarrow \quad m_{{\mathcal{C}}_\pm}^2
=\frac14\left(\varepsilon_\mu\pm\sqrt{\varepsilon_\mu^2+4m_{\mathcal{B}}^2}\right)^2
\label{cuadr2}
\end{equation}
reproducing the result we obtained in (\ref{cuadr}). Hence, after the breaking to $\cn=1$, one can associate
to each set of irreducible unitary representations a tower of massive $\mathcal{N}=1$
vector multiplets and two towers of $\mathcal{N}=1$ chiral multiplets, with their masses given
by eq.(\ref{cuadr2}).

Regarding the massless modes, for each stack of magnetized branes we have
two neutral $\mathcal{N}=1$ chiral multiplets $(\mathcal{C}^\alpha_\pm)_{0}$ and one
$\mathcal{N}=1$ vector multiplet $(\mathcal{A}^\alpha)_0$, while for each pair of factors
$U(n_\alpha)\times U(n_\beta) \subset G_{unbr}$ we have
$|I^1_{\alpha\beta}I^2_{\alpha\beta}|$ charged $\mathcal{N}=2$ hypermultiplets
$(\mathcal{H}^{\alpha\beta})_{0}^{(j_1,j_2)} = (\cc^{\a\b})_0^{(j_1,j_2)}
+ (\bar{\cc}^{\a\b})_0^{(j_1,j_2)}$ in the bifundamental representation.
Thus, of the three originally present neutral $\mathcal{N}=1$ chiral multiplets in flat space,
we see that only two remain massless in the presence of the closed string fluxes, whereas
$\mathcal{C}^\alpha_3$ gets a mass equal to $\varepsilon_\mu^2$. As we will see in
the next section, this is also what is expected from the four dimensional effective
supergravity analysis.

\begin{table}[!h]
\begin{center}
\begin{tabular}{|c||c|c|}
\hline
Multiplets& $(\textrm{Mass})^2$ & Degeneracy\\
\hline \hline
$({\mathcal{A}}^\alpha)_{k_1,k_2,k_4,k_5}$&$\Delta^2_{k_1,k_2,k_4,k_5}$&$1$\\
\hline
$({\mathcal{A}}^\alpha)_{n,k_3,k_4,k_5}^{(\delta)}$&
$|\varepsilon_\mu|\Delta_{k_3}(2n+1)+\Delta^2_{k_3,k_4,k_5}$&$|k_3M_3|$\\
\hline
$({\mathcal{A}}^\alpha)_{n,k_2,k_4,k_6}^{(\delta)}$&
$|\varepsilon_\mu| \Delta_{k_6} (2n+1)+\Delta^2_{k_2,k_4,k_6}$&$|k_6M_6|$\\
\hline
$({\mathcal{A}}^\alpha)_{n,k_3,k_4,k_6}^{(\delta_2,\delta_5)}$&
$|\varepsilon_\mu| \Delta_{k_3,k_6}(2n+1)+\Delta_{k_3,k_6}^2$&
$\textrm{l.c.m.}(|k_3M_3|,|k_6M_6|)$\\
\hline
$({\mathcal{C}}^\alpha_\pm)_{k_1,k_2,k_4,k_5}$&
$\frac14\left(\varepsilon_\mu\pm\sqrt{\varepsilon_\mu^2
+4\Delta^2_{k_1,k_2,k_4,k_5}}\right)^2$&$1$\\
\hline
$({\mathcal{C}}^\alpha_\pm)_{n,k_3,k_4,k_5}^{(\delta)}$&
$\frac14\left(\varepsilon_\mu\pm\sqrt{\varepsilon_\mu^2+\ 4|\varepsilon_\mu| \Delta_{k_3}(2n+1)
+4\Delta^2_{k_3,k_4,k_5}}\right)^2$&$|k_3M_3|$\\
\hline
$({\mathcal{C}}^\alpha_\pm)_{n,k_2,k_4,k_6}^{(\delta)}$&
$\frac14\left(\varepsilon_\mu\pm\sqrt{\varepsilon_\mu^2
+ 4|\varepsilon_\mu| \Delta_{k_6} (2n+1)+4\Delta^2_{k_2,k_4,k_6}}\right)^2$&$|k_6M_6|$\\
\hline
$({\mathcal{C}}^\alpha_\pm)_{n,k_3,k_4,k_6}^{(\delta_2,\delta_5)}$&
$\frac14\left(\varepsilon_\mu\pm\sqrt{\varepsilon_\mu^2
+4\varepsilon_\mu \Delta_{k_3,k_6}(2n+1)+4\Delta_{k_3,k_6}^2}\right)^2$&
$\textrm{l.c.m.}(|k_3M_3|,|k_6M_6|)$\\
\hline
$({\mathcal{A}}^\alpha)_0$&$0$&$1$\\
\hline
$(\mathcal{C}^\alpha_\pm)_0$&$0$&$1$\\
\hline
${\mathcal{C}}^\alpha_3$&$\varepsilon_\mu^2$&$1$\\
\hline
\end{tabular}
\end{center}
\caption{Spectrum of neutral $\mathcal{N}=1$ multiplets for
D9-brane fields in the model with non-vanishing $\mu$-term of subsection \ref{nvmu}.}
\label{table2}
\end{table}

\begin{table}[!h]
\begin{center}
\begin{tabular}{|c||c|c|}
\hline
Multiplets& $(\textrm{Mass})^2$ & Degeneracy\\
\hline \hline
$({\mathcal{A}}^{\alpha\beta})_{n,k_3}^{(\delta)}$&$\rho(n+1)+\Delta^2_{k_3}$&$|I^1_{\alpha\beta}I^2_{\alpha\beta}|$\\
\hline
$({\mathcal{C}}^{\alpha\beta}_\pm)_{n,k_3}^{(\delta)}$&$\frac14\left(\varepsilon_\mu\pm\sqrt{\varepsilon_\mu^2+4\rho(n+1)+4\Delta^2_{k_3}}\right)^2$&$|I^1_{\alpha\beta}I^2_{\alpha\beta}|$\\
\hline
$(\mathcal{H}^{\alpha\beta})_{0}^{(j_1,j_2)}$&$0$&$|I^1_{\alpha\beta}I^2_{\alpha\beta}|$\\
\hline
\end{tabular}
\end{center}
\caption{Partial spectrum of charged $\mathcal{N}=2$ and $\mathcal{N}=1$ multiplets
for D9-brane fields in the non-vanishing $\mu$-term model of subsection
\ref{nvmu}, for SUSY open string fluxes ($\sigma_+=0$).}
\label{table3}
\end{table}

Finally, let us point out that in the above discussion we have not included the effect
of the $\IZ_{2n}$ orbifold needed for the consistency of the construction. In principle,
this effect could partially project out the spectrum above, as it is known to happen for
the massless sector. This projection will however depend on the particular choice of
orbifold action,\footnote{Indeed, for some choices of, e.g., $\IZ_2$ orbifold the massless
chiral multiplets $(\cc^\a_\pm)_0$ in Table \ref{table2} are projected out, while for some
other choices like in \cite{gp96} it remains in the spectrum.} and it can be implemented
in our framework along the lines of \cite{ako08}. We defer a more detailed analysis of the
different possibilities to \cite{wip}.


\subsection{Comparison with 4d effective supergravity}
\label{sugra}

When analyzing the 4d effective theory of type I flux vacua we only
need to keep a small set of light modes in order to describe the low energy
dynamics. Such dynamics can then be encoded in terms of a 4d effective
K\"ahler potential and a superpotential which, at least at tree-level,
can be expressed as integrals over the internal space $\cam_6$. As a result,
finding vacua in the 4d effective theory can be translated into certain 10d
conditions which, if our effective theory is accurate, should describe 10d vacua.

The main caveat in the above approach is whether the appropriate set of light modes
has been chosen. Since in the presence of closed string fluxes the internal manifold
$\cam_6$ is no-longer Calabi-Yau, it is in general not known how to perform
the light mode truncation. A popular ansatz is to take the set of massless modes
of the Calabi-Yau $\cam_6^{\tiny \text{CY}}$ that is obtained from $\cam_6$ by `turning off'
the background fluxes. This procedure is well-defined when the fluxes are weak
compared to the KK scales in $\cam_6^{\tiny \text{CY}}$, but far from reliable beyond
this regime.
For instance, considering type IIB flux vacua on warped Calabi-Yau manifolds,
non-dilute fluxes in general lead to strong warping effects, which could in principle
lower the mass of an $\a'$ state below the flux scale.

Clearly, the same kind of observations apply to open strings
and, in particular, to the type I spectra analyzed above.
Since we have followed a well-defined prescription when dimensionally reducing
our flux vacua, comparing the 10d approach with the standard
4d effective supergravity analysis can be made manifest, and it can be checked
explicitly under which circumstances both approaches agree.
This will be the purpose of the present subsection.

\subsubsection{10d versus 4d approach}

For SU(3)-structure compactifications with O9/O5-planes, one can write
the 4d K\"ahler potential and superpotential in terms of integrals over the
internal manifold as \cite{kahler1,kahler2}
\begin{align}
\hat{K} &=-\textrm{log}\left[-i\int_{\mathcal{M}_6}\Omega\wedge\Omega^*\right]-\textrm{log}[2e^{-\phi}]-2\textrm{log}\left[\int_{\mathcal{M}_6} J\wedge J\wedge J\right]
\label{kahler}\\
W&=\int_{\mathcal{M}_6} \Omega\wedge (F_3+ie^{-\phi/2}dJ)
\label{super}
\end{align}
with $J$ and $\Omega$ the SU(3)-invariant 2-form and 3-forms of $\cam_6$, respectively.
In addition we can write $F_3 = F_3^{\text{cl}} + \om_3$, where $F_3^{\text{cl}}$ depends on the
RR closed string fields and
\begin{equation}
\omega_3=\textrm{Tr}\left(A\wedge dA+\frac23 A\wedge A\wedge A\right)
\end{equation}
is the 10d Chern-Simons 3-form, containing the open string degrees of freedom.

Now, when the internal manifold is not Calabi-Yau, as occurs in the presence of
closed string background fluxes, a prescription to expand $J$ and $\Omega$, and
$\om_3$ in terms of closed and open string light fields is in general not known.
In that case, one usually
proceeds by expanding them in a base of harmonics for the Calabi-Yau manifold
$\cam_6^{\tiny \text{CY}}$ which results in the limit of vanishing fluxes.

In our case, this prescription amounts to take either $\cam_6^{\tiny \text{CY}} = T^6$
or a toroidal orbifold, and so the wavefunctions used in our dimensional reduction
should look like those that arise from an unwarped $T^6$. From our results on open string
wavefunctions,  it is clear that this will be the case as long as ${\it i)}$ the warping can be
neglected and ${\it ii)}$ the light modes of the compactification do not contain any KK mode
excited along the fiber. Whether neglecting the warping is a good approximation can be
read from eq.(\ref{NSNSt1}). Using the  conditions it can be rewritten as
\beq
\nabla^2_{T^4}Z^2\, =\, - \varepsilon^2 + \dots
\eeq
where $\varepsilon$ is the flux mass scale of our compactification, and the dots
stand for $F_2$ and $\d$-function contributions. Thus, away from localized
sources and setting $F_2 =0$ for simplicity, the warp factor can be taken
constant for $m^{\text{KK}}_{\text{base}} \gg \varepsilon$.
It is easy to see \cite{Schulz04} that this is guaranteed if we take
$\textrm{Vol}_{B_4}^{1/2} \gg \textrm{Vol}_{\Pi_2}$, which in turn implies that
$m^{\text{KK}}_{\text{fib}} \gg m^{\text{KK}}_{\text{base}}$ and hence that
no fiber KK mode will be a light field of the theory.

Indeed, as we will show below, under the assumption
$\textrm{Vol}_{B_4}^{1/2} \gg \textrm{Vol}_{\Pi_2}$ the 4d effective supergravity
succeeds in describing the spectrum of light modes that we have obtained by
dimensional reduction. On the contrary, in the regime where the volume of the fiber
is of the same order of magnitude than the volume of the base, the mass of the
fiber KK modes will be comparable to the mass of the base modes and lifted open
string moduli, and they cannot be omitted from the 4d effective supergravity description.
As discussed around figure \ref{fig1}, the wavefunctions of these fiber KK modes present
interesting localization properties, which should be added to the standard localization
effects due to the strong warping effects. It would be very interesting to see how their combined
effect may affect standard dimensional reduction.

Let us then take the limit $R_{\text{base}} \gg R_{\text{fib}}$ and truncate the theory to
the lightest neutral and charged modes, denoted in the following by $\varphi^{\alpha,k}$
and $\varphi^{\alpha\beta,k}$, respectively. In terms of the notation of Section \ref{susyspect},
the scalar component of these fields are
\begin{equation}
(\xi_\pm^\alpha)_{0}\equiv \varphi^{\alpha,1}\pm i\varphi^{\alpha,2}\quad \quad (\xi_3^\alpha)_{0}\equiv \varphi^{\alpha,3}\quad \quad (\Phi_\pm^{\alpha\beta})_0\equiv \varphi^{\alpha\beta,1}\pm i\varphi^{\alpha\beta,2}
\label{lights}
\end{equation}
where the subscript $0$ denotes the lightest KK mode of each tower.
In the following we will analyze the two and three-point couplings for this set of light fields.

\subsubsection{2-point couplings}

In supersymmetric compactifications to 4d Minkowski, the only source for scalar masses are
$\mu$-terms in the superpotential. In terms of these, the 2-point couplings in the
4d effective action read\footnote{In this section we will be working in 4d Planck
mass units.}
\begin{equation}
-S=Z_{i\bar j}(M,M^*)\partial_\mu \varphi^i \partial^\mu (\varphi^i)^*+e^{\hat K(M,M^*)}\mu_{ik}\bar\mu_{\bar l\bar j}Z^{k\bar l}(M,M^*)\varphi^i(\varphi^j)^* + \ldots\label{ssugra}
\end{equation}
where we have expanded the effective superpotential and the full K\"ahler potential
in powers of the light open string fields $\varphi^i$ as
\begin{align}
K(M,M^*,\varphi,\varphi^*)&=\hat K(M,M^*)+Z_{i\bar j}(M,M^*)\varphi^i(\varphi^{\bar j})^*+\ldots \\
W(M,\varphi)&=\hat W(M)+\frac12 \mu_{ij}(M)\varphi^i\varphi^j+\frac{1}{3!}\tilde Y_{ijk}\varphi^i\varphi^j\varphi^k+\ldots
\end{align}
and $M$ stands for the full set of closed string moduli/light fields, whose K\"ahler
potential $\hat{K}$ is given by (\ref{kahler}). The standard procedure in the 4d supergravity
 approach is then to approximate (\ref{kahler}) by the K\"ahler potential of a factorizable $T^6$.
For $\mathcal{N}=2$ configurations of the open string flux $F_2$, this is given to quadratic
order in the fields by \cite{tkahler1,tkahler2,lerda1,lerda2,diveccia}
\begin{equation}
K=-\textrm{log }(2s)+\sum_{k=1}^3\left[-\textrm{log}(4t_k u_k)+\sum_\alpha \frac{|\varphi^{\alpha,k}|^2}{4t_ku_k}\right]+\sum_{\alpha,\beta}\frac{|\varphi^{\alpha\beta,1}|^2+|\varphi^{\alpha\beta,2}|^2}{16(t_1u_1t_2u_2)^{1/2}}
\end{equation}
where
\begin{equation}
2s=g_s^{1/2} \textrm{Vol}_{\cam_6}\quad \qquad 2t_a=4\pi^2g_s^{-1/2}R_aR_{a+3}
\quad \qquad 2u_a=\frac{R_{a+3}}{R_a}\quad \qquad a=1,2,3
\label{moduli}
\end{equation}
are the real parts of the moduli in a toroidal orientifold with O5/O9-planes  \cite{louisiib}.

Under these assumptions, the integration of the superpotential (\ref{super}) was performed in
\cite{geosoft} for toroidal compactifications, obtaining the following expressions for the gravitino
mass and for the effective $\mu$-term of the lightest neutral modes\footnote{We have corrected a normalization factor $t_I$ in eq.(3.46) of \cite{geosoft} and expressed the result in terms of the conventions used in this paper.}
\begin{equation}
m_{3/2}=e^{\hat K/2}\langle \hat W\rangle=\frac{3}{4\sqrt{2s}}f^k_{\bar i\bar j}\quad
\qquad \mu_{kk}=\frac{e^{-\hat K/2}Z_{k\bar k}}{\sqrt{2s}}f^{\bar k}_{\bar i\bar j}
\end{equation}
where $f^{\bar i}_{\bar j\bar k}$ are the (moduli dependent) structure constants of the
algebra (\ref{torsion}) expressed in the complex basis.
These equations, which depend
only on the NSNS part of the background, assume that the on-shell conditions
(\ref{rel1})-(\ref{rel2}) are satisfied.

Note that when the manifold is complex $f^k_{\bar i\bar j}=0$, the gravitino is massless
and the background preserves $\mathcal{N}\geq 1$ supersymmetry in four dimensions
\cite{geosoft,lawrence}. In that case, from (\ref{ssugra}) we get
\begin{multline}
-S=\frac{1}{4u_it_i}\partial_\mu\varphi^{\alpha,i}\partial^\mu(\varphi^{\alpha,i})^*
+\frac{1}{16(t_1u_1t_2u_2)^{1/2}}(\partial_\mu\varphi^{\alpha\beta,1}
\partial^\mu(\varphi^{\alpha\beta,1})^*+\partial_\mu\varphi^{\alpha\beta,2}
\partial^\mu(\varphi^{\alpha\beta,2})^*)\\
-\sum_{i\neq k\neq j}\frac{1}{8st_ku_k}
|f^{\bar k}_{\bar i\bar j}|^2|\varphi^{k,\alpha}|^2
\end{multline}
and so, making use of the moduli definitions (\ref{moduli}) we have
\begin{multline}
-S=\frac{g_s^{1/2}}{(2\pi R_{i+3})^2}\partial_\mu\varphi^{\alpha,i}\partial^\mu(\varphi^{\alpha,i})^*
+\frac{g_s^{1/2}}{16\pi^2R_4R_5}(\partial_\mu\varphi^{\alpha\beta,1}\partial^\mu
(\varphi^{\alpha\beta,1})^*+\partial_\mu\varphi^{\alpha\beta,2}\partial^\mu
(\varphi^{\alpha\beta,2})^*)\\
-\sum_{i\neq k\neq j}\frac{1}{\textrm{Vol}_{\cam_6}}
\frac{1}{(2\pi R_{k+3})^2}|f^{\bar k}_{\bar i\bar j}|^2|\varphi^{k,\alpha}|^2
\label{2point}
\end{multline}

Let us see how this expression applies to the two classes of type I flux vacua that have
been analyzed in this paper. First, note that in the example of subsection \ref{vmu} with
vanishing $\mu$-terms, the structure constants $f^{\bar k}_{\bar i\bar j}$ are all zero.
From (\ref{2point}) we see that then all the lightest scalars remain massless, in agreement
with the 10d result that there are no flux-generated $\mu$-terms in this case. The open string
massless content is therefore the same than in a fluxless toroidal (or toroidal orbifold)
compactification, as we have also concluded from direct dimensional reduction of the
10d supergravity background.

On the other hand, for the example of subsection \ref{nvmu} we see from
(\ref{com}) that the only non-vanishing structure constant
whose all indices are anti-holomorphic is given by
$f^{\bar 3}_{\bar 1\bar 2}=\varepsilon_\mu$.
Hence, as expected from the 10d analysis all the light scalars (\ref{lights}) are massless
except for $\varphi^{3,\alpha}$. Moreover, after the rescaling
$\varphi^{3,\alpha}\to 2\pi R_6 g_s^{-1/4}\varphi^{3,\alpha}$ in order
to have canonically normalized kinetic terms, one obtains a 4d mass given by
\begin{equation}
m^2_{\varphi^{3,k}}=(g_{YM}\varepsilon_\mu)^2
\label{massmu}
\end{equation}
where $g_{YM}=(g_s^{1/2}\textrm{Vol}_{\cam_6})^{-1/2}$ is the gauge coupling constant.
Again, this matches the result obtained in Section \ref{nonvanish} by means of
dimensional reduction.\footnote{There is a factor
$g_{YM}^2$ with respect of the expressions in Section \ref{conmu} which can
be explained from the fact that the results in the previous sections have been obtained in
the 10d Einstein frame, whereas in this section we are working in the 4d Einstein frame.}


\subsubsection{3-point couplings}

Let us now turn to the 3-point couplings between the lightest modes and compare again
with the effective supergravity results. We will focus on those Yukawa couplings of the
form
\begin{equation}
S=\int dx^4\, Y_{ijk} \bar \psi^{\alpha\beta,i}\psi^{\beta\alpha,j}\varphi^k
\label{yuki}
\end{equation}
where $\psi^{\alpha\beta,i}$ is some massless fermion in the bifundamental representation
of $U(n_\alpha)\times U(n_\beta)$, and $\varphi^k$ a complex scalar in the adjoint
representation. Recall that in the specific closed string background at hand, one cannot
turn on a magnetic flux $F_2$ such that $\int_{\cam_6} F_2^3 \neq 0$ since, in
particular, such $F_2$ cannot be turned on the elliptic fiber $\Pi_2$ wrapped by the
D5-branes. As a result, (\ref{yuki}) is the only possible class of Yukawa couplings involving
the light modes of these constructions.

As usual, the coupling $Y_{ijk}$ can be obtained by dimensional reduction of the kinetic term
of the 10d gaugino, given in eq.(\ref{accion}), resulting in the expression \cite{yukawa,diveccia}
\begin{equation}
Y_{ijk}\,=\, g_s^{-1/4}\int_{\mathcal{M}_6}(\Psi^{\beta\alpha}_i)^\dagger \tilde \gamma^m
 \Psi^{\beta\alpha}_j(\xi_k)_m
\end{equation}
with $\Psi^{\alpha\beta,i}$ and $\xi^k$ the corresponding wavefunctions for the 4d
modes $\psi^{\alpha\beta,i}$ and $\varphi^k$, respectively.

More precisely, in the two examples of flux compactifications considered above, the only
non-vanishing Yukawa coupling involving the two fermionic superpartners of
$\varphi^{\alpha\beta,1}$ and $\varphi^{\alpha\beta,2}$, denoted as
$\psi^{\alpha\beta,1}$ and $\psi^{\alpha\beta,2}$ respectively, are given by
\begin{equation}
Y_{123}\,=\, =\, \frac{1}{g_s^{1/4}\textrm{Vol}_{\cam_6}^{1/2}}
\int_{\mathcal{M}_6}(\Psi^{\beta\alpha,(j_1,j_2)}_1)^\dagger \tilde \gamma^3
\Psi^{\beta\alpha,(j_1',j_2')}_2=
-ig_{YM}\delta_{j_1j_1'}\delta_{j_2j_2'}
\end{equation}
where we have normalized the wavefunction of $\varphi^3$ such that
\begin{equation}
\int_{\mathcal{M}_6}(\xi_3)^\dagger \xi_3=1
\end{equation}
The computation then exactly follows the one carried out in \cite{diveccia} for fluxless toroidal compactifications. In terms of the moduli definitions (\ref{moduli}) we have
\begin{equation}
Y_{123}=-\frac{i\delta_{j_1j_1'}\delta_{j_2j_2'}}{\sqrt{2s}}
\end{equation}
which can be compared with the standard expression for the physical
Yukawa couplings in 4d effective supergravity
\begin{equation}
Y_{ijk}=e^{\hat K/2}\tilde Y_{ijk}(Z_{i\bar i}Z_{j\bar j}Z_{k\bar k})^{-1/2}
\end{equation}
where $\tilde Y_{ijk}$ is the holomorphic Yukawa coupling appearing in the superpotential.
We then obtain $\tilde Y_{123}=-i\delta_{j_1j_1'}\delta_{j_2j_2'}$, as in
standard toroidal compactifications.


\subsection{Comparison with T-dual type IIB vacua}\label{D7dual}

An interesting feature of the type I flux vacua analyzed in this paper is that
they have a simple dual description in terms of standard type IIB
flux compactifications. Indeed, if we take type I theory in an elliptically fibered
manifold of the form (\ref{mansatz}) and we perform two T-dualities along
the fiber coordinates $a \in \Pi_2$, we will obtain type IIB string theory compactified
on the direct product $\cam_6' = B_4 \times \Pi_2$ (up to an overall warp factor)
and threaded by an NSNS 3-from flux $H_3$. Regarding the open string sector,
the type I gauge theory analyzed in Section \ref{diraclap} will be mapped to a
set of O7-planes and D7-branes wrapped on $B_4$, while O5-planes and
D5-branes wrapped on $\Pi_2$ will be taken to O3-planes and D3-branes,
respectively.

This fact applies, in particular, to the twisted tori examples of Section
\ref{subsec:twisted}, for which $B_4 = T^4/\IZ_{2n}$. Following \cite{kstt02}
and ignoring the presence of the orbifold for simplicity, we have that the
type IIB T-dual of these twisted tori is given by the
following closed string background
\bes
\label{bgiib}
\begin{align}
\label{bgiib1}
&ds^2=Z^{-1}ds^2_{\IR^{1,3}}+Z\, ds^2_{T^4 \times T^2} \\
\label{bgiib2}
&ds^2_{T^4 \times T^2}=(2\pi)^2 \left[ \sum_{m=1,2,4,5}(R_mdx^m)^2
+ \sum_{m=3,6} \left(\frac{dx^m}{R_m}\right)^2 \right] \\
\label{bgiib3}
& F_5  =  (1 + *_{10})\, d{\rm vol}_{M_4} \wedge dh \\
\label{bgiib4}
& \tau  =  ie^{-\phi_0} = \text{const.}
\end{align}
\ees
with $e^{\phi_0} = g_s/R_3R_6$, and $h-Z^{-2} e^{-\phi_0}$= const.
In addition, the internal $T^6$ will be threaded by RR and NSNS 3-form
fluxes, which depend on the particular choice of T-dual type I flux
vacuum. In particular, the type IIB NSNS flux $H_3$ is related to the choice
of structure constants in the type I elliptic fibration, while the RR flux
$F_3$ comes from the type I quantity $F_3^{\text{bg}}$ defined in Appendix
\ref{ap:warp}. In particular, the type IIB duals of the vacua in subsection
\ref{vmu} contain the fluxes
\bes
\label{iibflux1}
\begin{align}
& H_3  =  (2\pi)^2\, N\, (dx^1 \wedge dx^2 + dx^4 \wedge dx^5) \wedge dx^6 \\
& F_3  =  - (2\pi)^2\, M\, (dx^1 \wedge dx^2 + dx^4 \wedge dx^5) \wedge dx^3
\end{align}
\ees
that impose the supersymmetry conditions $NR_6 = M R_3 e^{\phi_0}$ and
$R_1R_2 = R_4 R_5$, on the closed string moduli of the compactification,
identical to the ones obtained in the type I side.\footnote{In the type IIB picture
these conditions come from imposing that $G_3 = F_3 - \tau H_3$ is a (2,1)-form
\cite{granapolcho}.
For general choices of complex structure $dz^i  = dx^i + \tau_idx^{i+3}$
in $\prod_i (T^2)_i$ they read $M\tau^3 = N\tau$ and $\tau_1\tau_2 = -1$.}

The type IIB duals to the vacua in subsection \ref{nvmu} contain, on the other hand,
the 3-form fluxes
\bes
\label{iibflux2}
\begin{align}
& H_3  =  (2\pi)^2\,  (M_3\, dx^2 \wedge dx^3 + M_6\, dx^5 \wedge dx^6) \wedge dx^1 \\
& F_3  =  (2\pi)^2\,  (N_6\, dx^2 \wedge dx^3 + N_3\, dx^5 \wedge dx^6) \wedge dx^4
\end{align}
\ees
that impose the SUSY conditions $N_6 R_1 e^{\phi_0} = M_3 R_4$,
$N_3 R_1 e^{\phi_0} = M_6 R_4$ and $M_3 R_3R_5 = M_6 R_2R_6$,
again identical to the dual type I conditions.

Rather than analyzing the closed string sector of these type IIB vacua, we
would like to understand the dynamics governing the open string sector.
In particular, we would like to translate the type I open string spectrum
to the present picture, and interpret the open string wavefunctions of
Sections \ref{sec:wgauge} to \ref{sec:wmatter} in terms of type IIB quantities.
In this sense, note that the initial $G_{gauge} = U(N)$ gauge theory considered
in Section \ref{diraclap} will now arise from a stack of $N$ D7-branes, and that
the gauge group will be broken to $G_{unbr} = \prod_i U(n_i) \subset U(N)$
via the presence of a magnetic open string flux $F_2$ on them. The analysis
of the open string Dirac and Laplace equations could then in principle be
carried out via a dimensional reduction of the D7-brane 8D U(N) twisted SYM theory,
along the lines of \cite{quevedo}. Extracting our wavefunction information from
the type I T-dual setup, however, has the advantage of automatically including the
coupling of the D7-brane open strings to the warp factor and to the background
fluxes, which is in general only known for $U(1)$ theories \cite{dirac,fershiu}.

In the set of type IIB vacua at hand, the stack of N D7-branes under analysis
will wrap $T^4 = (T^2)_1 \times (T^2)_2 = \{x^1, x^4, x^2, x^5\}$ and sit at a
particular point in the transverse space $(T^2)_3$. Setting $F_2=0$ and
neglecting the effect of closed string fluxes, we obtain at the massless level
three 4d $\cn=1$ chiral multiplets $\Phi^i$ in the U(N) adjoint representation,
which are nothing but the D7-brane moduli and modulini. More precisely,
the bosonic components of these multiplets are given by two complex Wilson
line moduli $\phi^i$ arising from dimensional reduction of the 8D gauge boson
$A_M$ on $(T^2)_i$, $i=1,2$, and by the D7-brane geometric modulus $\phi^3$
in the $(T^2)_3$ transverse space. In the absence of background fluxes it is easy
to see that these D7-brane moduli are mapped to the type I Wilson line moduli
via the dictionary
\begin{center}
\begin{tabular}{lcl}
{D7-brane}& \qquad \qquad &{D9-brane}\\
{\bf Wilson line\ } $\phi^1$ $\phi^2$  & &
{\bf Wilson line\ } $(\xi^{1,2})_0 \equiv \varphi^{1,2}$ \\
{\bf Geom. modulus\ } $\phi^3$ & & {\bf Wilson line\ } $(\xi^3)_0 \equiv  \varphi^3$ \\
\end{tabular}
\end{center}
where we are defining our type I fields as in (\ref{standcom}) and (\ref{lights}).
Turning on the closed string background fluxes, it is easy to see that the same
dictionary will still apply. Indeed, using the results of \cite{ciu04,lustf,osl} one expects
the D7-brane Wilson line moduli $\phi^i$ to remain massless
in the presence of background fluxes, and the geometric modulus $\phi^3$ to
generically gain a mass. This latter point will of course depend on the choice
of background fluxes and, by construction, we expect it to differ for both set of
fluxes (\ref{iibflux1}) and (\ref{iibflux2}). Indeed, applying the analysis of \cite{ciu04}
to the background fluxes above, it is easy to check that for the choice (\ref{iibflux1})
$\phi^3$ remains massless, while for (\ref{iibflux2}) a $\mu$-term is generated
which exactly reproduces (\ref{massmu}).

In terms of wavefunctions, a more interesting sector is given by massive open string
modes. Again, in the absence of closed string fluxes one has the dictionary
\begin{center}
\begin{tabular}{lcl}
{D7-brane}& \qquad \qquad &{D9-brane}\\
{\bf KK mode on} $(T^2)_1 \times (T^2)_2$ & &
{\bf KK mode on} $B_4 \simeq (T^2)_1 \times (T^2)_2$ \\
{\bf Winding mode on} $(T^2)_3$ & & {\bf KK mode on} $\Pi_2 \simeq (T^2)_3$ \\
\end{tabular}
\end{center}
between D7-brane and D9-brane massive modes. Let us now turn on
background fluxes and translate our type I open string wavefunctions
to the type IIB setup via the above dictionary. For simplicity, we will
first focus on the gauge boson wavefunctions of Section \ref{sec:wgauge}.
A general result is then that a D9-brane KK mode along the base $B_4$
will never feel the effect of the fluxes, while the KK modes along the
fiber $\Pi_2$ could indeed have a distorted wavefunction. More precisely,
a KK mode on the fiber will behave as an open string charged under a
magnetic flux $F_2^{\text{cl}}$ that depends on the $\Pi_2$ KK momenta.

\begin{figure}[!h]
\begin{center}
\includegraphics[width=15cm]{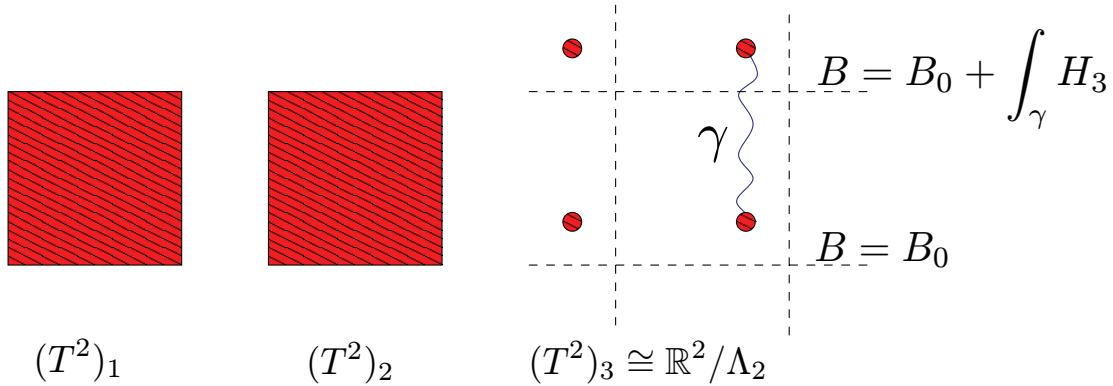}
\end{center}
\caption{\label{figw} Open string wavefunction for a D7-brane winding mode
in the T-dual type IIB flux picture. Even if both ends of the open string sit on the
same point in the internal space, they feel a different B-field due to the presence
of the NSNS flux $H_3$ and the extended nature of the winding mode. As a result,
D7-brane winding modes behave as open strings that end on D7-branes with
different magnetizations, and so do their wavefunctions.}
\end{figure}

In terms of D7-brane modes, we thus obtain that KK modes are unaffected
by the presence of type IIB $G_3$ fluxes, while winding modes behave as
magnetized open strings. Indeed, it is not hard to convince oneself that
a D7-D7 string winded around the closed path $\g \subset (T^2)_3$ can
in principle feel different B-fields on both ends, and that their difference is
given by
\beq
\Delta B|_{\text{D7}}\, =\, \int_\g H_3
\label{Bdif}
\eeq
as illustrated in figure \ref{figw}.
Moreover, for a closed $H_3$ (\ref{Bdif}) will only depend on the winding
numbers of $\g$, which upon T-duality translate into the KK-modes $(k_3, k_6)$
on the elliptic fiber $\Pi_2$. Finally, one can check that computing (\ref{Bdif})
for the examples (\ref{iibflux1}) and (\ref{iibflux2}) and mapping the result to
the T-dual type I setup one indeed obtains the closed string magnetic flux
$F_2^{\text{cl}}$. Hence, we can summarize the D7-brane winding mode
wavefunction as
\beq
\Psi^{\g}\, =\, \psi^{\Delta B}({\vec x_{B_4}}) \cdot e^{2\pi i (k_3 x^3 + k_6x^6)}
\label{wwinding}
\eeq
where $k_3, k_6$ are the winding modes of $\g$ in $(T^2)_3$,
$\vec x_{B_4} = \{x^1,x^4, x^2,x^5\}$, and $\psi^{\Delta B}$ is the
wavefunction of an open string in a magnetized D7-brane wrapping
$B_4$, and whose magnetic flux is given by (\ref{Bdif}). This clearly
matches our type I T-dual results.

Turning now to the wavefunctions for fermions and 4d scalars, it is easy to see
that D7-brane KK modes should be insensitive to the presence of the flux. Winding
modes, on the other hand, should feel the background flux in a more involved way
than their gauge boson counterparts, as it is manifest from the matrix $\mathbb{M}$
that appears in their equation of motion in the type I picture, and which contains off
diagonal terms proportional to the components of $F_2^{\text{cl}}$. In the case of
the example (\ref{iibflux1}) with vanishing $\mu$-terms on the D7, the off-diagonal
terms should correspond to those of (\ref{system1}), and they may be understood as
the mixing terms $G_m^{\ \, p}$ that usually appear in the equations of motion for magnetized
D-branes (see e.g., eq.(\ref{phieq})), with the substitution $F_2 \raw F_2^{\text{cl}}$.
The interpretation of these off-diagonal terms for the example with non-vanishing
$\mu$-term (\ref{iibflux2}) (given by those of (\ref{system2})) remain
however more obscure from the type IIB viewpoint. Note in particular that, according
to our first dictionary above, the eigenfunctions (\ref{muxi3}) and (\ref{muxipm})
obtained in the type I side, should correspond to a bound state of winding modes of
D7-brane Wilson lines and moduli. It would be interesting to understand how these
eigenstates arise from the type IIB side of the duality.\footnote{In view of the
non-commutative nature of (\ref{system2}), this could perhaps be naturally explained
in terms of a non-commutative field theory in the internal D7-brane coordinates.}

Finally, let us consider those matter field wavefunctions analyzed in Section
\ref{sec:wmatter}. From the type IIB side, the exotic W boson wavefunction
(\ref{setmagnetico}) and its generalization to non-vanishing D-term should arise
from a D7-brane winding mode which also feels a difference on the open string
magnetic flux $\Delta F_2 = (F_2^{\a\b})^{\text{op}}$.
Hence, in this picture the total difference in flux felt by such a D7$_{\a}$-D7$_{\b}$
 string is given by the gauge invariant quantity
\beq
\Delta \cf \, =\, \Delta B|_{\text{D7}} + 2\pi \a' \Delta F_2\,
=\, 2\pi F_2^{\text{cl}} + 2\pi (F_2^{\a\b})^{\text{op}}\, =\, 2\pi (F_2^{\a\b})_{\text{eff}}
\eeq
which is nothing but the open + closed effective flux entering the definition of the
wavefunction (\ref{wavematter}) and the more massive modes of this sector. Hence,
we find that the open string wavefunctions obtained in the type I flux vacua studied
in this paper fit nicely into our understanding of the D7-brane wavefunctions physics
in the type IIB T-dual setup.


\section{Conclusions and outlook}\label{sec:conclu}

In this work we have given a concrete prescription for performing dimensional reduction in flux compactifications. The procedure relies on the observation that in presence of closed string fluxes it is still possible to define some modified Dirac and Laplace-Beltrami operators in the internal manifold which account for the effect of the fluxes on the open string fluctuations. These operators are extracted from the type I supergravity action in the limit on which closed string fluctuations are frozen and the warping can be neglected.

To analyze the spectrum of eigenmodes of these operators, we have found very helpful some of the tools of non-commutative harmonic analysis and representation theory, which we have summarized in Section \ref{gener} and Appendix \ref{kirillov}. This formalism seems to point out towards a deep connection between the 4d spectrum of massive excitations, symplectic geometry and 4d gauged supergravity algebras. In particular, we have found that the spectrum of Kaluza-Klein excitations for neutral and charged modes in a stack of magnetized D9-branes is classified by irreducible unitary representations of the Kaloper-Myers gauge algebra \cite{kaloper}.\footnote{A similar observation has been made in \cite{douglas2} in the context of fluxless Calabi-Yau compactifications.} Notice that for sectors of the theory which preserve enough number of supersymmetries, one can in addition consider the global symmetries of the effective action and compute other massive excitations such as winding modes. Indeed, notice that the Kaloper-Myers algebra is only a portion of the full $\mathcal{N}=4$ gauged supergravity algebra. It is therefore natural to conjecture that irreducible unitary representations of the full algebra classify not only Kaluza-Klein modes, but also winding and non-perturbative modes associated to the $\mathcal{N}=4$ sectors of the theory. Following this philosophy we have conjectured the presence of some massive non-perturbative charged modes in the worldvolume of magnetized D9-branes.

We can extract several conclusions from the results of this paper. First, notice that generically there is always a set of fields which is insensitive to the background fluxes, and therefore their wavefunctions are the same than in a fluxless compactification. Moreover, the on-shell conditions usually ensure that these are the lightest modes in the limit of diluted fluxes and constant warping, which has two important consequences. On the one side, the lightest sector is usually not affected by the fluxes, up to possible flux induced mass terms. On the other, if one considers only this sector of the theory, it is enough to dimensionally reduce as if being in a fluxless compactification.\footnote{The same result was found in \cite{nearly} for the closed string sector of type IIA AdS$_4$ vacua.}

Thus, we find that fluxes mainly affect the structure of massive Kaluza-Klein replicas. In particular, for the class of vacua that we have considered, the resulting spectrum can be understood in terms of Landau degeneracies, mass shifts and mixings induced by the fluxes. We therefore expect that fluxes change in an important way the threshold corrections to the 4d low energy effective theory. The computation of gauge threshold corrections in flux compactifications will be addressed in a future publication \cite{wip}.

We have also observed that wavefunctions in the presence of closed string fluxes are not very different from wavefunctions in compactifications with only magnetized branes. This has been interpreted in the light of open/closed string duality, showing that in many cases the closed string fluxes can be interpreted as non diagonal magnetic fluxes in a dual background.

There are several possible further directions to explore, apart from the ones already mentioned. For example, it would be interesting to see how the warping fits in this picture, and in particular to try to combine these results with the ones e.g. in \cite{fershiu}. This is particularly important for applying these methods in the context of the AdS/CFT correspondence. Some recent applications of wavefunctions in this context include models of holographic gauge mediation \cite{holographic}, where Kaluza-Klein modes mediate the transmission of supersymmetry breaking between the hidden and visible sectors, and models for meson spectroscopy (see \cite{meson} for a review and references), where meson resonances are identified with Kaluza-Klein modes in a dual supergravity theory. We expect that the techniques introduced here will result useful in these contexts, once they are extended conveniently to account for the strong warping.

Also, one could similarly consider other vacua different than the no-scale solutions that we have analyzed. For instance, we could make use of the same methods for dimensionally reduce type IIA $\textrm{AdS}_4$ compactifications on nearly K\"ahler manifolds, in the same spirit than in \cite{kashani1,nearly}. This would be particularly relevant for computing the structure of massive modes in these backgrounds.

Finally, from the phenomenological point of view, the vacua considered here are not very appealing, since they are non-chiral. In this sense, it would be desirable to extend this computation to models including magnetized D5-branes and more realistic matter content. In particular, the T-duals of the chiral flux compactifications considered in \cite{marchesashiu} fall into this class. With that same aim, it would be also desirable to extend these techniques to general, non-parallelizable SU(3)-structure manifolds.

\section*{Acknowledgments}
{We would like to thank L. Alvarez-Gaum\'e, E. Dudas and A. Uranga
for useful discussions and comments. The work of P.G.C. is supported by
the European Union through an Individual Marie-Curie IEF. Additional support
comes from the contracts ANR-05-BLAN-0079-02, MRTN-CT-2004-005104,
MRTN-CT-2004-503369 and CNRS PICS \#~4172, 3747. Finally, we would like to thank the Ecole Polytechnique, CERN and
the Galileo Galilei Institute for Theoretical Physics for hospitality and the INFN
for partial support during the completion of this work.}

\appendix

\section{Fermion conventions}\label{ap:ferm}

In order to describe explicitly fermionic wavefunctions we take the following representation for $\G$-matrices in flat 10d space
\beq
\G^{\ul{\mu}}\, = \, \g^\mu \otimes \Id_2 \otimes \Id_2 \otimes \Id_2 \quad \quad  \quad \G^{\ul{m}}\, =\, \g_{(4)} \otimes \tilde{\g}^{m-3}
\label{ulG:ap}
\eeq
where $\mu = 0, \dots, 3$, labels the 4d Minkowski coordinates, whose gamma matrices are
\beq
\g^0\, =\,
\left(
\begin{array}{cc}
 0 & -\Id_2 \\ \Id_2 & 0
\end{array}
\right)
\quad
\g^i\, =\,
\left(
\begin{array}{cc}
 0 & \sig_i \\ \sig_i & 0
\end{array}
\right)
\eeq
$m = 4, \dots, 9$ labels the extra $\R^6$ coordinates
\beq
\begin{array}{lll}
\tilde{\g}^{1}\, = \, \sig_1 \otimes  \Id_2 \otimes \Id_2 & \quad \quad &  \tilde{\g}^{4}\, = \,  \sig_2 \otimes  \Id_2 \otimes \Id_2 \\
\tilde{\g}^{2}\, = \, \sig_3 \otimes  \sig_1 \otimes \Id_2 & \quad \quad &  \tilde{\g}^{5}\, = \,  \sig_3 \otimes  \sig_2 \otimes \Id_2 \\
\tilde{\g}^{3}\, = \,  \sig_3 \otimes \sig_3 \otimes \sig_1 & \quad \quad &  \tilde{\g}^{6}\, = \,  \sig_3 \otimes \sig_3 \otimes \sig_2
\end{array}
\label{tilgamma}
\eeq
and $\sig_i$ indicate the usual Pauli matrices. The 4d chirality operator is then given by
\beq
\G_{(4)}\, = \, \g_{(4)} \otimes  \Id_2 \otimes \Id_2 \otimes \Id_2
\eeq
where $\g_{(4)} = i \g^0\g^1\g^2\g^3$, and the 10d chirality operator by
\beq
\G_{(10)}\, = \, \g_{(4)} \otimes \g_{(6)}\, =\, \left(
\begin{array}{cc}
 \Id_2 & 0 \\ 0 & -\Id_2
\end{array}
\right) \otimes  \sig_3 \otimes \sig_3 \otimes \sig_3
\eeq
with $\g_{(6)} = -i \tilde{\g}^1\tilde{\g}^2\tilde{\g}^3\tilde{\g}^4\tilde{\g}^5\tilde{\g}^6$. Finally, in this choice of representation a Majorana matrix is given by
\beq
\label{ap:Maj}
\mathcal{B}\, =\, \G^{\ul{2}}\G^{\ul{7}}\G^{\ul{8}}\G^{\ul{9}}\, =\,
\left(
\begin{array}{cc}
 0 & \sig_2 \\ -\sig_2 & 0
\end{array}
\right)
\otimes \sig_2 \otimes i\sig_1 \otimes \sig_2 \,= \, \mathcal{B}_4 \otimes \mathcal{B}_6
\eeq
which indeed satisfies the conditions $\mathcal{B}\mathcal{B}^* = \Id$ and $\mathcal{B}\, \G^{\ul{M}} \mathcal{B}^* = \G^{\ul{M}*}$. Notice that the 4d and 6d Majorana matrices $\mathcal{B}_4 \equiv \g^2 \g_{(4)}$ and $\mathcal{B}_6 \equiv \tilde{\g}^4 \tilde{\g}^5 \tilde{\g}^6$ satisfy analogous conditions $\mathcal{B}_4\mathcal{B}_4^* = \mathcal{B}_6\mathcal{B}_6^* = \Id$ and $\mathcal{B}_4\, \g^{\mu} \mathcal{B}_4^* = \g^{\mu*}$, $\mathcal{B}_6\, \g^{m} \mathcal{B}_6^* = -  \g^{m*}$.

In the text we mainly work with 10d Majorana-Weyl spinors of negative chirality, meaning those spinors $\theta$ satisfying $\theta = - \G_{(10)} \theta = \mathcal{B}^*\theta^*$. In the conventions above this means that we have spinors of the form
\bes
\label{basisMW}
\begin{align}
\theta^0\, =\,
\psi^0 \,
\left(
\begin{array}{c}
\xi_+ \\ 0
\end{array}
\right) \otimes \chi_{---}
+ i (\psi^0)^*\,
\left(
\begin{array}{c}
0 \\ \sig_2\xi_+^*
\end{array}
\right) \otimes
\chi_{+++}\\
\theta^1\, =\,
\psi^1 \,
\left(
\begin{array}{c}
\xi_+ \\ 0
\end{array}
\right) \otimes \chi_{-++}
- i (\psi^1)^*\,
\left(
\begin{array}{c}
0\\ \sig_2\xi_+^*
\end{array}
\right) \otimes
\chi_{+--}\\
\theta^2\, =\,
\psi^2 \,
\left(
\begin{array}{c}
\xi_+ \\ 0
\end{array}
\right) \otimes \chi_{+-+}
+ i (\psi^2)^*\,
\left(
\begin{array}{c}
0 \\ \sig_2\xi_+^*
\end{array}
\right) \otimes
\chi_{-+-}\\
\theta^3\, =\,
\psi^3 \,
\left(
\begin{array}{c}
\xi_+ \\ 0
\end{array}
\right) \otimes \chi_{++-}
- i (\psi^3)^*\,
\left(
\begin{array}{c}
0 \\ \sig_2\xi_+^*
\end{array}
\right) \otimes
\chi_{--+}
\end{align}
\ees
where $\psi^j$ is the spinor wavefunction, $(\xi_+ \ 0)^t$ is a 4d spinor of positive
chirality and $\chi_{\epsilon_1\epsilon_2\epsilon_3}$ is a basis of 6d spinors of such that
\beq
\chi_{---}\, =\,
\left(
\begin{array}{c}
0\\1
\end{array}
\right) \otimes
\left(
\begin{array}{c}
0\\1
\end{array}
\right) \otimes
\left(
\begin{array}{c}
0\\1
\end{array}
\right)
\quad \quad
\chi_{+++}\, =\,
\left(
\begin{array}{c}
1\\0
\end{array}
\right) \otimes
\left(
\begin{array}{c}
1\\0
\end{array}
\right) \otimes
\left(
\begin{array}{c}
1\\0
\end{array}
\right)
\label{spinorbasis:ap}
\eeq
etc. Note that these basis elements are eigenstates of the 6d chirality operator $\g_{(6)}$, with eigenvalues $\epsilon_1\epsilon_2\epsilon_3$.

Finally, let us recall that to dimensionally reduce a 10d fermionic action, one has to simultaneously diagonalize two Dirac operators: $\slashed{\p}_{\IR^{1,3}}$ and $\slashed{D}^{\text{int}}$, built from $\G^{\ul{\mu}}$ and $\G^{\ul{m}}$, respectively. However, as these two set of $\G$-matrices do not commute, nor will $\slashed{\p}_{\IR^{1,3}}$ and $\slashed{D}^{\text{int}}$, and so we need instead to construct these Dirac operators from the alternative $\G$-matrices
\beq
\tilde{\G}^{\ul{\mu}}\, = \, \G_{(4)} \G^{\ul{\mu}}\, = \, \G_{(4)}\g^\mu \otimes \Id_2 \otimes \Id_2 \otimes \Id_2 \quad \quad  \quad  \tilde{\G}^{\ul{m}}\, =\, \G_{(4)} \G^{\ul{m}}\, =\, \Id_4 \otimes \tilde{\g}^{m-3}
\label{commG}
\eeq
following the common practice in the literature.


\section{Warped Dirac equation}
\label{ap:warp}

Let us consider the 6d Dirac equation deduced in eq.(\ref{dirac6d})
\beq
\left( \slashed{D}^{\cam_6}  + \frac{1}{4} e^{\phi/2} \slashed{F}_3 - \frac{1}{2} \slashed{\p} \ln Z \right) \chi_6 \, =\, Z^{1/4}  m_\chi\, \mathcal{B}_6^* \chi_6^*
\label{ap:dirac6d}
\eeq
where now all slashed quantities are constructed from the set of $\G$-matrices defined in
(\ref{commG}). Let us also consider a compactification ansatz of the form (\ref{mansatz}), where again $Z$ only depends on the coordinates of the base $B_4$.

Then, as in \cite{geosoft}, the 2-form $J$ splits as $J = J_{\Pi_2} + J_{B_4}$, and we can split $F_3$ accordingly. Indeed, let us define
\beq
e^{\phi/2} F_3^{\rm bg}\, \equiv \, e^{\phi/2} F_3 -  2  *_{\cam_6} (d \phi \wedge  J_{\Pi_2})
\label{F3bg}
\eeq
so that eq.(\ref{ap:dirac6d}) becomes
\beq
\left( \slashed{D}^{\cam_6}  + \frac{1}{4} e^{\phi/2} \slashed{F}_3^{\rm bg}  - \slashed{\p} \ln Z\, P_+^{\Pi_2} \right) \chi_6 \, =\, Z^{-1/4} m_4\, \cb_6^* \chi_6^*
\eeq
where we have introduced the projectors $P_\pm^{\Pi_2}$ defined in (\ref{ex1proj}). In addition, we have that the covariant derivative reads
\beq
\nabla^{\cam_6}_m\, =\, \p_m + \om^{B_4}_m - \frac{1}{8} \left(\p_m\ln Z -\G_{m}\slashed{\p}\ln Z \right)
- \frac{1}{4} \Lambda_m{}^{n} \left(\p_n\ln Z -\G_{n}\slashed{\p}\ln Z  - \slashed{f}_{n} \right)
\eeq
where $\om^{B_4}$ is the spin connection of $B_4$,  $f_{mnp}$ is defined by (\ref{metricflux}) and $\Lambda$ is a block-diagonal matrix specified by
\beq
\Lambda_{mn}\, =\, g_{mn} -2 e^{{a}}_m e_{{a}n},\quad \quad a \in \Pi_2
\label{Lam}
\eeq
Finally, (\ref{F3bg}) implies that
\beq
 e^{\phi/2} F_3^{\rm bg}\, =\, *_{\cam_6} \left[e^{\phi/2} d \left(e^{-\phi/2} J_{\Pi_2} \right) + e^{-3\phi/2} d \left(e^{3\phi/2} J_{B_4} \right) \right]
\eeq
and this, if the $B_4$ base is symplectic, implies that $e^{\phi/2} \slashed{F}_3^{\text{bg}} = i \slashed{f} \slashed{J}_{\Pi_2} \g_{(6)}$.

We thus obtain a 6d Dirac equation of the form
\beq
\left(\slashed{D}^{\Pi_2} + \slashed{D}^{B_4} +  \frac{1}{2}\slashed{f}P_+^{\Pi_2}  -  \slashed{\p} \ln Z  \left( P_+^{\Pi_2} -\frac{7}{8}\right)\right) \chi_6  \, =\, Z^{1/4} m_4\, \cb_6^* \chi_6^*
\label{ap:dirac6dsf}
\eeq
containing the coupling of fermions to the warping. Note that by taking $Z = 1$ we recover the unwarped equation (\ref{dirac6duw}) used in the main text.

Now, if we normalize the internal spinor as $\chi_6^\dagger\chi_6 =1$, then the warp factor dependence of the metric ansatz (\ref{mansatz}) will induce a non-standard 4d kinetic terms for $\chi_4$. In order to recover a canonical kinetic terms upon dimensional reduction we need instead to consider the rescaled Weyl fermion
\beq
\eta\, \equiv\, Z^{-7/8}\, \chi_6
\eeq
in terms of which the warped 6d Dirac equation reads
\beq
\left(\slashed{D}^{\Pi_2} + \slashed{D}^{B_4} +  \frac{1}{2}\slashed{f}P_+^{\Pi_2}  -  \slashed{\p} \ln Z   P_+^{\Pi_2} \right) \eta  \, =\, Z^{1/4} m_4\, \cb_6^* \eta^*
\label{ap:6dnorm}
\eeq

Note that the projector $P_+^{\Pi_2}$ is basically the chirality projector of the 4d base $B_4$. As in (\ref{split4+2}), let us split $\eta$ as
\beq
\eta\, =\, \eta_{\Pi_2} + \eta_{B_4}
\label{ap:split4+2}
\eeq
where $P_+^{\Pi_2} \eta_{\Pi_2} = \eta_{\Pi_2}$, $P_+^{\Pi_2} \eta_{B_4} = 0$. We can then split the Dirac equation (\ref{ap:6dnorm}) as
\beqa
\label{ap:6dsplit1}
{\slashed{D}}^{\Pi_2}_{\text{uw}}\eta_{B_4} + {\slashed{D}}^{B_4}_{\text{uw}} Z^{-1} \eta_{\Pi_2} & = & m_4 \cb_6^* \eta_{B_4}^*\\
\label{ap:6dsplit2}
{\slashed{D}}^{\Pi_2}_{\text{uw}}\eta_{\Pi_2} + Z^{-1} {\slashed{D}}^{B_4}_{\text{uw}} \eta_{B_4}  + \oh Z^{-2} {\slashed{f}}_{\text{uw}} \eta_{\Pi_2} & = & m_4 \cb_6^* \eta_{\Pi_2}^*
\eeqa
where we have extracted the warp factor dependence from the $\Gamma$-matrices contractions.

Note that a simple set of solutions is obtained by setting $\eta_{\Pi_2} = 0$, $m_4 = 0$ and  $\slashed{D}^{B_4} \eta_{B_4} = 0$, since neither the warp factor nor the fluxes play any role in this case. This simple zero mode equation does not come as a surprise if one compares eq.(\ref{ap:6dnorm}) with the Dirac equation for D7-branes in type IIB warped Calabi-Yau flux backgrounds. Indeed, by the results of \cite{fershiu} it is easy to identify the modes $\eta_{\Pi_2}$ in (\ref{ap:split4+2}) as those containing the gaugino and geometric modulini of a T-dual D7-brane, as well as their KK replicas, whereas $\eta_{B_4}$ are T-dual to the D7-brane Wilsonini.\footnote{In fact, such mapping can be made explicit by simply T-dualizing our type I compactification along the two fiber coordinates $a \in \Pi_2$ of the elliptic fibration (\ref{mansatz}), as done in Section \ref{D7dual}.} Now, since the Wilson line zero modes of a D7-brane do not feel the effect of the background fluxes \cite{ciu04,osl} nor that of the warping \cite{fershiu}, the same statement must apply to the open string zero modes arising from $\eta_{B_4}$, as is indeed the case.


\section{A non-supersymmetric example}\label{ap:N=0}

On the main text we have analyzed examples where the closed
string background fluxes preserve at least $\mathcal{N}=1$ supersymmetry
in four dimensions.
However, as we have treated bosons and fermions independently, our
techniques apply equally well to ${\cn = 0}$ vacua of the theory.
To illustrate this fact, in this appendix we apply them to one of such
examples, based on a compactification on the Heisenberg manifold.

Let us then consider the background
\bes
\begin{align}
&ds^2=Z^{-1/2}(ds^2_{\mathbb{R}^{1,3}}+ds^2_{\Pi_2})+Z^{3/2}ds^2_{T^4} \\
&ds^2_{T^4}=(2\pi)^2\sum_{m=1,2,4,5}(R_m dx^m)^2 \\
&ds^2_{\Pi_2}=( 2\pi)^2[ (R_3 dx^3)^2+(R_6
\tilde{e}^{6})^2] \\
&F_3=-(2\pi)^2N dx^1\wedge dx^2\wedge \tilde e^6-g_s^{-1}*_{T^4}dZ^2 \\
&e^\phi Z=g_s=\textrm{const.}
\end{align}
\ees
which is almost identical to (\ref{bg1}). In the present case, however,
$\tilde e^6$ stands for the left-invariant 1-form satisfying
\begin{equation}
d \tilde{e}^6=M dx^4\wedge dx^5
\end{equation}
so that $\mathcal{M}_6$ is given locally by $\mathbb{R}\times \mathcal{H}_3$.
The twisted derivatives are then
\begin{align*}
\hat\partial_1&=(2\pi R_1)^{-1}\partial_{x^1} & \hat\partial_4&=(2\pi R_4)^{-1}(\partial_{x^4}+\frac{M}{2}x^5\partial_{x^6})\\
\hat\partial_2&=(2\pi R_2)^{-1}\partial_{x^2} & \hat\partial_5&=(2\pi R_5)^{-1}(\partial_{x^5}-\frac{M}{2}x^4\partial_{x^6})\\
\hat\partial_3&=(2\pi R_3)^{-1}\partial_{x^3} & \hat\partial_6&=(2\pi R_6)^{-1}\partial_{x^6}
\end{align*}
Finally, the compact structure of $\mathcal{M}_6$ is produced by the following
 identifications which result from quotienting by $\Gamma = \G_{\ch_3} \times \IZ^3$
\begin{align*}
& x^4\to x^4+1 \qquad x^6\to x^6 - \frac{M}{2}x^5\\
& x^5\to x^5+1 \qquad x^6\to x^6 + \frac{M}{2}x^4\\
& x^i \to x^i + 1 \qquad \qquad \textrm{for }i\neq 4,5
\end{align*}
In addition, the equations of motion require the conditions $R_4R_5=4\pi^2 R_6^2R_1R_2$
and $g_sN=M$, with $N,\ M \in \mathbb{Z}$. This in particular ensures that the first torsion
class $\mathcal{W}_1$, defined as $J\wedge d\Omega = \mathcal{W}_1 J\wedge J\wedge J$,
 is non-vanishing and, hence, $\mathcal{M}_6$ is not a complex manifold. As the gravitino
 mass is proportional to $\mathcal{W}_1$ \cite{geosoft,lawrence,dwsb}, this reflects the fact
 that the background does not preserves any supersymmetry in 4d.

As in our previous examples, in order to cancel the RR charges and tensions, O5-planes
(and maybe also D5-branes) wrapping $\Pi_2$ are required, which again will be introduced
via the orbifold quotient $\mathcal{R}: x^m \mapsto -x^m$ on the $T^4$ base coordinates.
To simplify our discussion, in this section we will assume $F_2=0$, although one can easily
add the effect of a non-trivial $F_2$ along the lines of Section \ref{sec:wmatter}.


\subsection{Bosonic wavefunctions}

As usual, the wavefunction for the four dimensional neutral gauge bosons is given by the eigenfunctions of the corresponding Laplace-Beltrami operator of the manifold
\begin{equation}
\hat\partial_m\hat\partial^m B=-m_B^2B
\end{equation}
The solutions to this equations can be found using the techniques described in Section
\ref{sec:wgauge}. More precisely, we find two towers of KK modes associated to the four
dimensional gauge boson, which in a suitable polarization read
\begin{equation}
B_{k_1,k_2,k_3,k_4,k_5}(\vec x)=\textrm{exp}[2\pi i(k_1x^1+k_2x^2+k_3x^3+k_4x^4+k_5x^5)]\label{nosusy1}
\end{equation}
for the first tower, with mass eigenvalue
\begin{equation}
m_B^2=\sum_{i=1}^5\left(\frac{k_i}{R_i}\right)^2
\end{equation}
while for the second tower
\begin{multline}
B^{\delta}_{n,k_1,k_2,k_3,k_6}=\left(\frac{2\pi^2R_5|k_6 M|}{R_4\textrm{Vol}_{\mathcal{M}_6}}\right)^{1/4} \sum_{k_4\in
\delta+k_6M\mathbb{N}} \psi_{n}\left(\frac{\dot x^5}{\sqrt{2}}\right)
e^{2\pi i \left(k_1x^1+k_2x^2+k_3 x^3+k_4
x^4+k_6 \dot x^6\right)}
\label{nosusy2}
\end{multline}
with eigenvalue
\begin{equation}
m_B^2=\frac{
|k_6\varepsilon|}{R_6}(2n+1)+\sum_{m=1,2,3,6} \left(\frac{k_m}{R_m}\right)^2
\end{equation}
with $\delta=0\ldots k_6M-1$, $\varepsilon=MR_6/2\pi R_4R_5$ and
\begin{equation}
\dot x^5=\left(\frac{4\pi R_5}{R_4|k_6M|}\right)^{1/2}(k_4+k_6M)\quad \qquad \dot x^6\equiv x^6+\frac{M}{2}x^4x^5
\end{equation}
Similarly, we can work out the wavefunctions for the four dimensional scalars. Plugging the background into eqs.(\ref{xi+})-(\ref{xi-}) leads to an equation of the form (\ref{eigenval}) where, in complex coordinates (\ref{standcom}), the mass matrix now reads
\begin{equation}
\mathbb{M}=\begin{pmatrix}\hat\partial_m\hat\partial^m&-\varepsilon\hat\partial_6&-\frac{i\varepsilon}{2}\hat\partial_{z^2}&0&0&\frac{i\varepsilon}{2}\hat\partial_{z^2}\\
\varepsilon\hat\partial_6&\hat\partial_m\hat\partial^m&\frac{i\varepsilon}{2}\hat\partial_{z^1}&0&0&-\frac{i\varepsilon}{2}\hat\partial_{z^1}\\
-\frac{i\varepsilon}{2}\hat\partial_{\bar z^2}&\frac{i\varepsilon}{2}\hat\partial_{\bar z^1}&\hat\partial_m\hat\partial^m-\frac{\varepsilon^2}{2}&-\frac{i\varepsilon}{2}\hat\partial_{z^2}&\frac{i\varepsilon}{2}\hat\partial_{z^1}&\frac{\varepsilon^2}{2}\\
0&0&-\frac{i\varepsilon}{2}\hat\partial_{\bar z^2}&\hat\partial_m\hat\partial^m&-\varepsilon\hat\partial_6&\frac{i\varepsilon}{2}\hat\partial_{\bar z^2}\\
0&0&\frac{i\varepsilon}{2}\hat\partial_{\bar z^1}&\varepsilon\hat\partial_6&\hat\partial_m\hat\partial^m&-\frac{i\varepsilon}{2}\hat\partial_{\bar z^1}\\
\frac{i\varepsilon}{2}\hat\partial_{\bar z^2}&-\frac{i\varepsilon}{2}\hat\partial_{\bar z^1}&\frac{\varepsilon^2}{2}&\frac{i\varepsilon}{2}\hat\partial_{z^2}&-\frac{i\varepsilon}{2}\hat\partial_{z^1}&\hat\partial_m\hat\partial^m-\frac{\varepsilon^2}{2}
\end{pmatrix}\label{scalarnosusy}
\end{equation}
This is a non-commutative eigenvalue problem similar to the one found in Section \ref{nonvanish}.
Notice, however, that in the present case the mass matrix is not block diagonal, reflecting the
fact that the background does not preserves the complex structure of $T^6$, and in particular
the complex structure given by the choice (\ref{standcom}).

As in the supersymmetric case, the eigenvalues and eigenfunctions of (\ref{scalarnosusy})
can be found we the aid of the commutation relations of the twisted derivatives and the
Laplacian, which in the present case read
\begin{align*}
& [  \hat{\p}_{{z}^1},  \hat{\p}_{{z}^2}] \, =\,  [  \hat{\p}_{\bar{z}^1},  \hat{\p}_{\bar{z}^2}] \, =\, \varepsilon  \hat{\p}_{6} \quad \quad  [  \hat{\p}_{{z}^1},  \hat{\p}_{\bar{z}^2}] \, =\,  [  \hat{\p}_{\bar{z}^1},  \hat{\p}_{{z}^2}] \, =\, - \varepsilon  \hat{\p}_{6}\\
&- [ \hat{\p}_m\hat{\p}^m,  \hat{\p}_{{z}^1}] \, =\,  [ \hat{\p}_m\hat{\p}^m,  \hat{\p}_{\bar{z}^1}] \, =\,\varepsilon  \hat{\p}_{6} (\hat{\p}_{\bar{z}^2}-\hat{\p}_{z^2})\\
&[ \hat{\p}_m\hat{\p}^m,  \hat{\p}_{{z}^2}] \, =\, - [ \hat{\p}_m\hat{\p}^m,  \hat{\p}_{\bar{z}^2}] \, =\,\varepsilon  \hat{\p}_{6} (\hat{\p}_{\bar{z}^1}-\hat{\p}_{z^1})
\end{align*}
After some work, we find that the resulting spectrum is given by the two eigenvectors
\begin{equation}
\xi_3(\vec x)\equiv\begin{pmatrix}0\\ 0\\ 1\\ 0\\ 0\\ 1\end{pmatrix}B(\vec x)\quad
\qquad \xi_{3}^*(\vec x)\equiv \begin{pmatrix}\hat{\p}_{\bar{z}^1}\\ \hat{\p}_{\bar{z}^2}\\
2i\hat{\p}_6\\ \hat{\p}_{{z}^1}\\ \hat{\p}_{{z}^2}\\ 0\end{pmatrix}B(\vec x)
\end{equation}
with mass eigenvalues $m_{\xi_{3}}^2=m_{\xi_{3}^*}^2=m_B^2$, the two eigenvectors
\begin{equation}
\xi_+(\vec x)\equiv\begin{pmatrix}\hat{\p}_{\bar{z}^1}-i\hat{\p}_{{z}^2}\\
\hat{\p}_{\bar{z}^2}+i\hat{\p}_{{z}^1}\\ 0\\ -\hat{\p}_{{z}^1}+i\hat{\p}_{\bar{z}^2}\\
-\hat{\p}_{{z}^2}-i\hat{\p}_{\bar{z}^1}\\ 0\end{pmatrix}(\hat{\p}_4-i\hat{\p}_5)B(\vec x)
\quad  \qquad \xi_{-}(\vec x)\equiv \begin{pmatrix}-\hat{\p}_{\bar{z}^1}-i\hat{\p}_{{z}^2}\\
-\hat{\p}_{\bar{z}^2}+i\hat{\p}_{{z}^1}\\ 0\\ \hat{\p}_{{z}^1}+i\hat{\p}_{\bar{z}^2}\\
\hat{\p}_{{z}^2}-i\hat{\p}_{\bar{z}^1}\\ 0\end{pmatrix}(\hat{\p}_4+i\hat{\p}_5)B(\vec x)
\end{equation}
with mass eigenvalues $m_{\xi_{\pm}}^2=m_B^2\pm (\varepsilon k_6/R_6)$, and the two eigenvectors
\begin{equation}
\xi_{\pm}^*\equiv\begin{pmatrix}-(m^2_{\xi_{\pm}^*}-m^2_B)\hat{\p}_{z^2}+
\varepsilon\hat{\p}_{\bar{z}^1}\hat{\p}_6\\ (m^2_{\xi_{\pm}^*}-m^2_B)\hat{\p}_{z^1}+
\varepsilon\hat{\p}_{\bar{z}^2}\hat{\p}_6\\ i\frac{(m^2_{\xi_{\pm}^*}-m^2_B)^2}{\varepsilon}
+i\varepsilon\hat{\p}_6\\
-(m^2_{\xi_{\pm}^*}-m^2_B)\hat{\p}_{\bar{z}^2}+\varepsilon\hat{\p}_{{z}^1}\hat{\p}_6\\
(m^2_{\xi_{\pm}^*}-m^2_B)\hat{\p}_{\bar{z}^1}+\varepsilon\hat{\p}_{{z}^2}\hat{\p}_6\\
-i\frac{(m^2_{\xi_{\pm}^*}-m^2_B)^2}{\mu}+i\varepsilon\hat{\p}_6
\end{pmatrix}B(\vec x)
\end{equation}
with mass eigenvalue
\begin{equation}
m^2_{\xi_{\pm}^*}=\frac14
\left(\varepsilon\pm\sqrt{\varepsilon^2+4m_B^2-4\left(\frac{k_3}{R_3}\right)^2}\right)^2
+\left(\frac{k_3}{R_3}\right)^2
\end{equation}
where in all these expressions $B(\vec x)$ is a gauge boson wavefunction,
(\ref{nosusy1})-(\ref{nosusy2}), with mass eigenvalue $m_B$.


\subsection{Fermionic wavefunctions}

Regarding the fermionic wavefunctions, in this case we have that
\beq
\slashed{f}\, =\, (2\pi)^{-1} \frac{M R_6}{R_4R_5}   \tilde{\g}^{456} \, = \, (2\pi)^{-1} \frac{M R_6}{R_4R_5} \, i \sig_2 \otimes \sig_1 \otimes \sig_2\, =\, \varepsilon \cb_6
\eeq
and so the Dirac operator reads
\beq
{\bf D} + {\bf F}\, =\,
\left(
\begin{array}{cccc}
i \frac{\varepsilon}{2} & \hat{\p}_{{z}^1} & \hat{\p}_{{z}^2} & \hat{\p}_{{z}^3} \\
- \hat{\p}_{{z}^1} & 0 & - \hat{\p}_{\bar{z}^3} & \hat{\p}_{\bar{z}^2} \\
- \hat{\p}_{{z}^2} & \hat{\p}_{\bar{z}^3} & 0 & - \hat{\p}_{\bar{z}^1} \\
- \hat{\p}_{{z}^3} & - \hat{\p}_{\bar{z}^2} &  \hat{\p}_{\bar{z}^1} &  i \frac{\varepsilon}{2}
\end{array}
\right)
\eeq
from which we extract the following mass matrix
 \beq
-({\bf D} + {\bf F})^* ({\bf D} + {\bf F}) \, =\,
\left(
\begin{array}{cccc}
\hat{\p}_m\hat{\p}^m - \frac{\varepsilon^2}{4} & i\frac{\varepsilon}{2}  \hat{\p}_{{z}^1}
& i\frac{\varepsilon}{2}  \hat{\p}_{{z}^2} & 0 \\
i\frac{\varepsilon}{2}  \hat{\p}_{\bar{z}^1} & \hat{\p}_m\hat{\p}^m
& - \varepsilon  \hat{\p}_{6} & - i\frac{\varepsilon}{2}  \hat{\p}_{{z}^2} \\
i\frac{\varepsilon}{2}  \hat{\p}_{\bar{z}^2} & \varepsilon  \hat{\p}_{6}
& \hat{\p}_m\hat{\p}^m &  i\frac{\varepsilon}{2}  \hat{\p}_{{z}^1} \\
 0 & -i\frac{\varepsilon}{2} \hat{\p}_{z^2}
 & i\frac{\varepsilon}{2}\hat{\p}_{z^1} & \hat{\p}_m\hat{\p}^m - \frac{\varepsilon^2}{4}
\end{array}
\right)
\eeq
Since the background does not preserve any supersymmetry, it is natural to expect
the eigenfunctions and eigenvalues of this matrix to be rather different from their
bosonic counterparts. Indeed, after some algebra one can show  that
the fermionic wavefunctions are given by the eigenvectors
{\footnotesize
\begin{equation}
\Psi_{\pm}(\vec x)\equiv\begin{pmatrix}
i\varepsilon(m^2_{\Psi_\pm}+\hat{\p}_{z^3}\hat{\p}_{\bar{z}^3})\\
2(\hat{\p}_{\bar z^1}+i\hat{\p}_{z^2})(m^2_{\Psi_\pm}-m_B^2)\\
2(\hat{\p}_{\bar z^2}-i\hat{\p}_{z^1})(m^2_{\Psi_\pm}-m_B^2)\\
\varepsilon(m^2_{\Psi_\pm}+\hat{\p}_{z^3}\hat{\p}_{\bar{z}^3})
\end{pmatrix}B(\vec x)\quad
\qquad \Psi'_{\pm}(\vec x)\equiv\begin{pmatrix}
-i\varepsilon(m^2_{\Psi'_{\pm}}+\hat{\p}_{z^3}\hat{\p}_{\bar{z}^3})\\
-2(\hat{\p}_{\bar z^1}-i\hat{\p}_{z^2})(m^2_{\Psi'_{\pm}}-m_B^2)\\
-2(\hat{\p}_{\bar z^2}+i\hat{\p}_{z^1})(m^2_{\Psi'_{\pm}}-m_B^2)\\
\varepsilon(m^2_{\Psi'_{\pm}}+\hat{\p}_{z^3}\hat{\p}_{\bar{z}^3})
\end{pmatrix}B(\vec x)
\end{equation}}
with mass eigenvalues
\begin{align}
m^2_{\Psi_\pm}&=\frac{1}{16}\left(\varepsilon\pm
\sqrt{16m_B^2+\varepsilon^2-16\left(\frac{k_3}{R_3}\right)^2
-\frac{\varepsilon k_6}{2R_6}}\right)^2+\left(\frac{k_3}{R_3}\right)^2\\
m^2_{\Psi'_{\pm}}&=\frac{1}{16}\left(\varepsilon\pm
\sqrt{16m_B^2+\varepsilon^2-16\left(\frac{k_3}{R_3}\right)^2
+\frac{\varepsilon k_6}{2R_6}}\right)^2+\left(\frac{k_3}{R_3}\right)^2
\end{align}
%


\section{The orbit method}\label{kirillov}

In this appendix we summarize the notions of representation theory
required for solving the generalized Dirac and Laplace equations
in parallelizable manifolds. More precisely we consider the
orbit method developed mostly by A. Kirillov in the 60's, applied to
nilmanifolds.\footnote{See \cite{kirillovv} for a more rigorous introduction to the
orbit method and its application to general compact group manifolds, as well as
\cite{orbitst} and references therein for earlier applications of this method in the
context of CFT and string theory.}
Basically, the method relies  the existence of a connection between harmonic
analysis and symplectic geometry. The main objects are the orbits of a
coadjoint action, which we will define in brief. These orbits turn out to be in one to one correspondence with the irreducible unitary representations of the group.

More precisely,
consider a compact nilmanifold given by $\mathcal{M}=G/\Gamma$,
with $G$ a nilpotent group and $\Gamma$ a discrete subgroup. For matrix groups, we can introduce the hermitian product
\begin{equation}
\langle A,B \rangle \equiv \textrm{Tr}(AB)
\end{equation}
for $A,B\in \textrm{Mat}_n(\mathbb{R})$. We can then introduce the algebra $\mathfrak{g}^*$,
dual to the Lie algebra of $G$, $\mathfrak{g}=\textrm{Lie}(G)$, through the partition
\begin{equation}
\textrm{Mat}_n(\mathbb{R})=\mathfrak{g}^*\ \oplus\
\mathfrak{g}^\perp
\end{equation}
where
\begin{equation}
\mathfrak{g}^\perp=\{A\in \textrm{Mat}_n(\mathbb{R})\ |\ \langle
A,B\rangle = 0 \ \forall \ B\in \mathfrak{g} \}
\end{equation}

The coadjoint representation $K$ of
$\mathfrak{g}^*$ is then defined as
\begin{equation}
K(g): \ \mathfrak{g}^* \to \mathfrak{g}^*, \qquad
K(g)F=p_{\mathfrak{g}^*}(gFg^{-1})
\end{equation}
for $g\in G$ and $p_{\mathfrak{g}^*}$ the projector of $\textrm{Mat}_n(\mathbb{R})$ onto
$\mathfrak{g}^*$.

The central idea underlying the orbit method then states that there is a one to one correspondence between
the orbits $\Omega$ of the coadjoint action $K$, and the irreducible unitary representations of $\mathfrak{g}$ acting on $L^2(\mathbb{R}^{\frac{\textrm{dim }\Omega}{2}})$, given by
\begin{equation}
\pi_\Omega(g)u(\vec s)=e^{2\pi i \langle F,\textrm{log }h(\vec
s,g)\rangle}u(\vec s\cdot g)
\label{form}
\end{equation}
acting on $L^2(\mathbb{R}^{\frac{\textrm{dim }\Omega}{2}})$. This equation needs
some explanation. Here, $F$ is an arbitrary point in $\Omega$, whereas
$\textrm{log }h(\vec s,g)$ represents the Lie algebra element corresponding to
the group element $h(\vec s,g)$. The latter is a solution of the master equation
\begin{equation}
S(\vec s)g=h(\vec s,g)S(\vec s\cdot g)
\end{equation}
with $S$ a section $G/H \to G$, and $H\in G$ the subgroup corresponding to a subalgebra $\mathfrak{h}\in\mathfrak{g}$ of dimension
dim $\mathfrak{h}=\textrm{dim }\mathfrak{g}-\frac12\textrm{dim
}\Omega$\ \footnote{A general feature of coadjoint orbits, related
to their symplectic structure, is that they are always even
dimensional.} such that
\begin{equation}
\langle F,[\mathfrak{h},\mathfrak{h}]\rangle=0
\end{equation}
Each subalgebra of the right dimension satisfying this equation leads
to a different manifold polarization of the representation
associated to an orbit $\Omega$, and different polarizations are related among themselves by generalizations of the Abelian Fourier transform.

In order to illustrate this powerful procedure, in what follows we consider a couple
of examples relevant for the material presented in the main text.\\

{\bf Example 1. Irreducible unitary representations of
$\mathcal{H}_{2p+1}$}

Consider the $2p+1$ dimensional Heisenberg group. As already
mentioned in Section \ref{nili}, a suitable matrix representation
for the group is given by (\ref{heis})
\begin{equation}
G=\begin{pmatrix}1 & -\frac12\vec{y}^t &
\frac12\vec{x}^t & z\\
0 & 1 & 0 & \vec x \\
0 & 0 & 1 & \vec y\\
0 & 0 & 0 & 1\end{pmatrix}
\end{equation}
From here the matrix representations for $\mathfrak{g}$ and
$\mathfrak{g}^*$ are easily worked out
\begin{equation}
{g}=\begin{pmatrix}0 & -\frac12\vec{y}^t &
\frac12\vec{x}^t & z\\
0 & 0 & 0 & \vec x \\
0 & 0 & 0 & \vec y\\
0 & 0 & 0 & 0
\end{pmatrix}\quad \qquad {g}^*=\begin{pmatrix}0&0&0&0\\
-\vec g_y&0&0&0\\
\vec g_x&0&0&0\\
g_z&\frac12\vec{g}_x^t&\frac12\vec{g}_y^t&0
\end{pmatrix}\label{gg}
\end{equation}
where $\vec g_x$ and $\vec g_y$ are $p$-dimensional vectors. The
coadjoint representation then reads
\begin{equation}
K(G)(\vec{g}_x,\vec{g}_y,g_z)=(\vec{g}_x+\vec y\cdot\vec{g}_z,\
\vec{g}_y-\vec x\cdot\vec{g}_z,\ g_z)
\end{equation}
Observe that there are only two types of orbits: zero dimensional
orbits given by the points
$\Omega_{\mu,\nu}\equiv(\vec{\mu},\vec{\nu},0)$ with $\vec{\mu}$
and $\vec{\nu}$ constant vectors, and two dimensional orbits given
by the hyperplanes $\Omega_\lambda\equiv(*,*,\lambda)$, with
$\lambda\neq 0$.

The irreducible unitary representations associated to zero
dimensional orbits, $\Omega_{\mu,\nu}$, can be worked out very
easily. The corresponding subalgebra is $(2p+1)$-dimensional, and
therefore it is the full Heisenberg algebra. The master equation
becomes trivial, and the corresponding irreducible unitary
representations are given by
\begin{equation}
\pi_{\mu,\nu}=e^{2\pi
i\langle\left.\mathfrak{g}^*\right|_{\Omega_{\mu,\nu}},\mathfrak{g}\rangle}=e^{2\pi
i(\vec \mu\cdot \vec x+\vec \nu\cdot \vec y)}
\end{equation}

For the irreducible unitary representations associated to the
$2p$-dimensional orbits, $\Omega_\lambda$, we have to select a
$p+1$-dimensional subalgebra $\mathfrak{h}$ such that
\begin{equation}
\langle\left.\mathfrak{g}^*\right|_{\Omega_\lambda}, \
[\mathfrak{h},\mathfrak{h}]\rangle=0
\end{equation}
Different choices correspond to different manifold polarizations.
Here, for concreteness, we focus in the subalgebra generated by
$\vec x=0$ in (\ref{gg}). A suitable section in $G/H$ is then
given by
\begin{equation}
S(\vec s)=\begin{pmatrix}1 & 0 &
-\frac12\vec{s}^t & 0\\
0 & 1 & 0 & \vec s \\
0 & 0 & 1 & 0\\
0 & 0 & 0 & 1\end{pmatrix}
\end{equation}
and the solution to the master equation reads
\begin{equation}
h(\vec s, g)=\begin{pmatrix}1&-\frac12 \vec y&0&z+\frac12\vec
s\cdot \vec y\\
0&1&0&0\\
0&0&1&\vec y\\
0&0&0&1\end{pmatrix} \qquad \vec s\cdot g = (\vec x+\vec s, \vec y, z)
\end{equation}
Plugging into eq.(\ref{form}) we finally get the irreducible
unitary representations associated to the orbits $\Omega_\lambda$
\begin{equation}
\pi_\lambda u(\vec s)=e^{2\pi i\lambda [z+\vec y\cdot \vec s+\vec
x\cdot\vec y/2]}u(\vec x+\vec s)
\end{equation}
In this way we have rederived the Stone - von Neumann theorem,
discussed in eqs.(\ref{rep1})-(\ref{rep2}), by means of the orbit
method. Let us now consider a more involved example.\\

{\bf Example 2. Irreducible unitary representations of the algebra
(\ref{nil2})}

Consider the nilpotent group associated to the nilmanifold defined
by eq.(\ref{nil2}). Matrix representations for the group, the algebra, and the dual
algebra, can be easily worked out, resulting in
\begin{equation*}
G=\begin{pmatrix}1&-\frac{M_6x^1}{2}&0&0&0&\frac{M_6x^5}{2}&x^6\\
0&1&0&0&0&0&x^5\\
0&0&1&0&0&0&x^4\\
0&0&0&1&-\frac{M_3x^1}{2}&\frac{M_3x^2}{2}&x^3\\
0&0&0&0&1&0&x^2\\
0&0&0&0&0&1&x^1\\
0&0&0&0&0&0&1
\end{pmatrix}
\end{equation*}
\begin{equation*}
\mathfrak{g}=\begin{pmatrix}0&-\frac{M_6x^1}{2}&0&0&0&\frac{M_6x^5}{2}&x^6\\
0&0&0&0&0&0&x^5\\
0&0&0&0&0&0&x^4\\
0&0&0&0&-\frac{M_3x^1}{2}&\frac{M_3x^2}{2}&x^3\\
0&0&0&0&0&0&x^2\\
0&0&0&0&0&0&x^1\\
0&0&0&0&0&0&0
\end{pmatrix}\quad \quad \mathfrak{g}^*=\begin{pmatrix}0&0&0&0&0&0&0\\
-\frac{2g_1}{M_6}&0&0&0&0&0&0\\
0&0&0&0&0&0&0\\
0&0&0&0&0&0&0\\
0&0&0&0&0&0&0\\
\frac{2g_5}{M_6}&0&0&\frac{2g_2}{M_3}&0&0&0\\
g_6&g_5&g_4&g_3&g_2&g_1&0
\end{pmatrix}
\end{equation*}
The projector into $\mathfrak{g}^*$ is given by
\begin{equation*}
P_{\mathfrak{g}^*}(A)=\begin{pmatrix}0&0&0&0&0&0&0\\
\frac{A_{21}}{2}-\frac{A_{76}}{M_6}&0&0&0&0&0&0\\
0&0&0&0&0&0&0\\
0&0&0&0&0&0&0\\
0&0&0&0&0&0&0\\
\frac{A_{61}}{2}+\frac{A_{72}}{M_6}&0&0&\frac{A_{64}}{2}+\frac{A_{75}}{M_3}&0&0&0\\
A_{71}&\frac{A_{72}}{2}+\frac{M_6A_{61}}{4}&A_{73}&A_{74}&\frac{A_{75}}{2}+\frac{M_3A_{64}}{4}&\frac{A_{76}}{2}-\frac{M_6A_{21}}{4}&0
\end{pmatrix}
\end{equation*}
From these expressions, the coadjoint representation reads
\begin{multline*}
K(G)(g_1,g_2,g_3,g_4,g_5,g_6)=\left(g_1-\frac14 g_3M_3x^2-\frac12 g_6M_6x^5,\ g_2
+\frac12M_3g_3x^1,\ g_3,\right.\\
\left.g_4,\ g_5+\frac12M_6g_6x^1,\ g_6\right)
\end{multline*}
We observe, therefore, four classes of orbits, one 0-dimensional,
and three 2-dimensional, given by
\begin{align}
\Omega_{\mu,\nu,\sigma,\rho}&=(\mu,\nu,0,\sigma,\rho,0)\\
\Omega_{\mu,r,p}&=(*_1,M_3r*_2,r,\mu,M_6p*_2,p)\\
\Omega_{\mu,\nu,p}&=(*_1,\mu,0,\nu,*_2,p)\\
\Omega_{\mu,r,\nu}&=(*_1,*_2,r,\mu,\nu,0)
\end{align}
with $p,r\neq 0$. Proceeding as in the previous example, we arrive
to the following set of irreducible unitary representations
\begin{align}
\pi_{\mu,\nu,\sigma,\rho}&=e^{2\pi i(\mu x^1+\nu x^2+\sigma x^4+\rho x^5)}
\label{pi1}\\
\pi_{\mu,r,p}u(s_1)&=e^{2\pi i(\mu
x^4+ra_3+pa_6)]}u(s_1+x^1)\\
\pi_{\mu,\nu,p}u(s_1)&=e^{2\pi i(\mu
x^2+\nu x^4+pa_6)}u(s_1+x^1) \\
\pi_{\mu,r,\nu}u(s_1)&=e^{2\pi i(\mu x^4+\nu x^5+ra_3)}u(s_1+x^1)
\label{pi6}
\end{align}
where
\begin{equation}
a_3\equiv x^3-M_3x^2\left(s_1+\frac{x^1}{2}\right)\ ,\qquad
a_6\equiv x^6-M_6x^5\left(s_1+\frac{x^1}{2}\right)
\end{equation}
and for simplicity we have taken the same polarization for all the
representations.


\section{Scalar wavefunction matrix} \label{ap:matrix}

Let us rewrite eqs.(\ref{xi+}) and (\ref{xi-}) in matrix notation, and more precisely as
\begin{equation}
\left[\mathbb{M}+m_b^2\ \mathbb{I}_{6\times
6}\right]\mathbb{V}=0
\label{ap:eigenval}
\end{equation}
where
\beq
\mathbb{V}=
\begin{pmatrix}\xi^1\\
\xi^2\\ \xi^3\\ \xi^{*1}\\ \xi^{*2}\\ \xi^{*3}
\end{pmatrix}
\quad \quad \quad
\begin{array}{c}
\xi^1\, \equiv\,  \xi_{B_4}^1+i\xi_{B_4}^4\\
\xi^2\, \equiv\,  \xi_{B_4}^2+i\xi_{B_4}^5\\
\xi^3\, \equiv\,  \xi_{\Pi_2}^3+i\xi_{\Pi_2}^6
\end{array}
\quad \quad \quad
\begin{array}{c}
\xi^{*1}\, \equiv\,  \xi_{B_4}^1-i\xi_{B_4}^4\\
\xi^{*2}\, \equiv\,  \xi_{B_4}^2-i\xi_{B_4}^5\\
\xi^{*3}\, \equiv\,  \xi_{\Pi_2}^3-i\xi_{\Pi_2}^6
\end{array}
\label{ap:standcom}
\eeq
and
\beq
\mathbb{M}\, =\,
 \hat{\p}_m\hat{\p}^m\, \mathbb{I}_6 +
\left(
\begin{array}{cc}
A & B \\ -B^\dag & A^*
\end{array}
\right)
\eeq
with
\beq\nonumber
A \, =\, \left(
\begin{array}{ccc}
0 & - (G_-)^3_{\bar 1 2} \p_3 - (G_-)^{\bar 3}_{\bar 1 2} \p_{\bar 3} & - (G_+)^{\bar 3}_{\bar 1 \bar 2} \p_{2} - (G_+)^{\bar 3}_{\bar 1 2} \p_{\bar 2} \\
(G_-)^3_{1 \bar 2} \p_3 + (G_-)^{\bar 3}_{1 \bar 2} \p_{\bar 3} & 0 & (G_+)^{\bar 3}_{\bar 1 \bar 2} \p_{1} + (G_+)^{\bar 3}_{1 \bar  2} \p_{\bar 1} \\
(G_+)^{3}_{1 \bar  2} \p_{2} + (G_+)^{3}_{12} \p_{\bar 2} & - (G_+)^{3}_{\bar 1  2} \p_{1} - (G_+)^{3}_{12} \p_{\bar 1} & - a
\end{array}
\right)
\eeq
\beq\nonumber
B \, =\, \left(
\begin{array}{ccc}
0 & - (G_-)^3_{\bar 1 \bar 2} \p_3 - (G_-)^{\bar 3}_{\bar 1 \bar 2} \p_{\bar 3} & - (G_+)^{3}_{\bar 1 \bar 2} \p_{2} - (G_+)^{3}_{\bar 1 2} \p_{\bar 2} \\
(G_-)^3_{\bar 1 \bar 2} \p_3 + (G_-)^{\bar 3}_{\bar 1 \bar 2} \p_{\bar 3} & 0 & (G_+)^{3}_{\bar 1 \bar 2} \p_{1} + (G_+)^{3}_{1 \bar  2} \p_{\bar 1} \\
(G_+)^{3}_{\bar 1 \bar  2} \p_{2} + (G_+)^{3}_{\bar 1 2} \p_{\bar 2} & - (G_+)^{3}_{\bar 1 \bar 2} \p_{1} - (G_+)^{3}_{1\bar 2} \p_{\bar 1} & - b
\end{array}
\right)
\eeq
and where
\beq
(G_\pm)^a_{bc}\, \equiv\, f^a_{bc} \pm g^{a\bar{a}} F_{\bar a bc}
\eeq
is constructed without imposing the on-shell condition (\ref{jj}). Finally, we have
defined
\beqa
a & = & (G_+)^3_{12}f^{\bar 3}_{\bar 1 \bar 2} + (G_+)^3_{\bar 1 \bar 2} f^{\bar 3}_{12}
+ (G_+)^3_{\bar 1 2} f^{\bar 3}_{1 \bar 2} + (G_+)^3_{1\bar 2}f^{\bar 3}_{\bar 1 2} \\
b & = & (G_+)^3_{12}f^{3}_{\bar 1 \bar 2} + (G_+)^3_{\bar 1 \bar 2} f^{3}_{12}
+ (G_+)^3_{\bar 1 2} f^{3}_{1 \bar 2} + (G_+)^3_{1\bar 2}f^{3}_{\bar 1 2}
\eeqa
In general, for non-supersymmetric backgrounds the matrix $B$ is different from zero, reflecting the fact that in that case the internal manifold is not complex. In that case, holomorphic and anti-holomorphic indices label different elements in a complex basis of 1-forms. The fact that the internal manifold is not complex manifests in a spectrum of wavefunctions for which some of the ``holomorphic'' scalars have different mass eigenvalues than their ``anti-holomorphic'' counterparts.


\section{General magnetic fluxes and Riemann $\vartheta$-function}
\label{riem}

As emphasized in \cite{yukawa}, in the presence of general magnetic fluxes $F_2$ on
a $T^{2n}$, the zero modes of the Dirac  and Laplace operators are given in terms
of Riemann $\vartheta$-functions, instead of the more familiar Jacobi $\vartheta$-functions
that appear for the factorizable case of a $(T^2)^n$ with a magnetic flux
$F_2 = \sum_i F_2|_{(T^2)_i}$. As we have seen in the main text, for type I open string
wavefunctions in flux compactifications this non-factorizable case is quite natural,
and in particular for those matter field wavefunctions analyzed in
Section \ref{sec:wmatter} that feel closed and open string fluxes simultaneously.
The purpose of this appendix is thus to extend the discussion of \cite{yukawa} on
Riemann $\vartheta$-functions and non-factorizable magnetic fluxes, in order
to accommodate the wavefunctions of Section \ref{sec:wmatter} into the general
scheme of \cite{yukawa}. See also \cite{akp09} for some recent similar results on
this topic.

Let us then consider a general ${T}^{2n}$ and a magnetic $U(1)$ flux $F_2 = dA$ of the form
\beq
F_2\, =\, \pi \sum_{ij} q_{ij}\, dx^i \wedge dx^j
\label{flux2}
\eeq
where $q_{ij} \in \IZ$ and $x^i \in [0,1]$ label the $T^{2n}$ coordinates.
This means that we can write the vector potential as
\beq
A\, =\, \pi \sum_{ij} q_{ij}\, x^i dx^j\, = \, \pi\,  \vec{x}^{\, t} {\bf Q} d\vec{x}
\label{potential}
\eeq
with ${\bf Q}^t = - {\bf Q}$. Let us now define some complex coordinates in $T^{2n}$ as
\beq
\vec{z}\, =\, \vec{\xi} + {\bf \Omega} \cdot \vec{\eta}
\label{cpx}
\eeq
with $\vec{\xi}, \vec{\eta}$, two $n$-dimensional real vectors in which we split the components
of $\vec x$. The matrix {\bf Q} then splits as
\beq
{\bf Q}\, =\,
\left(
\begin{array}{cc}
{\bf Q^{\xi\xi}} & {\bf Q^{\xi\eta}} \\
{\bf Q^{\eta\xi}} & {\bf Q^{\eta\eta}}
\end{array}
\right)
\label{blocks}
\eeq

In practice, computing open string wavefunctions greatly simplifies if the
magnetic flux $F_2$ can be written as a (1,1)-form for some choice of complex
structure (\ref{cpx}). In that case we can express (\ref{potential}) as
\beq
A\, = \, \pi \, \im \left( \ov{\vec{z}}^{\, t} \, {\bf C}\, d \vec{z} \right)
\label{ansatz}
\eeq
Direct comparison reveals that the matrices {\bf C} and {\bf Q} are related as
\beqa
{\bf Q^{\xi\xi}} & = & \im {\bf C}
\label{integerxx}\\
{\bf Q^{\xi\eta}} & = & \im {\bf C}\, \re {\bf \Om} + \re  {\bf C}\, \im {\bf \Om}
\label{integerxy}\\
{\bf Q^{\eta\eta}} & = & \re {\bf \Om}^t \, \re {\bf C}\,  \im {\bf \Om}
- \im {\bf \Om}^t \, \re {\bf C}\, \re {\bf \Om} +
\nonumber \\
& & \re {\bf \Om}^t \, \im {\bf C}\,  \re {\bf \Om} + \im {\bf \Om}^t \, \im {\bf C}\, \im {\bf \Om}
\label{integeryy}
\eeqa
and that ${\bf Q}^t = - {\bf Q}$ implies ${\bf C}^\dag = {\bf C}$. It turns out that
the Dirac and Laplace zero mode wavefunctions can be easily expressed as
Riemann $\vartheta$-functions if we also impose the constraint
 ${\bf Q}^{\xi\xi} = \im {\bf C} = 0$. Indeed, the system (\ref{integerxx})-(\ref{integeryy})
  is then solved by taking
\beqa
\re {\bf C} & = & {\bf Q^{\xi\eta}} \, (\im {\bf \Om})^{-1}
\label{solfxy1} \\
{\bf Q^{\eta\eta}} & = &  \re {\bf \Om}^t \, {\bf Q^{\xi\eta}} - {\bf Q^{\xi\eta}}^{\, t} \, \re {\bf \Om}
\label{solfyy2}
\eeqa
where we have assumed that $\im {\bf \Om}$ is invertible.
If we now define the $n\times n$ integer matrix ${\bf N} = {\bf Q^{\xi\eta}}^{\, t}$,
we can express the above solution as
\beqa
\re {\bf C} & = & {\bf N}^t \, (\im {\bf \Om})^{-1}
\label{solfxya} \\
- {\bf Q^{\eta\eta}} & = &  {\bf N}\, \re {\bf \Om} - \left({\bf N}\, \re {\bf \Om}\right)^t
\label{solfyyb}
\eeqa

In terms of {\bf N}, the antiholomorphic covariant derivative reads
\begin{equation}
 \hat{D}_{\bar a}= \frac{1}{2\pi R_a} \left( \nabla - i A\right)_{\bar{a}} = \left(\nabla+\frac{\pi}{2}[{\bf N}\cdot \vec z]^t\cdot(\textrm{Im }{\bf \Omega})^{-1}\right)_{\bar a}
\end{equation}
and it is easy to check that it annihilates the wavefunction \cite{yukawa}
\begin{equation}
\psi^{\vec{j}, {\bf N}}(\vec z, {\bf \Om})\, =\, \mathcal{N}\ e^{i\pi [{\bf N}\, \vec{z}]^t (\text{Im}\, {\bf \Om})^{-1} \text{Im}\, \vec{z}}\,
\vartheta
\left[
\begin{array}{c}
\vec{j} \\ 0
\end{array}
\right]
\left({\bf N} \cdot \vec{z} \ ; {\bf N} \cdot {\bf \Om} \right)
\label{ap:wavematter}
\end{equation}
which is of the form (\ref{wavematter}) up to fiber-dependent phases.
The normalization constant is given by
\begin{equation}
\mathcal{N}=\left(2^n|\textrm{det}({\bf N}\, \im {\bf \Om})|\,\textrm{Vol}^{-2}_{T^{2n}}\right)^{1/4}
\end{equation}
whereas $\vartheta$ stands for the Riemann $\vartheta$-function, defined as
\begin{equation}
\vartheta\left[{\vec a \atop \vec b}\right](\vec \nu \, ; \, \mathbf{\Omega})=
\sum_{\vec m\in\mathbb{Z}^n}e^{i\pi(\vec m-\vec a)^t
\mathbf{\Omega}(\vec m-\vec a)}e^{2\pi i(\vec m-\vec a)\cdot(\vec\nu-\vec b)}
\label{thetariem}
\end{equation}
with $\vec a,\ \vec b\in \mathbb{R}^n$. Under lattice shifts $\vec n\in\mathbb{Z}^n$,
$\vartheta$ undergoes the transformations
\begin{align}
\vartheta\left[{\vec a \atop \vec b}\right](\vec \nu+\vec n\, ; \, \mathbf{\Omega})&=
e^{-2\pi i\vec a\cdot\vec n}\cdot\vartheta\left[{\vec a\atop \vec b}\right](\vec\nu\ ; \ \mathbf{\Omega}) \\
\vartheta\left[{\vec a \atop \vec b}\right](\vec \nu+\mathbf{\Omega}\, \vec n\ ; \ \mathbf{\Omega})&=
e^{-i\pi\vec n^t\mathbf{\Omega} \vec n-2\pi i\vec n\cdot (\vec \nu-\vec b)}\cdot
\vartheta\left[{\vec a\atop \vec b}\right](\vec\nu\, ; \, \mathbf{\Omega})
\end{align}
which implies that the wavefunction (\ref{ap:wavematter}) transforms as
\beqa
\psi^{\vec j, {\bf N}} (\vec z + \vec n, {\bf \Om}) & = &
e^{i \pi \vec n^t {\bf Q}^{\xi\eta} \vec \eta} \, \psi^{\vec j, {\bf N}} (\vec z, {\bf \Om})
\label{trans1} \\
\psi^{\vec j, {\bf N}} (\vec z + {\bf \Om}\, \vec n, {\bf \Om}) & = &
e^{i \pi \vec n^t \left({\bf Q}^{\eta\xi} \vec \xi + {\bf Q}^{\eta\eta} \vec \eta \right)}
\, \psi^{\vec j, {\bf N}} (\vec z, {\bf \Om})
\label{trans2}
\eeqa
provided that ${\bf N}^t \vec j = {\bf Q}^{\xi\eta} \vec j \in \IZ^n$. Here
we have used the fact that ${\bf N}\, \im \Om$ is symmetric, which is implied
by (\ref{solfxy1}) and that ${\bf C}$ is Hermitian. The transformations
 (\ref{trans1}) and (\ref{trans2}) are indeed those of a particle coupled with unit
 charge to a vector potential (\ref{potential}) that satisfies ${\bf Q}^{\xi\xi} =0$.

The above result, however, does not imply that for any potential (\ref{potential})
such that ${\bf Q}^{\xi\xi} =0$ for some choice of $\vec \xi$, $\vec \eta$, we can
find a zero mode wavefunction of the form (\ref{ap:wavematter}). First, recall that
$F_2 = dA$ should correspond to a (1,1) form for a choice of ${\bf \Om}$ compatible
with the $T^{2n}$ metric, and second we should guarantee the convergence of the
$\vartheta$-function in (\ref{ap:wavematter}), which requires the positive definiteness
condition
\begin{equation}
\mathbf{N}\cdot\textrm{Im }\mathbf{\Omega} >0
\label{conv}
\end{equation}

In Section \ref{sec:wmatter} we have provided some examples of wavefunctions
satisfying all these constraints for certain families of non-factorizable fluxes $F_2$
on $T^4$, more precisely for (\ref{totalflux}) and (\ref{totalflux2}). One can there check
that the complex structure ${\bf \Omega_U}$ is rotated by an $SO(2)$ matrix ${\bf U}$.
Let us see how these kind of solutions arise in the context of the above discussion.
For that aim, let us write the $T^{2n}$ metric as
\beq
ds^2 \, = \,
(d\vec{\xi}^t \quad d\vec{\eta}^t) \cdot {\bf G} \cdot
\left(
\begin{array}{c}
\vec{\xi} \\ \vec{\eta}
\end{array}
\right)
\, =\,
(\begin{array}{cc}
d\vec{z}^{\, t} & d\ov{\vec{z}}^{\, t}
\end{array})
\cdot
\left(
\begin{array}{cc}
0 & {\bf h} \\
{\bf \bar{h}} & 0
\end{array}
\right)
\cdot
\left(
\begin{array}{c}
d\vec{z} \\ d\ov{\vec{z}}
\end{array}
\right)
\label{metricm}
\eeq
with ${\bf h}$ an hermitian matrix. We then have that
\beq
{\bf G}\,=\, 2
\left(
\begin{array}{cc}
\re {\bf h} & \re ({\bf \bar{h} \cdot \Om}) \\
\re ({\bf \bar{h} \cdot \Om})^t & \re ({\bf \Om}^t\cdot {\bf h} \cdot {\bf \bar{\Om}})
\end{array}
\right)
\label{G}
\eeq
and so we would like to characterize those deformations of ${\bf \Omega}$ that
leave {\bf h} and {\bf G} invariant. Note that since ${\bf h}$ is Hermitian we can write
it as ${\bf h} \, = \, {\bf B}^\dag {\bf B}$, with ${\bf B}$ invertible. This allows to
parameterize a deformation of  ${\bf \Om}$ as
\beq
{\bf \Om_U}\, =\, {\bf \bar{B}}^{-1} \cdot {\bf \bar{U}} \cdot {\bf \bar{B}} \cdot {\bf \Om}
\eeq
with ${\bf U}$ an arbitrary matrix. Then we have that
\beq
\re ({\bf \Om_U}^t {\bf h} {\bf \bar{\Om}_U})\, =\, \re ({\bf \Om}^t {\bf B}^\dag  {\bf U}^\dag {\bf U}  {\bf B} {\bf \bar{\Om}})
\eeq
so this term remains invariant if ${\bf U} \in U(n)$.
The off-diagonal terms of (\ref{G}), on the other hand, remain invariant if
\beq
\re ({\bf B}^\dag {\bf U} {\bf B}{\bf \bar{\Om}})\, =\, \re ({\bf B}^\dag {\bf B} {\bf \bar{\Om}})
\eeq
which, for ${\bf B}$ real and ${\bf \Om}$ pure imaginary, is satisfied by simply imposing
that ${\bf U}$ is also real. Together with the above constraint this implies that
${\bf U} \in O(n)$.

The wavefunctions of Section \ref{sec:wmatter} precisely fall in the category of
wavefunctions (\ref{ap:wavematter}) with rotated complex structure ${\bf \Om_U}$.
Indeed, note that for the factorized $T^4$ metric of the form (\ref{bg12}), ${\bf \Om}$
is indeed pure imaginary and so, by the discussion above, ${\bf U}$
is an orthogonal matrix. In addition, if we take ${\bf N}$ definite positive (as we do in
the examples of Section \ref{sec:wmatter}) we need to constrain ${\bf U} \in SO(n)$
as a requirement for the convergence condition (\ref{conv}). The precise choice
of ${\bf U}$ is then given by the condition that $F_2$ is a (1,1)-form for the complex
structure ${\bf \Om_U}$.

Note, however, that the above setup clashes with the degree of freedom
${\bf Q}^{\eta\eta} \neq 0$ which in principle we have for our magnetic flux $F_2$.
Indeed, (\ref{solfyy2}) requires that $\re {\bf \Om_U} \neq 0$ if ${\bf Q}^{\eta\eta} \neq 0$,
while $\re {\bf \Om_U} \neq 0$ is not allowed by a rotation ${\bf U} \in SO(n)$.
Hence, at least naively, the wavefunctions (\ref{ap:wavematter}) apply directly to
those magnetic fluxes (\ref{flux2}) such that ${\bf Q}^{\xi\xi} = {\bf Q}^{\eta\eta} = 0$
for some choices of $\vec \xi, \vec \eta$. Note that this is not the case for the flux
(\ref{totalflux2}) in the more general situation $k_3, k_6 \neq 0$, and this is
the reason why in Section \ref{sec:wmatter} no explicit wavefunctions have been provided
for such sector of the theory.


\end{document}